\documentclass[12pt,a4paper,oneside]{article}
\usepackage{graphicx,amssymb,amsmath,color}
\usepackage[linkcolor={blue},citecolor={red},colorlinks=true]{hyperref}
\usepackage{cite}
\usepackage{relsize}
\usepackage{enumerate}
\usepackage[margin=1.5 cm]{geometry}
\usepackage{slashed}
\usepackage{multirow}
\usepackage[normalem]{ulem}
\usepackage[dvipsnames, x11names]{xcolor}
\usepackage[compat=1.0.0]{tikz-feynman}
\usepackage{comment}
\usepackage{soul}
\usepackage{dsfont}
\usepackage{mathtools}

\usepackage{cleveref}
\usepackage{cancel}

\def\be{\begin{equation}}
\def\ee{\end{equation}}

\def\beq{\begin{equation}}
\def\eeq{\end{equation}}
\def\bea{\begin{eqnarray}}
\def\eea{\end{eqnarray}}

\def\bit{\begin{itemize}}
\def\eit{\end{itemize}}

\def\baa{\begin{array}}
\def\eaa{\end{array}}

\def\simgt{\mathrel{\lower2.5pt\vbox{\lineskip=0pt\baselineskip=0pt
           \hbox{$>$}\hbox{$\sim$}}}}
\def\simlt{\mathrel{\lower2.5pt\vbox{\lineskip=0pt\baselineskip=0pt
           \hbox{$<$}\hbox{$\sim$}}}}

\def\bfc{\begin{figure}\begin{center}}
\def\efc{\end{center}\end{figure}}

\newcommand{\red}[1]{{\color{red} #1}}

\definecolor{chromeyellow}{rgb}{1.0, 0.65, 0.0}
\definecolor{darkcoral}{rgb}{0.8, 0.36, 0.27}
\definecolor{cadmiumgreen}{rgb}{0.0, 0.42, 0.24}

\usepackage{caption}
\begin{document}

\begin{flushright}
\hspace{3cm} 
\end{flushright}
\vspace{.6cm}
\begin{center}

\hspace{-0.4cm}{\Large \bf 
 DW-genesis: \\
 baryon number from domain wall network collapse\\}
\vspace{1cm}{Alberto Mariotti$^{a}$, Xander Nagels$^{a}$, A\"aron Rase$^{a}$,   Miguel Vanvlasselaer$^{a}$  }
\\[7mm]

\end{center}

\vskip 0.2cm
\begin{center}
\vskip0.05cm
{\it $^a$ Theoretische Natuurkunde and IIHE/ELEM, Vrije Universiteit Brussel,\\
\& The International Solvay Institutes, Pleinlaan 2, B-1050 Brussels, Belgium }
\vskip1.cm
\end{center}

\bigskip \bigskip \bigskip

\centerline{\bf Abstract} 
\begin{quote}

Axionic domain walls, 
as they move through the early universe plasma during their collapse, can generate a net baryon and lepton number through the mechanism of spontaneous baryogenesis, provided that there is a coupling between the axion and the lepton or baryon current. In this paper, we study systematically the baryon asymmetry produced by these domain walls (DWs) at annihilation, within different realisations of the $L$- or $B$-violating sector, and refer to this process as \emph{DW-genesis}. 
We find that the baryon number is maximised when the DW network collapses  approximately
at the moment when the $L$- or $B$-violating interaction decouples.
We study a model of minimal leptogenesis, a model of cogenesis, a model of baryogenesis and finally the possibility that the baryon asymmetry is produced by 
electroweak sphalerons. 
As phenomenological consequences of DW-genesis, we discuss the expected gravitational wave signal from the DW network annihilation
and the prospects for detecting it. 
However, 
 we finally emphasize that in realisations of the 
DW-genesis in minimal post-inflationary scenarios, there is a suppression induced by the cancellation between the 
asymmetry created by ``opposite'' 
axionic domain walls attached to the string. We quantify the impact of this cancellation and discuss possible ways to avoid it.

\end{quote}

\vfill
\noindent\line(1,0){188}
{\scriptsize{ \\ E-mail:
\texttt{\href{Alberto.Mariotti@vub.be}{Alberto.Mariotti@vub.be}},
\texttt{\href{Xander.Staf.A.Nagels@vub.be}{Xander.Staf.A.Nagels@vub.be},
\texttt{\href{aaron.rase@vub.be}{aaron.rase@vub.be}},
\texttt{\href{miguel.vanvlasselaer@vub.be}{miguel.vanvlasselaer@vub.be}}}
}}

\newpage

\tableofcontents
\newpage
\section{Introduction}
\label{sec: intro}

The numerical value of the baryon asymmetry of the universe (BAU), obtained via big bang nucleosynthesis\cite{Fields:2019pfx} and the latest CMB measurements\cite{Planck:2015fie}, is given by
\bea 
\label{eq:observed_BAU}
Y_{\Delta B} \equiv \frac{n_{B}-n_{\overline{B}}}{s} \bigg|_{0} \approx (8.69 \pm 0.22) \times 10^{-11}
\eea 
where $n_{B, \overline{B}}$ and $s$ are the number densities of baryons, antibaryons and entropy evaluated at present time, respectively. This observed imbalance in matter and anti-matter is one of the most puzzling features of early universe physics.  
Indeed, within the inflationary paradigm, any initial conditions preexisting the period of inflation will unavoidably be washed out. Consequently, the explanation of the imbalance between baryons and anti-baryons very likely requires a dynamical production mechanism, a \emph{baryogenesis mechanism}, occurring after inflation. 

A successful baryogenesis mechanism 
typically 
would 
satisfy the 
well-known Sakharov conditions~\cite{Sakharov:1967dj}: the mechanism needs to contain i)
the 
violation of baryon number $B$, ii) the violation of
$C$ and $CP$ symmetry, and iii) a departure from equilibrium dynamics. Based on these
general requirements, various baryogenesis scenarios have been constructed (for reviews see for example
\cite{Buchmuller:2021,Riotto:1998bt, Bodeker:2020ghk}), including GUT baryogenesis\cite{PhysRevLett.47.391, PhysRevLett.42.850}, leptogenesis\cite{FUKUGITA198645} (possibly catalized by the decay of heavy particles\cite{Tong:2024lmi,Cataldi:2024bcs}, by first order phase transition (FOPT) bubble walls\cite{Shuve:2017jgj,Azatov:2021irb, 
Baldes:2021vyz,
Huang:2022vkf,Azatov:2022tii,
Chun:2023ezg,Vanvlasselaer:2024iqz, Cataldi:2024pgt} or by the presence of Q-balls in the plasma \cite{Bai:2021xyf}) and electroweak baryogenesis\cite{Kuzmin:1985mm,Shaposhnikov:1986jp}. 

The third Sakharov condition, however, relies on CPT invariance which 
could
be spontaneously broken in a specific period of the cosmological history and then restored today\cite{Dolgov:1991fr,Prokopec:1995gp,Brandenberger:1999je}. 
In this case, successful baryogenesis
 can also be achieved without a departure from thermal equilibrium.
This is exemplified by the case of \textit{spontaneous baryogenesis}\cite{PhysRevD.94.123501,Kaplan:1991ah,AFFLECK1985361, COHEN1988913} which relies on a non-vanishing coupling between a source $\mu$, acting as a chemical potential, and the lepton or baryon current, $\mu j_{L,B}^0 $  biasing the $L$ or $B$-violating interactions $\mathcal{L}_{\slashed{L}, \slashed{B}}$. An early example of an application of this mechanism is the \emph{Affleck-Dine} baryogenesis\cite{AFFLECK1985361}.
More modern examples include the mechanism of axiogenesis\cite{Co:2019wyp}, leptogenesis catalized by a rotating majoron \cite{Bossingham:2017gtm,Chun:2023eqc} and  \emph{wash-in} leptogenesis\cite{Domcke:2020quw} (see also \cite{Domcke:2020kcp}).
All the above-mentioned scenarios of spontaneous baryogenesis involve a \emph{long-lived source}, i.e. active on much longer timescales than the equilibration time of the $L$ or $B$-violating interactions.

Alternatively, the chemical potential $\mu$ could be active for a time much shorter than the actual equilibration time of the $L$- or $B$-violating interactions. In this case, the equilibrium point is not reached and the relevant Boltzmann equations (BEs) need to be explicitly solved.  A prototypical example of such a mechanism is spontaneous electroweak baryogenesis\cite{COHEN1988913} (see also e.g. \cite{Comelli:1993ne,COMELLI1994441,Jeong:2018jqe}), where the non-trivial bubble wall background, which propagates in the plasma, biases the sphalerons during a timescale much shorter than the timescale of equilibration of the sphalerons in the plasma.

\begin{figure}[h!]
    \centering
    \includegraphics[width=.3\linewidth]{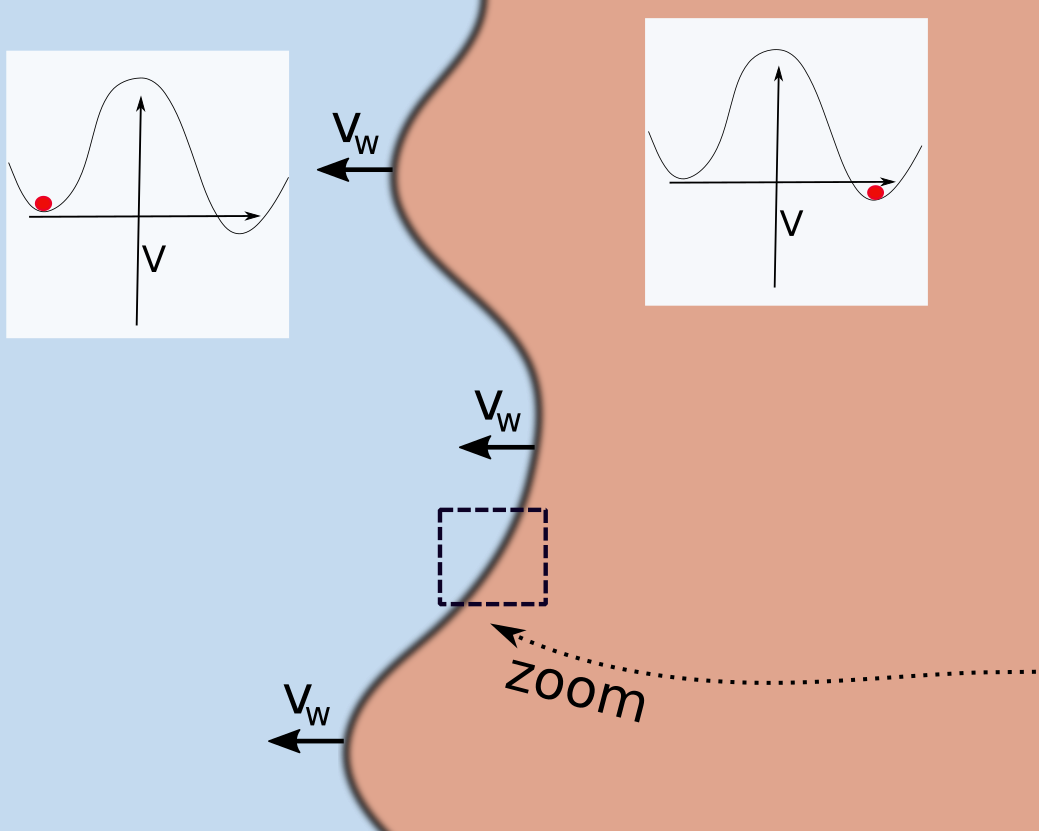}\includegraphics[width=.55\linewidth]{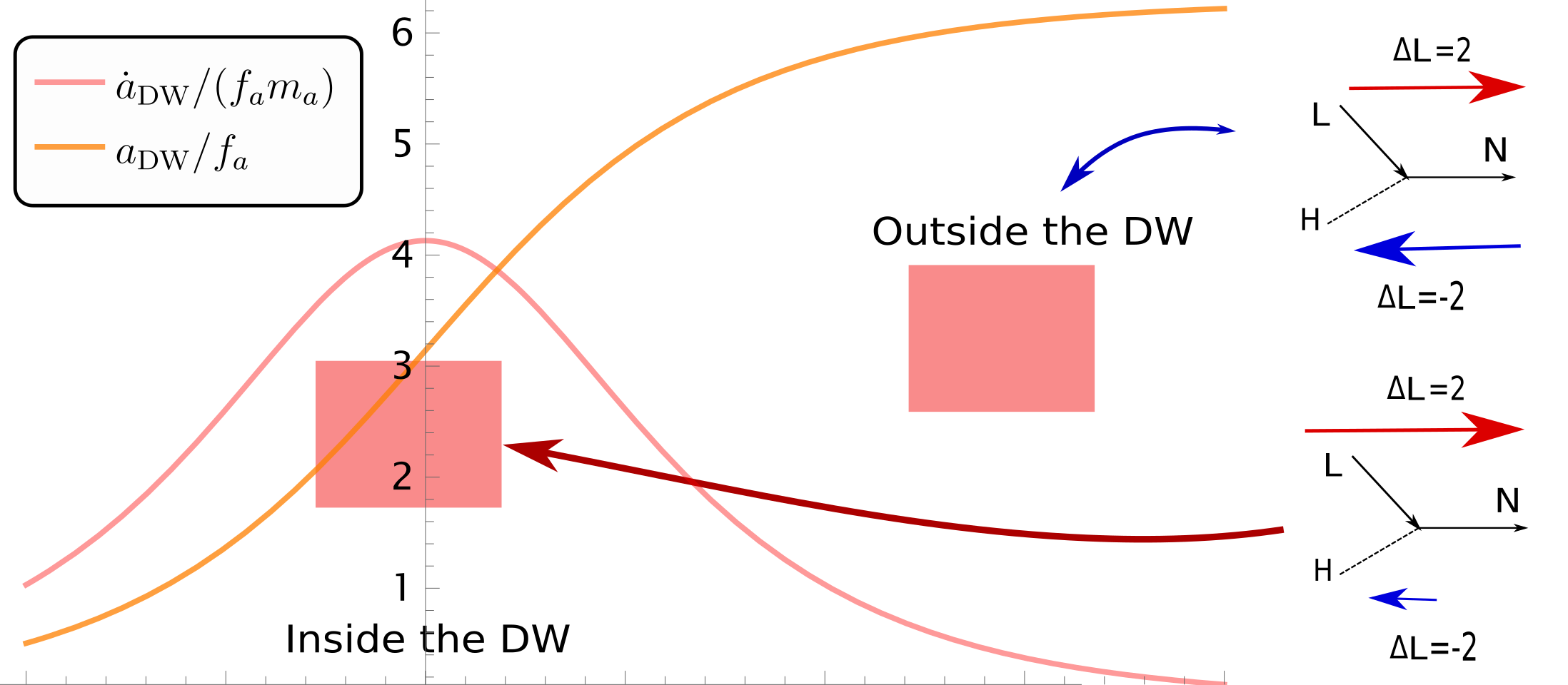}
    
    \caption{\textbf{Scheme of the DW-genesis}: Mechanism discussed in this paper. For concreteness, we show the case of DW leptogenesis. \textbf{Left panel:} Bias-induced collapse of the DW network where the DW moves in a preferred direction. \textbf{Middle panel}: Illustration of the kink DW profile and the induced chemical potential. \textbf{Right panel}: Sketch of the $L$-violating processes active when the DW collapses. Inside the DW, the interactions are biased and the $L$-number is created inside the wall.}
    \label{fig: illustration}
\end{figure}

In this paper we focus on an analogous possibility where baryogenesis is driven by the motion and collapse of axionic domain walls
\footnote{See
\cite{Cline:1998rc,Brandenberger:1994mq,Schroder:2024gsi} for previous alternative studies of defect-driven baryogenesis where EW symmetry is restored in the interior of the topological defect.}.
Indeed, similarly to the case of the bubble wall, an axionic domain wall (DW), which is a non-trivial profile of the $a/f_a$ background, also offers a short-lived source via the interaction $(\partial_\mu a/f_a) j^\mu_{L, B}$ , where $j^\mu_{L, B}$ is a lepton or baryon current, to bias some existing $L$ or $B$-violating interaction. An illustration of the idea behind this mechanism, which we call \emph{DW-genesis}, is provided in Fig.\ref{fig: illustration}. 
One can observe that, as long as the DW network motion does not prefer any direction, asymmetries induced with opposite signs will compensate each other. 

However, since the DW network tends to dominate the energy density of the universe\cite{Zeldovich1974CosmologicalCO}, which is cosmologically unacceptable, the DW network has to collapse before domination occurs. 
This collapse can for example happen because of a bias between the two nearly degenerate vacua. As soon as the axionic DW, initially of the size of the Hubble radius, starts to collapse, the preferential direction of the movement selects one sign for the dominant asymmetry and some lepton/baryon number is leftover after the end of  collapse phase of the DW. 

A realisation of this scenario was previously studied in \cite{Daido:2015gqa}, where it was shown that a DW could induce leptogenesis due to the presence of the $\Delta L=2$ effective Weinberg operator. 
However, the regime where decays and inverse decays of the right handed neutrino (RHN) could be relevant, were not considered.

In this paper, we systematically investigate 
the mechanism of the aforementioned scenario, the \emph{DW-genesis},
by studying in detail the general Boltzmann equations governing it,
and by
considering several concrete realisations: 
i) By considering the decays, the inverse decays and the Weinberg operator, we study the model of DW leptogenesis, where we emphasize that in this case, CP-breaking phases in the Yukawa couplings are not necessary.  ii) We study a model of DW cogenesis where dark matter is simultaneously produced in a dark sector with an abundance that can differ by orders of magnitude with respect to the baryon number. This scenario shows interesting interplay between the two sectors, displaying a \emph{rescuing} mechanism hiding the lepton number in the dark sector and allowing to match the observed abundance of dark matter (DM) for $m_{\rm DM} \in [0.1, 10^5]$ GeV. iii) We study a model of DW baryogenesis where the domain wall collapse biases an interaction violating the baryon number by two units. We observe that the scale necessary to match the observed abundance is however too high to induce detectable $n-\bar n$ oscillations. iv) We finish our overview of model realisations with the study of a model where the DW biases the electroweak sphalerons themselves. Unfortunately this last case cannot accommodate the observed BAU, consistent with the general upper limit that we find for the DW-genesis mechanism.

Here we anticipate a few relevant general results from our study: 
\begin{itemize}
\item Firstly, we observe that the maximum of the asymmetry is produced when the DW network collapses at the temperature $T_{\rm ann}$ at which the $L$- or $B$-violating interaction decouples, at $T_{\rm ann} \sim T_{\rm dec}$. This can be understood intuitively by considering the fact that at $T_{\rm dec}$, wash-out is mild while the  asymmetry production from the DW is still efficient.
    \item Secondly, because the maximal asymmetry that can be produced decreases linearly with the annihilation temperature, $Y^{\rm max}_{\Delta B} \propto T_{\rm ann}$, there is an absolute lower bound on the $T_{\rm ann}$ for matching the observed baryon abundance of around  $T_{\rm ann} \sim 10^{9}$ GeV. 
     \item Gravitational wave (GW) signals are copiously emitted during the evolution and the collapse of the DW network\cite{Hiramatsu:2010yz,Hiramatsu:2013qaa,Saikawa:2017hiv,Saikawa:2020duz, Gruber:2024pqh} (see e.g. \cite{Ferreira:2022zzo,NANOGrav:2023hvm,Gouttenoire:2023ftk,Blasi:2023sej} for attempts to explain the PTA signal with DWs). 
     However the axions produced at DW collapse tend to induce a period of matter domination accompanied by the dilution of the baryon and GW signal.  Accounting for this dilution, we observe that in our concrete realisations DW-genesis can not be probed with the expected sensitivity of the Einstein Telescope \cite{Hild:2008ng,Hild:2010id,Sathyaprakash:2012jk, Maggiore:2019uih}. In this respect, we emphasize as a generic result that detectable GWs from DWs and successful DW-genesis appear to be mutually in tension.     
     \item Importantly, 
     in the typical post-inflationary scenario where axion DWs are attached to strings, there are different types of DWs, leading to a partial cancellation in the generated asymmetry.
     This can be clearly noted
     e.g. in the simplest models of DWs, with $N_{\rm DW}=2$, 
     where there are two types of DWs: the DWs interpolating from $0 \to \pi$ and $\pi \to 2 \pi$. We emphasize that those two types of DWs produce opposite baryons/lepton number when they collapse. If the induced asymmetry abundances were exactly opposite they would cancel. We will observe that this is however not the case, as difference in the tension of different DWs will lead to slightly different collapse temperatures and velocities (as already observed in \cite{Daido:2015gqa}). 
     However this implies a significant suppression in the final resulting asymmetry, that we will quantify. 
     These conclusions naturally extend to larger $N_{\rm DW}$.
     Let us however emphasize that such a suppression can be avoided if a breaking of the $U(1)$ symmetry at very high energies, as this is necessary for Affleck-Dine (AD) baryogenesis and axiogenesis mechanisms, induces a population bias among the two types DWs (see e.g.  \cite{Gonzalez:2022mcx,Kitajima:2023kzu}), favoring one type of DW over the other.

     \item 
     Finally, in all the parameter space allowing DW-genesis, the bias necessary to induce the DW annihilation spoils the solution to the strong CP problem. We conclude that the axion responsible for the DW-genesis cannot be a \emph{heavy QCD axion}
     (see \cite{Marsh:2015xka,DiLuzio:2020wdo} for axion reviews).

\end{itemize}

The remainder of this paper is organised as follows: In Section \ref{sec:spont_baryo},
we introduce the axionic domain walls and the mechanism of DW-genesis in generality.
In Section \ref{sec:spont_lepto}, we focus on a 
model 
of domain wall induced leptogenesis:
we compute in detail the Boltzmann equations necessary for our analysis
and explore the viable parameter space of the model.
In Section \ref{sec:spont_cogenesis}, we extend the previous scenario 
with a dark sector containing  asymmetric dark matter which is also populated by the DW network, realizing DW-cogenesis.
As further interesting realisations,
we present in Section \ref{sec:spont_baryogenesis} a model of baryogenesis where we introduce an effective operator violating the baryon number by two units, and in Section \ref{Sec:with_spha}, we consider the case in which the electroweak sphalerons induce the necessary $B$-violating interaction. In Section \ref{sec:dynamics_DW_GW}, we discuss the collapse of axionic domain walls and the related gravitational wave signal, and we explore the parameter space of DW-genesis,
taking also into account possible suppression factors induced by the annihilation of different domain walls. Finally we summarize and conclude in Section \ref{sec:conclusions}. 

\section{Spontaneous baryogenesis via domain walls}
\label{sec:spont_baryo}

In this first section, we describe the basics of the DW network formation, evolution, collapse and the basics of the mechanism of DW-genesis. 

\subsection{The axionic domain walls}

Domain walls are topological defects that arise when a discrete symmetry is spontaneously broken in the plasma of the early universe (see e.g. \cite{Vilenkin:2000jqa, Saikawa:2017hiv} for an introduction on DWs)
through the Kibble mechanism \cite{Kibble:1976sj}. 
In this work, we focus on discrete symmetries emerging from anomalous global symmetries as in the case of axion models and the
generalization of the Peccei-Quinn solution to the strong CP problem \cite{Peccei:1977hh, Peccei:1977ur, Weinberg:1977ma, Wilczek:1977pj}.
To this end, 
we consider the following 
Lagrangian, where the axion arises as the pseudo Nambu-Goldstone boson from the spontaneous breaking of an anomalous $U(1)$ symmetry,
\bea 
\label{eq:Lagaxion}
\mathcal{L} =  \partial_\mu \Phi^\dagger \partial^\mu \Phi - \lambda \bigg(|\Phi|^2 - v_a^2/2 \bigg)^2 - V(a) \, .
\eea 
Here we denote $\Phi = \rho/\sqrt{2} \exp\left(ia/v_a\right)$ with $\rho$ the radial mode and $a$ the axion. The $U(1)$ symmetry is spontaneously broken by the first potential term.
For concreteness,
in the following we consider a post-inflationary axion scenario where the $U(1)$ phase transition occurs at $T\sim v_a$, after the end of inflation. In this case, at the phase transition, cosmic strings around which the axion field winds by multiples of $2\pi v_a$
are created (on which the axion domain walls will attach). 
Alternatively, one can consider pre-inflationary (or intermediate-inflationary \cite{Redi:2022llj,Harigaya:2022pjd}) scenarios as considered in \cite{Daido:2015gqa}, 
where the Peccei-Quinn symmetry is broken during inflation.
\footnote{
Domain walls can emerge also in multi-axion models with mixed inflationary conditions (see e.g. \cite{Lee:2024xjb,Lee:2024toz}).
} We emphasise that our analysis about the mechanism controlling DW-genesis does not depend on the details of the domain wall formation.

The last term in Eq.\eqref{eq:Lagaxion} is the axion potential induced by the mixed anomaly between the global $U(1)$ and a confining dark gauge group (e.g. a dark $SU(N)$), which explicitly breaks the $U(1)$ symmetry to a $\mathbb{Z}_{N_{\rm DW}}$ discrete symmetry,
and leads to the formation of domain walls.
The axion potential reads  
\bea 
\label{eq:potential_axion}
V(a) = \Lambda^4 \bigg(1- \cos \left(\frac{ a N_\text{DW}}{v_a}\right)\bigg)= m_a^2 f_a^2\bigg(1- \cos \left(\frac{ a }{f_a}\right)\bigg) , \qquad m_a = \Lambda^2/f_a\,,
\eea 
with the axion defined in the range $a \in [0,2\pi v_a)$,
and with the axion decay constant given by $f_a \equiv v_a/N_\text{DW}$.
The axion potential depends on the 
strong scale $\Lambda \simeq \sqrt{m_a f_a}$ of the dark gauge group
and on the anomaly coefficient which sets $N_{\rm{DW}}$.
It is temperature dependent and 
we assume that it turns on at $T \simeq \Lambda$.
\footnote{For $T>\Lambda$ the axion potential is expected to go to zero with a power-law in $T$, see e.g. \cite{DiLuzio:2020wdo} and references therein for details
 on the case of the QCD axion.}
The potential possesses $N_\text{DW}$ discrete vacua to which the axion field can locally roll. This rolling is however initially prevented by the Hubble friction, until $m_a \sim H[T = T_{\rm form}]$, and the DWs can form and attach to the strings.  The temperature at which the DW network forms is approximately given by
\beq
T_{\rm form} \simeq {\rm Min}\left[
\left(\frac{90}{\pi^2 g_*}\right)^{1/4}
\sqrt{m_a M_{\rm pl}}\,, \Lambda
\right] \, ,
\eeq
where we assumed a radiation dominated universe with $H = \sqrt{\pi^2 g_*/90}\ T^2/M_\text{pl}$,  $M_\text{pl}$ is the reduced Planck mass and $g_*$ is the number of relativistic degrees of freedom.
Considering that $\Lambda \approx \sqrt{m_a f_a}$, we thus observe that for $f_a \ll M_\text{pl}$
the DW formation naturally occurs at $\Lambda$.

If $N_\text{DW} = 1$, the string-domain wall network is unstable and decays quickly at the DW formation temperature \cite{Barr:1986hs, Vilenkin:1982ks, Shellard:1986in, Chang:1998tb}. On the other hand,
for $N_\text{DW} > 1$,
which we will assume in this work,
the network is stable and the DW energy density rapidly overcomes the one of the strings \cite{Vilenkin:2000jqa,Hiramatsu:2012sc}.
The system approaches quickly an attractor solution, called the scaling regime \cite{Hiramatsu:2012sc, Ryden:1989vj, Hindmarsh:1996xv, Garagounis:2002kt, Oliveira:2004he, Avelino:2005pe, Leite:2011sc}, where  
the average number of DWs per Hubble patch is a constant of order one and the average velocity of the individual DW is mildly relativistic. 

For the potential in Eq.\eqref{eq:Lagaxion},
the DW profiles are  
kink solutions \cite{Vilenkin:2000jqa} along the $z$-direction (chosen arbitrarily as the direction perpendicular to the wall), which 
in the plasma frame take the form 
    \bea 
    \label{eq:axionbkg}
    \frac{a_{\rm DW}(t, z)}{f_a} = 2\pi k + 4 \text{tan}^{-1} \bigg[e^{m_a\gamma_w (z- v_w t)}\bigg], \qquad \gamma_w \equiv 1/\sqrt{1-v_w^2} \,, 
    \eea 
with $k = 0,1, \ldots, N_\text{DW}-1$,
and $v_{w}$ the wall velocity.
The region along the $z$-direction along which the axion field significantly varies is the width of the wall and in the DW frame is of order $L_{w} \sim 1/m_a$. The energy stored in the DW is characterized by the energy density per unit surface, i.e. the tension,
\beq
\sigma = \int dz T_0^0 \simeq 8 m_a f_a^2 \, ,
\eeq
where $T_0^0$ is the 00-component of the energy-momentum tensor of the ALP field evaluated on the DW solution. 

In the scaling regime, the energy density of the DW network scales like $ \rho_{\rm DW} \sim \sigma H$ and tends to dominate the energy budget of the early universe \cite{Zeldovich1974CosmologicalCO, Sikivie:1982qv}. This domination of the universe energy occurs at the temperature  when
$\rho_{\rm DW} \approx \rho_{\rm rad}$ and is given by 
\beq
\label{eq:T_dom}
T_{\rm dom} = \left( 
\frac{10}{\pi^{2} g_*} \right)^{1/4} \sqrt{\frac{\sigma}{M_{\rm pl}}} \, .
\eeq
This problematic overclosing of the universe has often been invoked as a crucial argument against the possible existence of DWs. However, DW domination can be avoided if a small energy difference $\Delta V$ biases the vacua mapped by the discrete symmetry \cite{Sikivie:1982qv, Gelmini:1988sf}. The resulting pressure  $\sim \Delta V$ will induce the collapse of the DW network when it is of the order of the tension force induced by the DW curvature $\sim \sigma H $. Equating these two forces gives us an estimate of the annihilation temperature of the network,
\begin{equation}\label{eq:T_ann}
    T_\text{ann} = \left(\frac{90}{\pi^2 g_*}\right)^{1/4}\sqrt{\frac{\Delta V M_\text{pl}}{\sigma}}\,.
\end{equation}
Requiring that the DW network never dominates the energy density of the universe means that the network has to annihilate at a temperature 
\bea 
T_{\rm ann} \gtrsim T_{\rm dom} \,.
\eea 
As we will see, the necessary collapse of the DW network implies that any axionic DW network with coupling to the lepton or the baryon current would induce lepto/baryogenesis, if there is a lepto/baryon violating operator 
which is effective at the time of the collapse. In the next section, we quantify the abundance produced in this process.

\subsection{Spontaneous DW-genesis}

In this section we explain in a nutshell the mechanism of spontaneous DW-genesis. In the previous section, we have discussed how a network of axionic DWs 
is created in the early Universe, and that it reaches rapidly a scaling evolution with $\mathcal{O}(1)$ DWs per hubble volume.
After the DWs have been produced,  they oscillate in the plasma and then eventually collapse. The particles from the plasma, entering into the DWs, undergo interactions with the wall and among themselves. The consequences of these interactions and the formalism to describe them depend mostly on the thickness and velocity of the DWs. In this paper, we will put ourselves in the regime in which the relevant particles (here leptons and quarks) can efficiently thermalise within the DWs, implying the following requirement:
     \bea 
     \label{eq:condthermeq}
    L_w/\gamma_w v_w > t_{\rm thermalisation}  \approx 1/\Gamma_{2 \to 2}\qquad  \Rightarrow \qquad T_{\text{ann}}\gtrsim \mathcal{O}(0.1)\frac{\gamma_w v_wm_a}{\alpha^2}\, ,
    \eea 
   where $\Gamma_{2 \to 2}$ is the rate of the \emph{fast} gauge mediated scatterings which enforce kinetic equilibrium within the DW,
   and which we can estimate to scale as
   $\Gamma_{2 \to 2} \sim  g_{\rm dof}\alpha^2 T_{\text{ann}}$ where $g_{\rm dof}$ counts the number of target particles and range from $20-50$. Depending on the types of particles involved, leptons (quarks), one can have $\alpha = \alpha_w (\alpha_s)$.
   The condition of Eq.\eqref{eq:condthermeq}
   can be easily satisfied with the typical mildly relativistic DW velocity and for thick axionic DWs
   (also in the annihilation phase, as we will study later).
In analogy with the studies on baryogenesis with a time-varying background\cite{COHEN1988913,COMELLI1994441,Bossingham:2017gtm, Domcke:2020kcp}, we dub this regime the  
\emph{spontaneous DW-genesis}.

In general, for spontaneous DW-genesis to occur, three ingredients are necessary:
\begin{enumerate}
\item
The presence of a time-dependent condensate. This is given by the fact that the DW motion implies a time-dependent background for the axion field (see Eq.\eqref{eq:axionbkg});
Note that this time-dependent motion should have a preferred direction. 
This implies that the scaling regime of the DW network cannot generate any net asymmetry, 
since the asymmetry generated at every wall passage will cancel out and average to zero.
On the other hand,
the DW-genesis mechanism can work at the annihilation of the network at $T_{\text{ann}}$, where the bias induces a preferred direction in the DW motion\footnote{For more details about the CP and the CPT breaking in the presence of a condensate, see\cite{Brandenberger:1999je}.}.

\item The presence of a term in the Lagrangian that couples the time-dependent condensate to the current of leptons or baryons,
such as
\bea 
\label{eq:coupling_DW_ax}
\mathcal{L} \supset
c_X\frac{\partial_\mu a}{f_a} j^\mu_X, \qquad X = L, B \qquad \text{(Coupling between $X$ and DW)}
\eea 
and thus inducing a spontaneous violation of the CPT invariance
by generating an effective chemical potential, distinguishing particles from anti-particles.

\item The presence of an interaction violating the quantum number of the current coupling to the condensate, 
which we can denote schematically  
with an operator $\mathcal{O}_X$ in the Lagrangian
\bea 
\mathcal{L} \supset\mathcal{L}_{\slashed{X}} = \mathcal{O}_X   \qquad \text{such that} \qquad \partial_\mu j_X^\mu \propto \mathcal{O}_X  \, . 
\eea 
\end{enumerate}
If the previous three conditions are satisfied, the Boltzmann equation governing the $X$ asymmetry  
takes the schematic form
\bea 
\label{eq:BEschematic}
\frac{dY_{\Delta X}}{dt} \simeq \:
\langle \Gamma_{\slashed{X}} \rangle (Y_{\Delta X} + Y^{\rm eq}_{\Delta X}(t))  \, , \qquad Y^{\rm eq}_{\Delta X}(t) \equiv \frac{n_{X}^{\rm eq}}{s} \frac{2c_X\dot a}{f_a T}, \qquad \text{(Schematic Boltzmann equation)} \,. 
\eea 
where $Y^{}_{\Delta X}$ represents the yield of the $X$ asymmetry,
and $n_{X}^{\rm eq}$ is the equilibrium number density of the species $X$. We also denote with $\langle \Gamma_{\slashed{X}} \rangle$ the thermally averaged rate of the $X$-violating interaction corresponding to the operator $\mathcal{O}_X$, which is characterized by a decoupling temperature $T_{\text{dec}}^X$ (identifying the temperature below which this rate becomes smaller than Hubble).

Eq.\eqref{eq:BEschematic} encloses all the relevant aspects of the mechanism.
In the presence of the source $\dot a/f_a$ and of an efficient $X$-violating interaction, the asymmetry is tracked to the 
non-vanishing equilibrium value 
$Y^{\rm eq}_{\Delta X}$.
When the source is turned off, the 
$X$ violating interaction, if still 
active, washes out the asymmetry and tends to set $Y^{\rm eq}_{\Delta X}$ back to zero. This clarifies also the reason why the mechanism is more effective when $T_{\text{ann}} \sim T_{\text{dec}}^X$
as we mentioned in the introduction.

It is interesting to compare the mechanism presented here with spontaneous baryogenesis in different contexts \cite{Ibe:2015nfa,Kusenko:2014lra,Cohen:1988kt,Cohen:1987vi,Jeong:2018ucz,Jeong:2018jqe,Craig:2010au,Harigaya:2023bmp,Co:2020jtv,Sarkar:2022odh,Chun:2023eqc,deCesare:2014dga,Bossingham:2017gtm,Mavromatos:2018map,Sarkar:2022odh}.
In the literature \emph{spontaneous baryogenesis} can refer both to the generation of a baryon asymmetry at a first order electroweak phase transition (FOEWPT) due to the bubble wall passage in the adiabatic regime or to the generation of an asymmetry from a long-lasting source of a time varying background field.
In the second case, the asymmetry reaches the equilibrium value set by the induced chemical potential.
In the DW-genesis scenario, the time-dependent axion background is active during a short period of time (at the wall passage), which, as we will see, typically does not allow the asymmetry to reach the equilibrium value. 

In this perspective, the DW mechanism is more similar to the spontaneous baryogenesis at the FOEWPT, albeit with several important differences.
In spontaneous baryogenesis at the FOEWPT, the source of the baryon chemical potential is a phase varying along the Higgs wall profile, and the relevant $B$-violating interactions are the sphalerons. The sphaleron rate is also a function of the wall profile, and rapidly shuts off in the interior of the bubble wall. This differs from our scenario where the $B$ (or $L$)-violating interaction is active along the entire domain wall width, and outside.
A simple consequence of this difference is that washout effects are important for DW-genesis and the mechanism can be successful only if the decoupling temperature of the relevant rate is close to the DW annihilation temperature.
In addition, in spontaneous baryogenesis at the FOEWPT, the width of the wall is typically of the same order of the temperature, making the requirement of thermal equilibrium inside the wall at the edge of validity (depending on the wall velocity). In the case of spontaneous DW-genesis, instead, the width of the wall is controlled by a separate mass scale (the axion mass) which can be parametrically different from the annihilation temperature, hence allowing a large room to  fulfill the condition in Eq.\eqref{eq:condthermeq}.

In the following sections, we will describe in detail how the 
requirements for a successful spontaneous DW-genesis
can be fulfilled in 
several concrete realisations, and we will derive the Boltzmann equations (the detailed version of Eq.\eqref{eq:BEschematic}) 
controlling the production of baryon/lepton number after the introduction of a baryon/lepton violating operator.

\section{A model of domain wall leptogenesis}
\label{sec:spont_lepto}

In the previous section, we described the basics of the DW-genesis mechanism.  From now on, we will apply it to different baryon and lepton number violating models. We start with a model of DW leptogenesis. 

\subsection{Derivation of the Boltzmann equations}
\label{sec:derivation_BE}
In this section we consider a minimal model where the left handed leptons couple to the axion only via 
\be
\label{eq:inte}
\mathcal{L}_{a-L} =\frac{c_L \partial_{\mu} a}{f_a}  j^\mu_L   \, , \qquad j^\mu_L \equiv \bar L \gamma^{\mu}  L 
\ee
where 
$L$ 
could designate any lepton flavor. We will proceed with only one flavor to simplify the analysis, and comment later on possible generalizations. 
We do not include couplings to right-handed (RH) particles (lepton or neutrinos) since they will not change qualitatively our conclusions.

On its own, the interaction in Eq.\eqref{eq:CPT_viol_1} cannot have any impact on the abundances of particles. On top of the CPT breaking induced by the axion time-dependent background, one needs to consider an $L$ (or $B$)-violating interaction.
For the purpose of illustration, we consider the following Lagrangian 
\bea 
\label{eq:L_violating_1}
\mathcal{L}^{}_{\slashed{L}} = y_N (  \tilde H \bar L)N_R +  \frac{1}{2}M_N \bar N^c_R N_R + \text{h.c.} \; , 
\eea 
where $\tilde H \equiv i\sigma^2 H^\star$, responsible for  the neutrino masses via the standard type-I see-saw mechanism~\cite{Minkowski:1977sc,Yanagida:1979as,Mohapatra:1979ia,Schechter:1980gr}. The full Lagrangian of our model is then \footnote{In this model, the axion couples to the left-handed (LH) leptons as a \emph{Majoron}\cite{CHIKASHIGE1981265, GELMINI1981411}, which was studied in the context of spontaneous baryogenesis\cite{Chun:2023eqc}.}
\bea 
\mathcal{L} = \mathcal{L}_{a-j}  + \mathcal{L}^{}_{\slashed{L}} \, . 
\eea

When the DW is sweeping through the plasma, the interaction of Eq.\eqref{eq:inte} in the plasma frame inside the DW takes the form 
\bea
\label{eq:CPT_viol_1}
\qquad \mathcal{L}_{a-L} = \dot \theta j_L^0  \simeq \mu j_L^0 \, , \qquad \qquad 
\dot \theta  \equiv \frac{\dot a_{\rm DW}(t, z)}{f_a} = 
\frac{2 m_a \gamma_w v_w}{\cosh\left[m_a \gamma_w v_w\left(t-t_\text{passage}\right)\right]} \, ,
\eea 
where we used that $j_\mu = j_0(1, 0, 0, 0)$ and introduced $\dot \theta \equiv \dot a/ f_a$. The equality emphasises that the presence of the background acts as a chemical potential $\mu$ for the lepton number. $t_\text{passage}$ is the time when the region of the plasma under consideration is in the center of the DW. 
In the following we will take $t_{\rm passage} \sim t_{\rm ann}$ assuming that the DW annihilation happens instantaneously in all the Hubble volume. We  discuss in Appendix \ref{app:finite_time_collapse} the corrections one obtains by taking into account the time duration of the DW collapse.

One can understand the consequences of the coupling in Eq.\eqref{eq:inte} in two different ways. First, one might notice that the existence of the coupling in Eq.\eqref{eq:inte}, in presence of a varying background field for the axion ($\dot a \neq 0$), can modify the chemical potential for the 
particle thermal distributions
of the leptons, as presented in Appendix \ref{app:spon_baryo} (see however \cite{Arbuzova:2016qfh} for subtleties about this line of reasoning). Second, one might reabsorb the operator  $\mu j_L^0$ into the vertices by rotating the fermions, as presented in \cite{Arbuzova:2016qfh}. This is the method we will follow in this section.

To eliminate the coupling between the axion and the left-handed current\footnote{
This transformation also induces an axion phase in front of the SM lepton Yukawa coupling, that will not play any role in the following (and which is anyway degenerate with the derivative axion RH lepton coupling that we have not specified here).
Similar reasoning applies to the extra contribution to the axion non-derivative couplings to the gauge field strengths resulting from this transformation.
See Appendix \ref{app:axionEFT} for details.
}, we can now perform the following rotation, 
\begin{align} 
\label{eq:rotation_1}
L \to e^{i c_L a/f_a} L \qquad  \Rightarrow\qquad  i\bar L \slashed{\partial} L    \to i\bar L \slashed{\partial} L  -  c_L\frac{\partial_\mu a}{f_a} j^\mu_L \, ,
\qquad  y_N ( \tilde H \bar L)N_R \to e^{-i c_L a/f_a}y_N ( \tilde H \bar L)N_R\,.
\end{align} 
This rotation indeed eliminates the interaction in Eq.\eqref{eq:inte} and the Lagrangian becomes
\bea 
\label{eq:L_violating_1}
\mathcal{L}^{}_{\slashed{L}} = e^{-i c_L a/f_a} y_N (\tilde  H \bar L)N_R + \frac{1}{2}M_N \bar N^c_R N_R + \text{h.c.}  \,. 
\eea 
which we will now use in the computation of the collision term of the Boltzmann integrals. 

Before proceeding, few comments are in order. First, the phase in the term $y_N  N_R(\tilde  H \bar L)$ will appear in the computation of the decay as well as the $L$-violating scatterings $HL \to H^cL^c$, but not in the $\Delta L = 0$ scatterings $HL \to HL$. In the remainder of this paper, we can ignore this interaction, since it is not biased by the DW passage. Secondly, one could rotate away
the phase in the yukawa coupling by a rotation of the RHN, $N_R \to e^{ic_L a/f_a}N_R$, and conclude that the decay and the inverse decay are not biased by the source. This is not the case, as the phase will reappear in the mass term $M_N \bar N^c_R N_R$, leaving at the end the same amplitudes for the $L$-violating interactions.  This fact is better understood in the \emph{unsubtracted} scheme, as explained in Appendix \ref{app:un_sub}. 

In this section, when computing the rates and the BEs, we will use the \emph{narrow width approximation} (another route is presented in Appendix \ref{app:un_sub}), in which
\bea 
|\mathcal{M}_{LH \to H^c L^c}|^2 = |\mathcal{M}'_{LH \to H^c L^c}|^2+ |\mathcal{M}^{\rm res}_{LH \to H^c L^c}|^2 \text{Br}_{N \to HL} \, , \qquad \text{Br}_{N \to HL} = 1/2 \, ,
\eea 
where the resonant piece corresponds to the on-shell decay and inverse decay of the intermediate RHN\cite{Buchmuller:2000nq}
\bea
|\mathcal{M}^{\rm res}_{LH \to H^c L^c}|^2 \approx \mathcal{M}_{LH \to N}\mathcal{M}_{N \to HL}^\star = |\mathcal{M}_{N \to HL}|^2 \, . 
\eea 
Consequently the rate for the scatterings $\Gamma_{LH \to H^c L^c}$ will be computed keeping only the off-shell piece of the integral. This procedure avoids double counting the rates of the on-shell processes\cite{Davidson:2008bu}. 

Let us now move to the BEs themselves. The BEs in the plasma frame for the decay and the scatterings take the following form 
\be
\label{eq:boltzmann_eq}
  \dot n_L + 3 H n_L =\int  \frac{d^3 p_L}{(2\pi)^3} \bigg(\mathcal{C}^{\rm decay}[f_{L}] + \mathcal{C}^{\rm scatt}[f_{L}] \bigg)  \equiv I^{\rm decay} + I^{\rm scatt} \, ,
\ee
where $\mathcal{C}^{\rm decay}$ is the collision term which involves $N \to LH$ with rate
\bea
\label{eq:collision_int_1}
 I^{\rm decay} =  \int d \Pi_L d \Pi_N d \Pi_{H} (2 \pi)^4 \delta^3(p_N-p_L - p_H)\delta(E_N-E_L - E_H -c_L \dot a/f_a) 
\\ \nonumber
\times \left[|\mathcal{M}_{N \to LH}|^2 f_N  - |\mathcal{M}_{LH \to N}|^2 f_L f_H \right]
\eea
and  $\mathcal{C}^{\rm scatt}$ is the collision term which involves $LH \to H^c L^c$ and $H^c L^c \to LH$ with rate
\bea 
\label{eq:collision_int_2}
I^{\rm scatt} = \int d\Pi_{L}d\Pi_{H}d\Pi_{H^c}d\Pi_{L^c}\delta^3(\sum_i  p_i - \sum_f  p_f ) \delta (E_{H}+E_{L}-E_{H^c}-E_{L^c} + 2 c_L \dot a/f_a )
\\ \nonumber
\times \left[|\mathcal M'_{H^c L^c \to LH }|^2 f_{H^c}f_{ L^c}-|\mathcal M'_{LH \to H^c L^c}|^2 f_{H}f_{L} \right] \, , 
\eea 
where the phase space integrals are ($g_i$ keeps into account d.o.f.)
\be
d \Pi_i \equiv \frac{g_i}{(2 \pi)^3} \frac{d^3 p_i}{2 E_i} \, .
\ee
A similar set of equations holds for antiparticles, with opposite sign for the $\dot a$ term. In the former equations, we have assumed Maxwell-Boltzmann (MB) statistics. 

In the collision integrals, the delta Dirac, imposing the conservation of energy, contains a new unusual term $c_L \dot a/f_a$, which comes from the vertex obtained in Eq.\eqref{eq:L_violating_1}. In the derivation, following the lines of \cite{Arbuzova:2016qfh}, we approximated $a \approx \dot a t$ and then performed the fourier transform which lead to the delta Dirac with the new term containing $\dot a$. This approximation is valid if $T_{\rm ann} > \text{few}/L_wv_w \gamma_w$ as shown in Appendix \ref{app:validity_app}.
By assumption, the leptons reach kinetic equilibrium within the domain wall, due to \emph{fast} interactions with the photon bath and the gauge bosons and consequently the particles distributions can be captured with MB distributions. However, chemical equilibrium is ensured by \emph{slow} interactions, namely $L$-violating ones: $N \to HL$, $HL \to H^cL^c$ which have a rate $\Gamma_{\slashed{L}}^{-1} \gg L_w/v_w \gamma_w$,
i.e. the source is short lived. Consequently, chemical equilibrium typically will not be reached in our scenario.
To capture the asymmetry produced by the DW, we introduce a parameter $\xi_L$  and the distributions are thus parameterized by
\bea 
\label{eq:distributions}
f_{L} = \frac{1}{e^{\frac{E}{T} - \xi_L} +1 } \approx e^{-E/T + \xi_L} \, , 
\qquad 
f_{ L^c} \approx e^{-E/T -\xi_L} \, . 
\eea 
In this parameterization, $\xi_L$ can be related to the lepton number by subtracting the number density of leptons and anti-leptons
\bea 
n_{\Delta L} \equiv n_L - n_{L^c} = n^0_L (e^{\xi_L}- e^{-\xi_L}) \approx 2n^0_L \xi_L \, .
\eea 
 where we expanded the exponential in the last line by assuming $\xi_L \ll 1$ and we defined $n^0_L \equiv n_L(\xi_L = 0)$.

 We also neglect the chemical potential for the Higgs since it would only lead to a $\mathcal{O}(1)$ correction to the asymmetry\cite{Bertuzzo:2010et}.  Using the distributions in Eq.\eqref{eq:distributions} and noticing that 
\begin{align} 
&\text{decay}: \qquad  f_L f_{H} \approx e^{-\frac{E_{H}+ E_L}{T}+ \xi_L} = e^{-\frac{E_N - c_L \dot a/f_a}{T}+ \xi_L}  
\notag \\
&\text{scatterings}: \qquad f_{H}f_L \approx e^{-\frac{E_{H}+ E_L}{T} + \xi_L}, \qquad f_{H^c}f_{L^c} \approx e^{-\frac{E_{H^c}+ E_{ L^c}}{T} - \xi_L} = e^{-\frac{E_{H}+ E_{ L} + 2c_L\dot a/f_aT}{T} - \xi_L} \, .
\end{align} 
The scattering rates for $n_L$ simplify to the form
\begin{align}
\label{eq:collision_int_1_bis}
 I^{\rm decay}_{ L}(\xi_L) 
 = I^{\rm decay}_{ L}(\xi_L = 0)   \left(1  - e^{\frac{  c_L \dot a/f_a}{T}+ \xi_L} \right ) \, ,
\quad 
I^{\rm scatt}_{ L} (\xi_L) = I^{\rm scatt}_{ L} (\xi_L=0)   \bigg(  e^{- \xi_L - \frac{2c_L\dot a}{f_a T}}- e^{\xi_L}\bigg) \, ,
\end{align}  
 and for $n_{L^c}$ to
 \begin{align}
\label{eq:collision_int_1_bis}
 I^{\rm decay}_{ L^c} (\xi_L) &
 =I^{\rm decay}_{ L^c} (\xi_L=0)   \left(1  - e^{-\frac{  c_L \dot a/f_a}{T}- \xi_L} \right ) \, , \quad 
I^{\rm scatt}_{ L^c} (\xi_L) = I^{\rm scatt}_{ L^c} (\xi_L =0)   \bigg(  e^{ \xi_L + \frac{2c_L\dot a}{f_a T}}-e^{-\xi_L}\bigg) \, ,
\end{align}  
 at leading order in $\xi_L$. 
We can now write the BEs for the lepton number $n_{\Delta L}$
by using that $I^{\rm scatt}_{ L^c} (\xi_L =0) = I^{\rm scatt}_{ L} (\xi_L =0) $ and $I^{\rm decay}_{ L^c} (\xi_L=0)= I^{\rm decay}_{ L} (\xi_L=0)$ and the results above,
\begin{align} 
\dot n_{\Delta L} + 3 H n_{\Delta L} &= I^{\rm scatt}_{ L} (\xi_L=0)   \bigg( e^{- \xi_L - \frac{2c_L\dot a}{f_a T}}- e^{\xi_L}-  e^{ \xi_L + \frac{2c_L\dot a}{f_a T}}+e^{-\xi_L}\bigg) \, ,
\notag
\\
&
+ I^{\rm decay}_{ L} (\xi_L=0)   \bigg(  - e^{\frac{  c_L \dot a/f_a}{T}+ \xi_L}  + e^{\frac{ - c_L \dot a/f_a}{T}- \xi_L}   \bigg)  \, ,
\end{align}
which upon linearisation can be written in the form 
\begin{align} 
\dot n_{\Delta L} + 3 H n_{\Delta L} &= -4I^{\rm scatt}_{ L} (\xi_L=0) \times  \bigg(\frac{c_L \dot a}{f_a T} +\xi_L(t)\bigg) 
- 2I^{\rm decay}_{ L} (\xi_L=0) \times  \bigg(  \frac{  c_L \dot a}{Tf_a} + \xi_L(t)   \bigg)  \, . 
\end{align}

Switching to notations familiar to the leptogenesis literature, we obtain the following compact equation
\bea 
\label{eq:lepto_equation}
\boxed{
\frac{dY_{\Delta L}}{dt} = -\bigg( 
\frac{\gamma_D}{n_L^{\rm eq}} (Y_{\Delta L} + Y^{\rm eq}_{\Delta L}(t)) + 2 \frac{\gamma_{2 \to 2} }{n_L^{\rm eq}} \bigg( Y_{\Delta L} + Y^{\rm eq}_{\Delta L}(t)\bigg)\bigg) \, , \quad Y^{\rm eq}_{\Delta L}(t) \equiv \frac{n_{L}^{\rm eq}}{s} \frac{2c_L\dot a}{f_a T}, \quad Y_{\Delta L} \approx 2\frac{n_{L}^{\rm eq}}{s}  \xi_L}
\eea 
where we defined the yield via the usual $Y\equiv n/s$, where $s$ is the entropy density
\bea
\label{eq:basics_eq}
n^{\rm eq}_{L} =  \frac{g_{L}}{\pi^2} T^3 \, , \qquad s = \frac{2 \pi^2}{45} g_* T^3 \, ,
\eea 
and where $g_{L} = 2$ are the number of degrees of freedom of the lepton.
The rates in Eq.\eqref{eq:lepto_equation} can be obtained from \cite{Buchmuller:2002rq}(see also\cite{Davidson:2008bu, Buchm_ller_2005, Giudice:2003jh})\footnote{Notice that the decay rate $\Gamma_D$ is from time to time defined as $ \Gamma_D = \Gamma_{N \to HL} + \Gamma_{N \to H^cL^c} = y_N^2\frac{M_N}{8 \pi}$ in the literature. 
In this paper we do not use this definition and keep $\Gamma_D = \Gamma_{N \to HL}$. }.
\begin{align}
\frac{\gamma_D}{n_L^{\rm eq}} &= \frac{n^{(\rm eq)}_N}{n_L^{\rm eq}} \frac{K_1(z)}{K_2(z)} \Gamma_D = \frac{M_N^2}{ T^2} K_1(z)\Gamma_D \,  ,
\qquad \Gamma_D \equiv \Gamma_{N \to HL} = \Gamma_{N \to H^cL^c} = y_N^2\frac{M_N}{16 \pi} \, , 
\\ \notag
\frac{\gamma^{\Delta L = 2}_{2 \to 2} }{n_L^{\rm eq}} &\approx \Gamma^{\Delta L = 2}_{2 \to 2} \approx \Gamma_{HL \to H^cL^c}+ \Gamma_{LL \to H^cH^c}\,. 
\end{align}
where the rate for the $2 \to 2$ process above contains only the \emph{off-shell} piece of the integral, where the pole at  $s\to M_N^2$ has been subtracted.  
The same procedure can be followed by considering unsubtracted $2 \to 2 $ rates and is sketched in Appendix \ref{app:un_sub}. The two approaches give the same final result.
We observe that the derived BE indeed reflects the schematic form discussed around Eq.\eqref{eq:BEschematic}.

In the limit $T \ll M_N$, the scattering rate simplifies to \cite{Buchmuller:2002rq}
\bea 
\Gamma_{HL \to H^cL^c} = \Gamma_{LL \to H^cH^c}\approx \frac{T^3}{4\pi^3}  \frac{\sum m_{\nu_i}^2}{v_{\rm EW}^4}, \qquad \Gamma^{\Delta L = 2}_{2 \to 2} = \frac{T^3}{2\pi^3}  \frac{\sum m_{\nu_i}^2}{v_{\rm EW}^4}
\eea 
where $v_{\rm EW} \approx 174$ GeV is the EW scale  and $m_{\nu_i}$ are the masses of the light neutrino, $\sum m_{\nu_i}^2 \approx \Delta m^2_{\rm atm} \approx 2.4\times 10^{-3} \text{eV}^2$. This takes into account the $HL \to H^cL^c$ and the $LL \to H^cH^c$ interactions. We neglect the scattering processes with $\Delta L = 1$
since the effect of such scatterings is
subdominant to the inverse decay term as in thermal leptogenesis.  Moreover, these channels typically decouple at temperatures much higher than the inverse decay, and thus are not relevant for the \emph{final} value of the asymmetry. For the interested reader, we refer to papers where they do take them into account \cite{Buchmuller:2004nz, Giudice:2003jh, Hahn-Woernle:2009jyb}.

Eq.\eqref{eq:lepto_equation} will be our master equation for the remainder of this paper, with the necessary amendments in each section. It could actually be obtained without doing the rotation in Eq.\eqref{eq:rotation_1} but by considering the shift in the dispersion relations induced by the presence of the term $j^0_L\dot a/f_a$. We show that we reach the same final results in Appendix \ref{app:spon_baryo}. 

At this point, one comment is in order: in Eq.\eqref{eq:lepto_equation}, we assumed that the RHN abundances track perfectly its equilibrium abundance. In principle, the abundance of $N_R$ should be also solved for and plugged in the $\gamma_D$ expression all along the evolution. We however verified that this only brings a correction of less than 1\%, which we thus neglect in the remainder of this paper.

\subsection{Domain wall leptogenesis within a see-saw model}
\label{sec:numerics}

We now move to the study of the phenomenology of DW leptogenesis. We start from the same  model we considered in Eq.\eqref{eq:L_violating_1} and use directly the master BEs derived in Eq.\eqref{eq:lepto_equation} to compute the lepton yield. 

The analysis has been performed for a specific flavor of leptons. However, the same analysis would hold for any family as long as the condition for the thermalisation in the DW, Eq.\eqref{eq:condthermeq}, holds. We repeat here the Lagrangian with several families of leptons
\bea 
\label{eq:L_violating}
\mathcal{L}_{\slashed{L}} = \sum_{\alpha, i} y^{i\alpha}_N ( \tilde H \bar L_\alpha)N_{R,i} + \frac{1}{2}\sum_{ i} 
 M_{N_i} \bar N^c_{R,i} N_{R,i} + \text{h.c.}   \,, 
\eea 
where $i$ runs over the RHNs and $\alpha$ runs over the lepton families, and we choose the basis where the mass term is diagonal. For the simplicity of the analysis, we embrace a hierarchical scenario with $M_{N_1}\ll M_{N_2} $ and a peculiar texture, such that $y^{1 \tau} \gg y^{1 \mu} \gg y^{1 e}$ and $\sum_{\alpha} |y^{1 \alpha}_N|^2 \sim m_\nu M_N/v_{\rm EW}^2 \approx 1.6 \times 10^{-12} M_{N_1}/\text{TeV}$, such that effectively we can consider only one lepton family\footnote{Within a 2RHN model, considering the three families will only lead to an $\mathcal{O}(1)$ deviation from the prediction with only one lepton family.} and we can rename $M_{N} \equiv M_{N_1}$ and $y_N \equiv y^{1 \tau}$.

Such type of models, when embedded with enough RHN families, can account for the observed light SM neutrino mass and provide a model for leptogenesis\cite{FUKUGITA198645} via the CP-violating decay of RHNs. However, we emphasize that, in the DW-genesis scenario we consider, no explicit 
CP violation in the Yukawa couplings
is needed to create the baryon number. Thus in what follows, we will assume that the CP phase in the couplings are negligible and the decay is fully CP conserving. Of course, in principle, these two scenarios could combine.

\begin{figure}[h!]
    \centering
    \includegraphics[width=.52\linewidth]{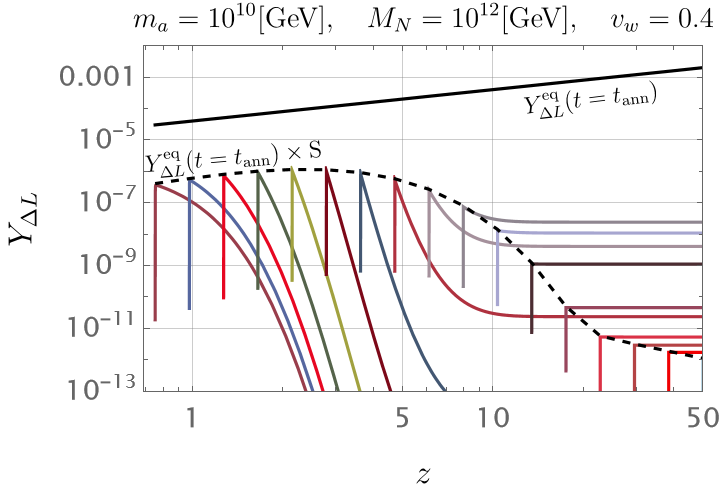}
    \includegraphics[width=.47\linewidth]{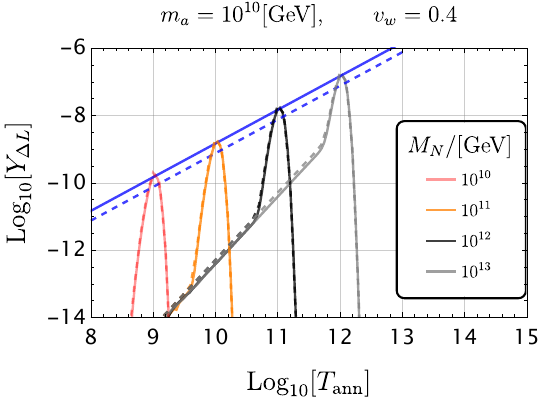}
    \caption{\textbf{Left panel}: trajectories in $z \equiv M_N/T$ of the asymmetry for $m_a = 10^{10}$ GeV and $M_N  = 10^{12}$ GeV, for values of $T_{\rm ann} \in[2\times 10^{10}, 10^{12}]$ GeV separated by a factor $1.3$. 
    The solid Black line represents the value of $Y^{\rm eq}_{\Delta L}(t = t_{\rm ann})$ and the dashed line represents $Y^{\rm eq}_{\Delta L}(t = t_{\rm ann}) \times S$ where $S = \pi \Gamma(T=T_{\rm ann})/(m_a\gamma_w v_w)$. \textbf{Right panel}: Final lepton asymmetry as a function of the temperature at which the wall collapses. The dashed lines represent the analytical approximation from Eq.\eqref{eq:final_analytics} and the solid lines the numerical solution of Eq.\eqref{eq:lepto_equation}. The Blue dashed line is the approximation of the height of the peak as obtained in Eq.\eqref{eq:analytics_max} for $c_L = 1$, $Y_{\Delta L} \approx 0.18 T_{\rm ann}/M_{\rm pl}$, while the Blue solid line is the numerical envelope of the peak.
    }
    \label{fig: T_ann}
\end{figure}

We have solved the master BE numerically (considering as a toy model one family of leptons and one RHN with $c_L=1$), and we now present the results. On Fig.~\ref{fig: T_ann}, using a $\tanh$ solution for the wall as in Eq.\eqref{eq:axionbkg}, 
we present on the Left panel the trajectories of the asymmetry $Y_{\Delta L}(z)$, with $z \equiv M_N/T$, for different annihilation temperatures of the DW network $T_{\rm ann}$. 
On the Right panel, we present the final value of the asymmetry as a function of $T_{\rm ann}$. 
The pattern of the final yield presents a peak structure which can be explained by inspecting 
in parallel the Left panel in
Fig.\ref{fig: T_ann}. 

The peak in the final lepton asymmetry occurs approximately at $T_{\rm ann} \sim M_N/10$, where the decay and inverse decay are close to decoupling. Solving the BEs numerically, we observe that the decoupling value $Y_{\Delta L}(T_{\rm ann} = T_{\rm dec})$ is at a slightly lower temperature than that of the peak value $Y_{\Delta L}^{\rm max}$: $T_{\rm peak} \sim 1.5 T_{\rm dec}$. When $T_{\rm ann}$ is much larger, the inverse decays are still strongly coupled and largely washes out the asymmetry produced by the DW. This produces the exponential suppression at larger $T_{\rm ann}$
in the Right panel.
It corresponds, in the Left panel, to the asymmetry trajectories with a steep bump, which is afterwards washed out for larger $z$.

At lower $T_{\rm ann}$, the interaction producing the asymmetry is already decoupled and the asymmetry produced is weaker. These correspond, in the Left panel, to trajectories where the asymmetry is suddenly created at the wall passage, and then remains constant for larger $z$.
In the Right panel, this regime corresponds to the slope at the left of the peak.

As a final remark, for reference, 
in the Left panel of Fig.~\ref{fig: T_ann},
the solid black line presents the value of $Y^{\rm eq}_{\Delta L}(t = t_{\rm ann})\sim z$, and the dashed black line presents $Y^{\rm eq}_{\Delta L}(t = t_{\rm ann}) \times S$ where $S = \pi \Gamma(T=T_{\rm ann})/(m_a v_w \gamma_w)$, with $\Gamma=(\gamma_D+2\gamma_{2\to 2})/n_L^{\rm eq}$. Interestingly, we observe that the equilibrium abundance at the passage of the wall $Y^{\rm eq}_{\Delta L}(t = t_{\rm ann})$ is never reached but is exactly suppressed by a factor $\pi \Gamma(T=T_{\rm ann})/(m_a v_w\gamma_w)$. 
This factor can be understood by simple analytical reasoning as we will investigate in the next subsection.

\subsection{Analytic approximation}
We would now like to derive an analytic approximation for the asymmetry from Eq.\eqref{eq:lepto_equation}. 
In the regime in which the wall is thin on the timescale of the $L$-violating interactions, the effective chemical potential can be approximated with a delta function as
\beq
\label{eq:delta_approx}
\frac{\dot a}{f_a} 
\simeq 2 \pi \delta(t-t_{\rm passage}) \, , 
\eeq 
We expect this approximation to be valid when the time of passage of the wall is small compared to the timescale of the lepton-violating interaction sourcing the asymmetry, so when 
\beq
\label{eq:regime_for_delta}
(L_w \gamma_D/n_L^{eq})/(\gamma_wv_w)  <1 
\qquad \text{(Validity condition for delta function approximation)}
\eeq
Employing this approximation, 
we can derive analytical predictions for the final lepton asymmetry, by splitting
 the integration of Eq.\eqref{eq:lepto_equation} into three parts.
 
i) \textbf{Scaling behaviour:} from $t \sim 0$ to $t_{\rm passage} - \epsilon/2$, where $\epsilon$ is the approximate timescale, a portion of the plasma resides inside the DW, the walls wiggle in arbitrary directions and create a small lepton number at every passage, that however statistically averages out. We hence assume that the asymmetry starts at $0$
and stays $0$ up to 
$t_{\rm passage} - \epsilon/2$.

ii) \textbf{Collapse}: when the bias, which favors one minimum over the other one in the axion potential, starts to dominate, the favored patches crunch the disfavored ones, accelerating the DW in a specific preferred direction. Each region of the plasma will be in the DW for a time $\sim \epsilon \ll 1/H$. 
We can estimate that the sourcing of the lepton asymmetry will occur in the time range 
from $t_{\rm passage} -\epsilon/2$ 
to $t_{\rm passage} +\epsilon/2$, 
and we can
integrate the equation assuming that the initial $Y_{\Delta L}$ term is irrelevant, getting
\beq
\label{eq:analytic_prod}
Y_{\Delta L}(t = t_{\rm passage}+ \epsilon/2) = 4 \pi  c_L \frac{n^{\rm eq}_L}{s} \bigg(\frac{ M_N^2}{T^3} K_1(M_N/T)\Gamma_D + \frac{T^2}{2\pi^3}  \frac{\sum m_{\nu_i}^2}{v_{\rm EW}^4}\bigg)\Big|_{T=T_{\rm ann}} \, .
\ee

iii) \textbf{Wash-out:} the subsequent evolution occurs in the absence of the source term and washes out the produced asymmetry. 
Assuming that all the regions in a Hubble volume will experience the wall passage at a time $t_{\text{passage}} \sim t_{\rm ann}$,
the evolution from 
$t_{\rm ann}$ to today is given by integrating the equation from $t_{\rm ann} $
to $\infty$ without the source term (setting $\dot a = 0$), using Eq.\eqref{eq:analytic_prod} as the initial condition for the asymmetry. This results in the following expression
\beq
Y_{\Delta L}^{\rm final} = Y_{\Delta L}(t = t_{\rm passage}+ \epsilon/2) e^{-\beta}\, , 
\eeq
where the exponent is a number obtained integrating
\beq
\label{eq:analytic_wash}
\beta = \int_{t_{\rm ann}}^{\infty} \frac{\gamma_D+ 2\gamma_{2\to 2}}{n^{\rm eq}_L} dt =
\int_0^{T_{\rm ann}} \frac{\gamma_D+ 2 \gamma_{2\to 2}}{HT n_L^{\rm eq}}dT \; .
\eeq
which measures the wash-out suppression.

Combining Eq.\eqref{eq:analytic_prod} with Eq.\eqref{eq:analytic_wash}, we obtain the final asymmetry
\bea 
\label{eq:final_analytics}
Y^{\rm final}_{\Delta L} \approx  4 \pi  c_L  \frac{n^{\rm eq}_L}{s}\bigg(\frac{ M_N^2}{T^3} K_1(M_N/T)\Gamma_D + \frac{T^2}{2\pi^3}  \frac{\sum m_{\nu_i}^2}{v_{\rm EW}^4}\bigg)\Big|_{T=T_{\rm ann}} e^{-\beta} \, .
\eea 
This approximation is consistent with the numerical results from Eq.\eqref{eq:lepto_equation} at typically 10 to 20\% accuracy,
as it can be seen from the dashed lines in the Right panel of Fig.~\ref{fig: T_ann}.
Note that from the analytic approximation in Eq.\eqref{eq:final_analytics} we can read the maximal asymmetry which can be obtained during the temperature evolution, simply by setting the washout to zero (i.e. $\beta=0$) in Eq.\eqref{eq:final_analytics}.
This indeed gives 
$Y^{\rm eq}_{\Delta L}(t_{\rm ann}) \times \pi \Gamma(T_{\rm ann})/(m_a v_w \gamma_w)$, with $\Gamma=(\gamma_D+2\gamma_{2\to 2})/n_L^{\rm eq}$,
which corresponds to the Black dashed line in the Left panel of Fig.\ref{fig: T_ann}, as we observed at the end of the previous subsection.

The analytic approximation allows us also to compare the decay and the scattering contributions to the final abundance.
Around the peak in Fig.\ref{fig: T_ann}, the dominant process is the decay (and inverse decay). For smaller $T_{\rm ann}$, instead, the most relevant processes are the $L$-violating scatterings.
This scattering regime is the one that has been previously considered in
\cite{Daido:2015gqa}. We verified that in this particular regime our results are consistent with theirs.

Let us now comment on the expression in Eq.\eqref{eq:final_analytics}. 
First of all, in this approximation there is no $m_a$ or $v_w$ dependence. 
This dependence
would be captured by considering deviations of the $\dot a/f_a$ profile with respect to a delta function. 
Moreover, 
we also know that baryon/lepton production from a standing wall, $v_w \to 0$ should go to zero. This is not accounted for in the expression of Eq.\eqref{eq:final_analytics}, which by construction is a valid approximation only in the regime given by Eq.\eqref{eq:regime_for_delta}
and cannot properly model the $v_w \to 0$ limit.

    In the approach described so far, we considered that the DW network collapse instantaneously and hence $t_{\rm passage} \sim t_{\rm ann}$ for every point of the Hubble volume.
    Let us however notice that realistically the DW network does not collapse instantaneously and that there can be a sizable change in the temperature between the beginning and the end of the collapse process, leading to a sizable difference in the scattering rates, and ultimately the produced asymmetries. 
    We study the correction of such an approximation in Appendix \ref{app:finite_time_collapse}, which depends on the assumed average velocity of the DW network. We found that varying the velocity of collapse in $v_w \in [0.1, 0.5]$, the suppression of $Y_{\Delta L}$ can range in $[0.1, 0.5]$.

\subsection{Phenomenology of the DW leptogenesis}
\label{sec:para_space_spon_lepto}

We now turn to the study of the parameter space where DW leptogenesis can successfully explain the observed matter abundance in the universe. The final lepton abundance will be transferred to the baryon sector via sphalerons. It can be computed that\cite{Davidson:2008bu}
\bea 
Y_{\Delta B} \approx \kappa_{\rm sph}  Y_{\Delta L} \, ,
\eea 
where $\kappa_{\rm sph} = 28/79 $ accounts for the transfer of the asymmetry in the leptons to the baryons.  This requires, using Eq.\eqref{eq:observed_BAU}, that $Y_{\Delta L} \approx 2.7\times 10^{-10}$ for matching the observed value. 

With the results obtained so far, we can explain some of the crucial features of DW baryogenesis with simple analytic reasonings. 
Fig.~\ref{fig: T_ann} (Right)  displays a peak structure for $Y_{\Delta L}$ at roughly $T^{\rm peak} \approx M_N/10$. 
Around the peak, the dominant process is the decay of the RHN.
Moreover, looking back at Fig.~\ref{fig: T_ann}, it seems that the height of the peak increases linearly with the value of $M_N$. 
These observations can be understood in the following way, finding also the maximal expected asymmetry obtainable through this mechanism. From the analytic expression in Eq.\eqref{eq:final_analytics}, and the numerical results in Section \ref{sec:numerics}, we can deduce that the produced lepton number will be maximised when the annihilation time of the DW and the decoupling time of the $L$-violating interaction coincide approximately. 
This corresponds to minimizing the wash-out effects, while still being able to create efficiently a lepton asymmetry, which indeed 
can be neglected if the interaction has decoupled right after the passage of the wall $\gamma_{D}/n^{\rm eq}_L \lesssim H(T_{\rm dec})$.
This regime can be captured in the analytic approximation by neglecting the factor $e^{-\beta}$ in Eq.\eqref{eq:final_analytics},
and 
one obtains the following estimate for 
the final asymmetry
\bea 
\label{eq:maximal}
Y_{\Delta L}^{\rm final}\approx  \frac{\gamma_D}{n^{\rm eq}_L} \frac{180}{4\pi^4 g_\star}\frac{4\pi  c_L }{ T }  \, .  
\eea 
where we used the relation between $s$ and $n_L^{\rm{eq}}$ (see Eq.\eqref{eq:basics_eq}).
 In this regime of weak wash-out, it appears that the final lepton asymmetry scales like $Y_{\Delta L}^{\rm final} \propto \Gamma/T|_{T = T_{\rm ann}}$.  
 Analogous arguments can be followed to conclude that  
when instead the production is dominated by the scatterings (off the peak, so for $T_{\rm ann}$ significantly smaller than $M_N$), we have thus $Y_{\Delta L}^{\rm final} \propto T_{\rm ann}^2$, as can be confirmed by looking at the slope on the Right panel of Fig.\ref{fig: T_ann}. 

The maximal lepton asymmetry for DW leptogenesis, i.e. the peaks of Fig.\ref{fig: T_ann} (Right), can thus be approximated by evaluating Eq.\eqref{eq:maximal} at decoupling, $\gamma_{D}/n^{\rm eq}_L \approx H(T_{\rm dec})$, 
\bea 
\label{eq:maximal_asymmetry}
Y_{\Delta L}\big|_{\text{max}}   \approx \frac{180}{\pi^3 g_\star}\frac{H(T)}{T}\bigg|_{\rm dec} 
\approx 10^{-10}\left(\frac{T_{\rm dec}}{10^9 \, \text{GeV}} \right)
\, .
\eea 
where in the second equality we have assumed radiation domination and the
number of degrees of freedom of the SM.
As mentioned, the maximum is obtained under the condition $T_{\rm ann} \sim T_{\rm dec}$.
The estimate above already implies that for successful leptogenesis one needs $T_{\rm ann} \sim T_{\rm dec} \gtrsim 10^9$ GeV.

In the DW leptogenesis scenario under study, the most important interactions are the decay and the inverse decay, $T_{\rm dec} \approx M_N/\text{few}$, where $\text{few} \sim \mathcal{O}(10)$.   This permits to write 
\bea
\label{eq:analytics_max}
Y_{\Delta L}\big|_{\text{max}} 
\approx
 10^{-10}\left(\frac{T_{\rm ann}}{10^9 \, \text{GeV}} \right)
\approx
10^{-10}
\left(\frac{M_N}{10^{10} \, \text{GeV}} \right)
\eea
This estimate is shown as a Blue dashed curve on the Right panel of Fig.\ref{fig: T_ann}.
We observe that the maximum lepton asymmetry produced in Eq.\eqref{eq:analytics_max} scales proportionally to $M_N$. Requiring that the produced asymmetry matches the observation in turn imposes a lower bound on $M_N \gtrsim 10^{10}$ GeV for successful leptogenesis.

\begin{figure}[h!]
    \centering
    \includegraphics[width=.5\linewidth]{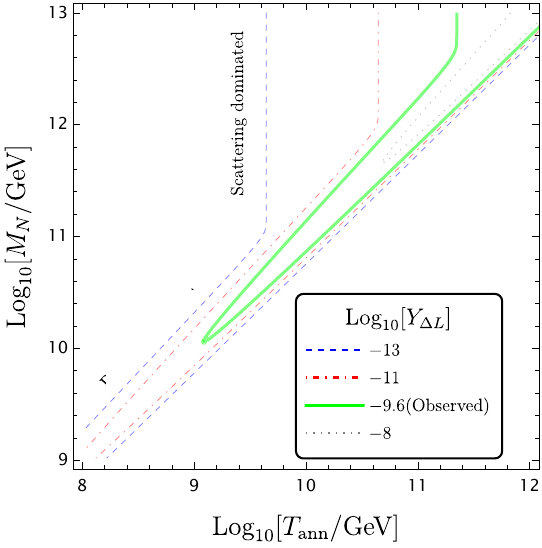}
    \caption{Contours of the final abundance $Y_{\Delta L}^{\rm final}$,  the lepton asymmetry in the parameter space $T_{\rm ann} \text{ vs } M_N$. The Green contour illustrates the region matching the observed asymmetry today. The region which is dominated by the scattering $LH \to H^c L^c$ interactions and by the decay are also illustrated. }
    \label{fig: succ_lep}
\end{figure}

To summarize our findings,
on Fig.\ref{fig: succ_lep} we present contours
of the lepton asymmetry abundance
in the $T_{\rm ann} - M_{N}$ plane.
The Green curve represents the region matching the observed abundance of baryons.
The ankle in the contour of equal asymmetry separates the region dominated by the decay contribution and the region dominated by the scattering operator. We observe, as expected from the Right panel of Fig.\ref{fig: T_ann}, that the scattering dominated regime is a vertical line and does not depend on the mass of the RHN $M_N$. 

Before concluding this section, we would like to make a further general remark.
As we will observe in the following examples, the upper limit of Eq.\eqref{eq:maximal_asymmetry} is a generic result independent on the baryon/lepton number violating interaction which is responsible for the DW-genesis mechanism. 
Indeed, in order to avoid strong wash-out from the very same interaction which is biased by the DW passage, the annihilation temperature which maximise the final asymmetry should be approximately the decoupling temperature of such interaction, which by definition is when $\gamma_X/n^{\rm{eq}} \sim H(T_{\rm{dec}})$. We can hence already anticipate that, independently on the specificity of the interaction, the DW-genesis mechanism can work for baryon/lepton violating interactions which decouple approximately at
$T \gtrsim 2.5 \times 10^{9}$ GeV.

\section{A model of domain wall cogenesis}
\label{sec:spont_cogenesis}

In the former section, we studied the minimal realisation of DW-catalized leptogenesis, where the chemical potential, induced by the presence of the DW, biases the $L$-violating interactions in the see-saw Lagrangian.  This leptogenesis scenario did not require any CP violation in the RHN decay. 

\begin{figure}[t]
    \centering
    \includegraphics[scale=0.8]{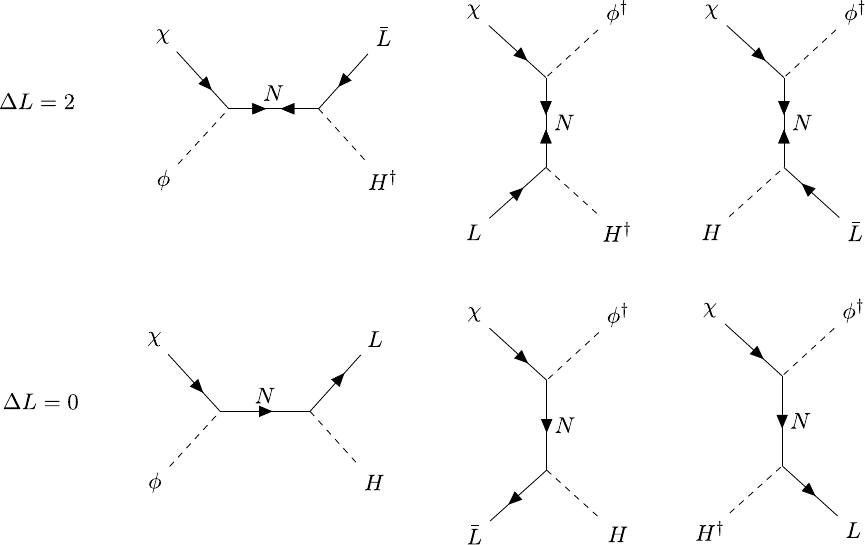}
    \caption{Feynmann diagrams of the $\Delta L=2$ and $\Delta L=0$ interactions for the processes 
    that share the asymmetry between the visible and the dark sector in the cogenesis scenario.
    }
    \label{fig:Feynmann_diag}
\end{figure}

In this section we present a model of cogenesis 
\cite{PhysRevLett.68.741}
of visible matter and dark matter asymmetry 
that employs the presence and the annihilation of the DWs.
 Models of cogenesis\cite{Cohen:2010kn, Shelton:2010ta,March-Russell:2011ang,Graesser:2011wi, Bell:2011tn, Co:2020xlh}, combined with asymmetric dark matter\cite{BARR1990387,PhysRevLett.68.741,Kaplan:2009ag,Zurek:2013wia,Petraki:2013wwa, Cheung:2011if}, offer the interesting avenue to address the \emph{coincidence problem}, $\Omega_{\rm DM} \sim 5 \Omega_{B}$. 
 
 In our setting, the CP \emph{conserving} RHN decay and $2 \to 2$ scatterings are biased by the DWs and produce the lepton number in the visible and in a new dark sector containing the DM candidate.\footnote{See instead \cite{Falkowski:2011xh,  Arina:2011cu,Arina:2012fb,Falkowski:2017uya,Ciscar-Monsalvatje:2023zkk, Herrero-Garcia:2024tyh} for models where cogenesis occurs through the CP-violating decay of the RHN.}.
 This is a realisation of \textit{spontaneous cogenesis}. 
Models of spontaneous cogenesis have already been proposed in previous works. 
 Different from the scenarios typically considered in the literature\cite{March-Russell:2011ang,Falkowski:2011xh,  Arina:2011cu,Arina:2012fb,Kamada:2012ht,Nagata:2016knk,Falkowski:2017uya,Biswas:2018sib,  Narendra:2018vfw,Ciscar-Monsalvatje:2023zkk,Mahapatra:2023dbr,Borah:2024wos, Bodas:2024idn, Chun:2023eqc,Chun:2024gvp}, for the situation we describe in this paper, the sharing operators are typically out-of-equilibrium
and the source  $\dot a$ is only active during a very short moment in time, i.e. when the DW passes.

To introduce the idea, we start with the following  minimal Lagrangian realisation 
\begin{align}
\label{Eq:Lag_for_cogenesis}
    \mathcal{L}\supset 
     y  (\tilde H \bar L)N_R+ \frac{1}{2}M_N \bar{N_R^c}N_R 
     + yR  (\phi^\dagger \bar \chi)N_R
     + m_{\chi}\bar \chi \chi + h.c.\, ,
\end{align}
 which is the same model as in Eq.\eqref{eq:L_violating}, with the addition of a dark sector made of a scalar field $\phi$ and a Dirac fermion DM candidate $\chi$.
 The possibility for $\chi$ to be a viable DM candidate is discussed in Section \ref{sec;cog_DM}.  We furthermore defined $R \equiv y_D/y$ as the ratio between the coupling of $\chi$ and $L$ to the RHN $N_R$. The $L$-number assignments are $ L(\chi) =1 $ and $L(N_R) = 1$. 
We also assume that fast interactions convert $\phi$ to $\phi^{\dagger}$ such that the asymmetry cannot be stored in the dark scalar sector.

We assume that the axion couples with the SM lepton and the dark matter via derivative 
 interactions:
\bea 
\mathcal{L}_{a-j} = c_L\frac{\partial_\mu a}{f_a} j^\mu_L + c_\chi\frac{\partial_\mu a}{f_a} j^\mu_\chi \, . 
\eea 
The passage of the axion DW can hence bias processes such as the RHN decay involving leptons as well as dark matter in the final state,
and also $2\to 2$ scatterings mediated by the RHN.
Depending on the coupling ratio $R$
and on the axion couplings $c_L$ and $c_{\chi}$, the DW annihilation can create a lepton and/or a DM asymmetry. In addition, a transfer of the asymmetry from the visible to the dark sector can occur if the relevant sharing interaction is active after the passage of the DW.
The resulting visible and dark asymmetry abundance can then be obtained by solving a coupled system of Boltzmann equations that we will now describe.

First of all, to get an intuition on the relevant processes involved in the DW cogenesis scenario, let us first focus on temperatures $T \ll M_N/10$, where the RHN can be safely integrated out. 
In this regime, the effective Lagrangian contains the following dimension-5 $\Delta L = 2$ operators

\bea 
\mathcal{L}^{\rm IR}_{\Delta L = 2}\supset 
y^2R\mathcal{O}^{\Delta L =2}_{(HL)(\phi \chi)} + y^2\mathcal{O}^{\Delta L =2}_{(HL)(HL)}  + y^2 R^2\mathcal{O}^{\Delta L =2}_{(\phi \chi)(\phi \chi)}  + h.c.
\eea 
where we have factorised the dependence on the coupling and $\mathcal{O}$ represent the effective operator after integrating out the heavy RHN. 
The effect of the coupling with the axion on these interactions can be understood by performing the axion dependent rotation of the lepton and the dark matter,
\begin{align} 
\label{eq:rotation}
 L \to e^{i c_L a/f_a} L \, , \qquad \chi \to e^{i c_\chi a/f_a} \chi \, ,
\end{align}
obtaining the non-derivative axion interactions
\footnote{As previously, the rotations also induce a non-derivative interaction of the axion to the SM lepton Yukawa coupling and possibly anomaly-type interactions with the gauge bosons, which will not play any role in the following. }
 
\bea 
\label{Eq:Lag_for_cogenesis_IR}
 \mathcal{L}^{\rm IR}_{\Delta L = 2}\supset 
    y^2 Re^{i (c_L+c_\chi) a/f_a}\mathcal{O}^{\Delta L =2}_{(HL)(\phi \chi)} + y^2e^{2i c_L a/f_a}\mathcal{O}^{\Delta L =2}_{(HL)(HL)} + y^2 R^2 e^{2i c_\chi a/f_a}\mathcal{O}^{\Delta L =2}_{(\phi \chi)(\phi \chi)} + h.c. \, .
\eea  
Note that the first of these operators is sharing the asymmetry between the visible and the dark sector
(the corresponding processes are depicted in Fig. \ref{fig:Feynmann_diag}).

On the top of these operators, there are however also $\Delta L=0$ interactions which do not violate the lepton number.
%
 The rates for the different operators $\Delta L =2$  and $\Delta L = 0$ have been computed in\cite{Falkowski:2011xh, Falkowski:2017uya}. 
 One important interaction of the second type is the sharing interaction via $\Gamma^{\Delta L = 0}_{(HL)(\phi \chi)} $ which are presented on Fig.\ref{fig:Feynmann_diag}, which we will keep in our analysis. There are also $\Delta L = 0$ interactions involving only the visible or the dark sector 
 which 
 will instead not play a role in the computation of the asymmetry.

To move to the Boltzmann equations involving the aforementioned interactions, we define the distribution functions with the chemical potential $\xi_{L, \chi}$
\bea 
f_\chi = e^{-E_\chi/T + \xi_\chi} \, , \qquad f_L = e^{-E_L/T + \xi_L} \,, \qquad f_{\chi^c} = e^{-E_\chi/T -\xi_\chi} \, , \qquad f_{L^c} = e^{-E_L/T -\xi_L} \, . 
\eea 
We  will also assume that the particles in the dark sector interact with each other through weak scale interactions and the constraint from thermalisation is the same as the one for the visible sector. 
Following a procedure similar to the one presented in Section \ref{sec:spont_baryo}, we obtain the set of equations (see Fig.\ref{fig:Feynmann_diag} for the diagrams involved)
\begin{align} 
\label{eq:full_cogenesis}
\frac{dY_{\Delta L}}{dt} & = - 
 2\text{Br}_{N \to LH}^2 \Gamma^{\Delta L = 2}_{(HL)(LH)} \bigg( Y_{\Delta L} + Y^{\rm eq}_{L}(t)\bigg)
 - \text{Br}_{N \to LH}\text{Br}_{N \to \phi \chi}\Gamma^{\Delta L = 2}_{(HL)(\phi \chi)} \bigg( Y_{\Delta L} +  Y_{\Delta \chi} + Y^{\rm eq}_{L+\chi}(t)\bigg)
 \notag
 \\
 &-   \text{Br}_{N \to LH}\text{Br}_{N \to \phi \chi}\Gamma^{\Delta L = 0}_{(HL)(\phi \chi)} \bigg( Y_{\Delta L} -  Y_{\Delta \chi} + Y^{\rm eq}_{L-\chi}(t)\bigg) \, ,
  \notag
 \\ 
 \frac{dY_{\Delta \chi}}{dt}& =
- 
 2 \text{Br}_{N \to \phi \chi}^2\Gamma_{(\phi \chi)(\phi \chi)}^{\Delta L = 2} \bigg( Y_{\Delta \chi} + Y^{\rm eq}_{\chi}(t)\bigg) 
 -  \text{Br}_{N \to LH}\text{Br}_{N \to \phi \chi} \Gamma_{(HL)(\phi \chi)}^{\Delta L = 2} \bigg( Y_{\Delta L}+ Y_{\Delta \chi} + Y^{\rm eq}_{L+\chi}(t)\bigg) 
 \notag
 \\
&  -   \text{Br}_{N \to LH}\text{Br}_{N \to \phi \chi}\Gamma^{\Delta L = 0}_{(HL)(\phi \chi)} \bigg( -Y_{\Delta L} +  Y_{\Delta \chi} - Y^{\rm eq}_{L-\chi}(t)\bigg) 
 \, ,
\end{align} 
where all the rates are the unsubtracted ones computed in the Appendix \ref{app:un_sub} and presented in \cite{Falkowski:2017uya}. 
We also defined 
\bea 
\text{Br}_{N \to \phi \chi} = \frac{R^2}{1+R^2},\qquad   \text{Br}_{N \to HL} = \frac{1}{1+R^2}\qquad \Gamma^{D}_{\rm tot} = \frac{y^2(1+R^2)M_N}{16\pi}. 
\eea

The expression of $Y^{\rm eq}_{...}(t)$ depends on the source induced by the coupling with the axion background,
\bea 
\label{eqn:Yeqsource}
Y^{\rm eq}_{\chi}(t) = \frac{n_{L}}{s} \frac{2c_\chi\dot a}{f_a T} \, , \quad Y^{\rm eq}_{L}(t) = \frac{n_{L}}{s} \frac{2c_L\dot a}{f_a T} \, , \quad Y^{\rm eq}_{L-\chi}(t) = 2\frac{n_{\chi}}{s} \frac{(c_L-c_\chi)\dot a}{f_a T} \, ,  \quad Y^{\rm eq}_{L+\chi}(t) = 2\frac{n_{\chi}}{s} \frac{(c_L+c_\chi)\dot a}{f_a T}\,.
\eea 
Several comments are in order concerning the equations in Eq.\eqref{eq:full_cogenesis}. First, one might notice the absence of the rate corresponding to the RHN decay. For compactness, we reabsorbed this piece in the $2 \to 2$ rates and followed the ``unsubtracted'' scheme as described in Appendix \ref{app:un_sub}. Secondly, as one can expect intuitively, if $c_L = c_\chi$, the source of the $\Delta L =0$ term vanishes, since the source pumps the same amount in each sector and does not bias the interaction.
In this case, the sharing 
operator will try to balance the asymmetry in the two sectors as it can \emph{transfer} one asymmetry abundance into the other.
Several consistency conditions related to these equations are discussed in Appendix \ref{app:un_sub}. 
We will now study the phenomenology of this model.

\subsection{Phenomenology of the DW cogenesis}
\label{sec:cogenesis_pheno}

Let us now study the phenomenology of the cogenesis model by solving numerically the two coupled equations in Eq.\eqref{eq:full_cogenesis}. 
We report here on the main results and features which emerge from our numerical analysis. 
We refer to the Appendix \ref{app:un_sub} 
and \ref{app:moreplots} for more details about the rates and the evolution of the yield.

\begin{figure}[t!]
\centering
\includegraphics[width=.42\linewidth]{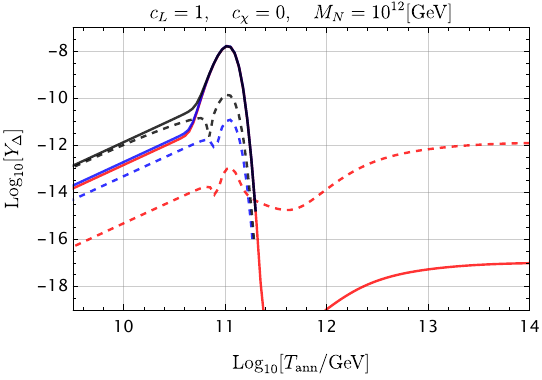}
\includegraphics[width=.42\linewidth]{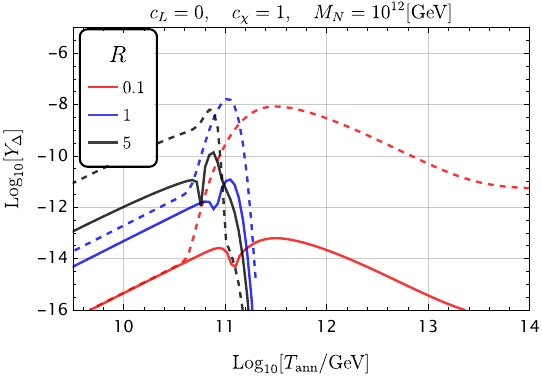}
    \caption{The final values of the asymmetries in the $L$ (solid) and $\chi$ (dashed) sectors for different values of $R= 0.1, 1, 5$ denoted by Red, Blue and Black lines, respectively. \textbf{Left panel}: The production in the SM  $c_L=1, c_\chi =0$ (SM active). \textbf{Right panel}: The production in the DS sector $c_L=0, c_\chi =1$ (DS active). Notice that Log$_{10}\left[|Y_{\Delta}^{\rm final}|\right]$ is plotted, so the sign of the asymmetry is not available on this plot.}
    \label{fig:TrajCog_On}
\end{figure}

In  Fig.\ref{fig:TrajCog_On} we display the evolution of $Y_{\Delta L, \Delta \chi}$ (solid and dashed lines) as a function of $T_{\rm ann}$  for different values of $R$. We display $R = 0.1$ (Red), $R = 1$ (Blue) and $R = 5$ (Black), assuming $c_L =1, c_\chi=0$ (SM active) on the Left panel \footnote{From now on, conventionally, we will call the sector with $c=1$, \emph{active} and the sector with $c=0$ passive.}, i.e. the source only couples to the leptons and not to $\chi$. The opposite case of $c_L =0, c_\chi=1$ (DS active) is displayed on the Right panel. 

A first important feature appearing is the dependence of the asymmetries on the $R$ parameter. This dependence varies across different temperature regimes, being $T_{\rm ann} \gg, \ll, \sim M_N/10$.  The different characteristics of the curves in Fig.\ref{fig:TrajCog_On} 
 are further elucidated in Appendix \ref{app:moreplots},
 while here we summarize the more important findings in the different regimes:
\begin{itemize}
\item $T_{\rm ann} \ll M_N/10$: In this case, the production comes from the $2 \to 2$ interaction with weak wash-out. If the DS is active (Right panel of Fig.\ref{fig:TrajCog_On}), we observe that  
$Y_{\Delta  L} \propto R^2$, because $Y_{\Delta  L}$ is produced from the off-shell sharing $\Gamma_{(HL)(\phi\chi)}$, while $Y_{\Delta  \chi} \propto R^4$ 
for $R\gg 1$, as the asymmetry in the DS is produced mainly via $\Gamma_{(\phi\chi)(\phi\chi)}\gg \Gamma_{(HL)(\phi\chi)} $.
In the opposite regime $R\ll 1$, $Y_{\Delta  \chi} \sim Y_{\Delta  L}\propto R^2$ since the sharing operator will be the dominant source of asymmetry. 
If the SM sector is active, one has that for large $R$ the sharing operator is effective and the same asymmetry is created in both sectors. In this case $Y_{\Delta L}\sim Y_{\Delta \chi}\propto R^2$. For low $R$, the asymmetry will be much less in the DS since it is proportional to $R^2$, while the SM asymmetry will be independent on $R$
(for small enough $R$) because the operator $\Gamma_{(HL)(HL)}$ will dominate.

\item At the position of the peak $T_{\rm ann} \sim M_N/10$, one has for the SM-active case that the lepton asymmetry is independent of $R$. This is because, as we show in Eq.\eqref{eq:BEs_sub_bis} of Appendix \ref{app:un_sub}, on the resonance the BEs of the SM and the DS decouple and evolve independently. This means that the passive sector is always only produced by the \emph{off-shell} piece of the rates. This incidentally explains why there is no sharp production peak at $T_{\rm ann} \sim M_N/10$ for the passive sector and the asymmetry scales as $R^2$.\footnote{It may appear that there is an observable peak, however, this observation is attributed to a change in the sign of the asymmetry.} A more precise estimate of the scaling can be made by considering the BE of the DS: for $R \ll 1$, one has the following relation
\bea 
\frac{dY_{\Delta \chi}}{dT} \approx  -R^2\frac{Y_{\Delta L}}{HT} \bigg(\Gamma^{\text{sub},\Delta L = 2}_{(HL)(\phi \chi)} - \Gamma^{\text{sub},\Delta L =0}_{(HL)(\phi \chi)}\bigg)  \, , 
\eea 
where the subscript ``$\text{sub}$'' emphasizes the fact that the rate only contains the \emph{off-shell} piece. We furthermore assume $Y_{\Delta L}$ to be constant throughout the evolution. 
The equation above can be approximately solved in the following way,
\begin{align}
Y^{\rm final}_{\Delta \chi} &\approx Y_{\Delta L}R^2\int dT \frac{\Gamma^{\text{sub},\Delta L = 2}_{(HL)(\phi \chi)} - \Gamma^{\text{sub},\Delta L =0}_{(HL)(\phi \chi)} }{HT}
\sim Y_{\Delta L}
R^2\frac{m^\dagger_{\nu}m_\nu}{2 \pi^3 v_{\rm EW}^4} \frac{M_{\rm pl} T_{\rm ann}}{3.4} \, .
\end{align} 
Notice that there might be a further suppression coming from a cancellation between the $\Delta L =0$ and the $\Delta L=2$ rates. 
Plugging in numerical values, one obtains 
\bea 
\label{eq:approximation_Ychi}
\frac{Y^{\rm final}_{\Delta \chi} }{Y_{\Delta L}}\approx 3\times 10^{-3} \times 
R^2 \times \left( \frac{T_{\rm ann}}{10^{11} \, \text{GeV}} \right) \, \qquad \qquad \text{(SM active)} \, .
\eea 
This reproduces the difference between the size of the peak on Fig.\ref{fig:TrajCog_On}.

\begin{figure}[t!]
\centering
\includegraphics[width=.42\linewidth]{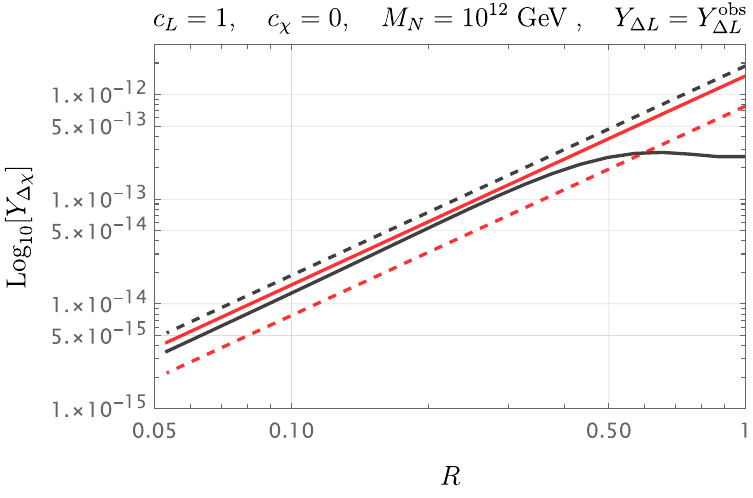}
    \caption{ Abundance of the DM $Y_{\Delta \chi}$ imposing that $Y_{\Delta L}$ matches the observed baryon abundance. The Red and Black lines represent two solutions to the equation $Y_{\Delta L}=Y^{\rm obs}_{\Delta L}$, with the Black line corresponding to the solution at larger temperatures. The solid line indicates the numerical value, while the dashed line provides the numerical approximation from Eq.\eqref{eq:approximation_Ychi}, with $M_N= 10^{12}$ GeV. }
    \label{fig:YchiEx}
\end{figure}

The estimate in Eq.\eqref{eq:approximation_Ychi} is shown against the numerical computation in Fig.\ref{fig:YchiEx} where we impose $Y_{\Delta L}= Y^{\rm obs}_{\Delta L}$ in order to fix $T_{\rm ann}$. Notice that there are two solutions satisfying $Y_{\Delta L}= Y^{\rm obs}_{\Delta L}$, which we show in Red and Black, where the latter corresponds to the solution at higher $T_{\rm ann}$. We observe that the approximation in Eq.\eqref{eq:approximation_Ychi} is precise within a factor of few. Moreover, one notices that the solution corresponding to the higher temperature displays an ankle for high $R$. This is because of the fact that the asymmetry will be washed out due to the $L$-violating operator in the dark sector becoming efficient.

In the case where the DS is active (Right panel) the RHN decay processes implies a peak in the DS asymmetry around $T_{\rm ann } \approx M_N/10$
(and an accompanying one in the lepton asymmetry).
One can also notice 
that in this case the position of the peak depends on $R$. Indeed, for lower $R$, the decay will be sooner out of equilibrium. As a consequence, the peak will shift to higher values of $T_{\rm ann}$. Furthermore, the broadening of the peak at low $R$ occurs because the decay process remains out of equilibrium, rendering the exponential suppression due to washout negligible.

\item At $T_{\rm ann} \gg M_N/10$, the wash-out is strong. In the absence of a DS, the asymmetry in the visible sector would be extremely suppressed, due to in-equilibrium wash-outs, as we observed in Section \ref{sec:spont_lepto}, namely in Fig.\ref{fig: T_ann}. We will now see that, in the presence of a weakly coupled DS, some asymmetry could be hidden in the DS and then leaked back in the SM via weak sharing operators, which will create a non-vanishing asymmetry in the visible sector. We call this process the \emph{``rescuing mechanism''}. 

\begin{figure}[h!]
    \centering
    \includegraphics[width=.4\linewidth]{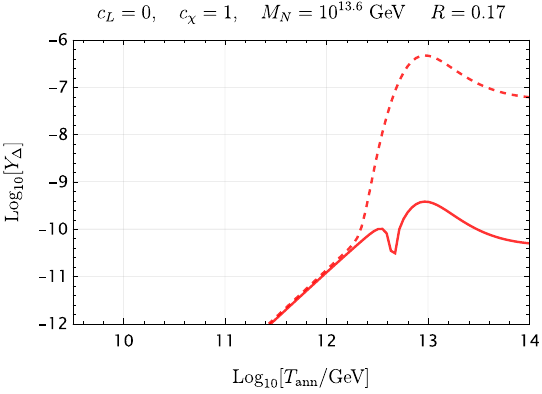}
    \includegraphics[width=.42\linewidth]{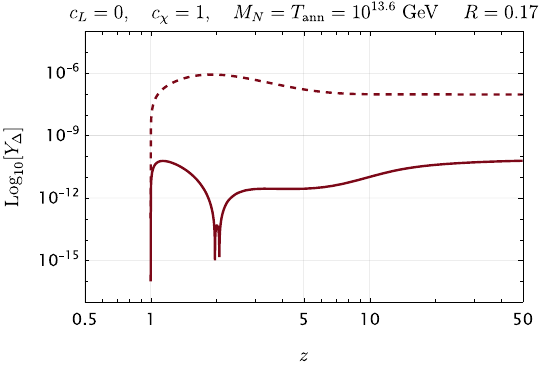}
    
    \caption{\textbf{Left panel}: Example of the rescuing mechanism producing the observed abundance of baryons at $T_{\rm ann} \gg M_N$. \textbf{Right panel}: Trajectory of the $Y_{\Delta L}$ (solid line) and $Y_{\Delta \chi}$ (dashed) line for $R = 0.1 $ in the specific case of the rescuing mechanism. The asymmetry in the DS $Y_{\Delta \chi}$ remains roughly constant while the $Y_{\Delta L}$ is washed out. The sharing operators then feed back $Y_{\Delta L}$ and this can set the observed abundance of baryons.}
    \label{fig:saving}
\end{figure}

This process can be understood from Fig.\ref{fig:saving}, where a final asymmetry is apparent in both sectors.
At large $T_{\rm ann} \gg M_N$ and for $R \ll 1$, the system of equations after production take the form
\begin{align} 
\label{eq:full_cogenesis_3}
\frac{dY_{\Delta L}}{dt}  \approx  -  
 2 \Gamma^{\Delta L = 2}_{(HL)(LH)}   Y_{\Delta L}  +  R^2  Y_{\Delta \chi}\bigg( \Gamma^{\Delta L = 0}_{(HL)(\phi \chi)} -  \Gamma^{\Delta L = 2}_{(HL)(\phi \chi)} \bigg) \qquad  
 \frac{dY^{\chi}_{\Delta \chi}}{dt} \approx 0  \, .
\end{align} 
 where we kept only the leading operators.
The asymmetry in the SM is however strongly washed out when the $HL\to H^cL^c$ recouples, while the asymmetry in the DS remains mostly constant. 
After the SM wash-out decouples again at $T_{\rm dec} \sim M_N/10$, the abundance in the SM freezes out to a value 
approximately
\bea
\label{eq:saving_scaling}
\frac{Y^{\rm final}_{\Delta L} }{Y_{\Delta \chi}}  \sim 3\times 10^{-3}  \times R^2 \times  \left( \frac{T_{\rm dec}}{10^{11} \,\text{GeV}} \right)  
\qquad \qquad\text{(rescuing mechanism)}
\eea 
This scales like $R^2$ as long as the wash-out in the DS \emph{never} reaches equilibrium.
 This relation can be verified on Fig.\ref{fig:TrajCog_On}. 
In general, for the rescuing mechanism to be operative, we need the following conditions to be fulfilled: i) there is a non-zero initial asymmetry in the DS, ii) the wash-outs in the DS never enter in equilibrium, iii) there is a weak sharing operator between the DS and the SM. 

\end{itemize}

\begin{figure}[h!]
    \centering
    \includegraphics[width=.45\linewidth]{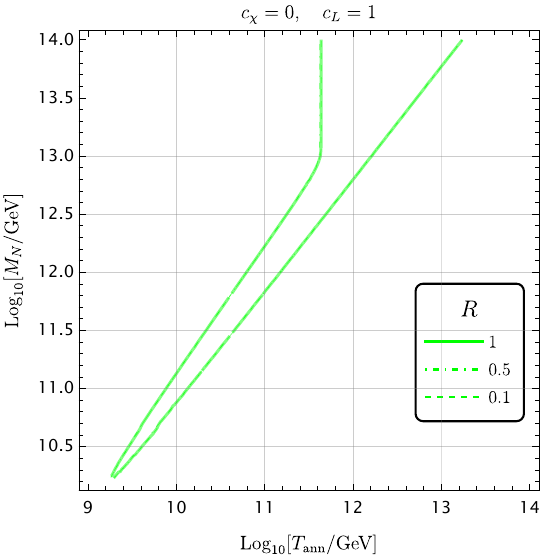}
    \includegraphics[width=.45\linewidth]{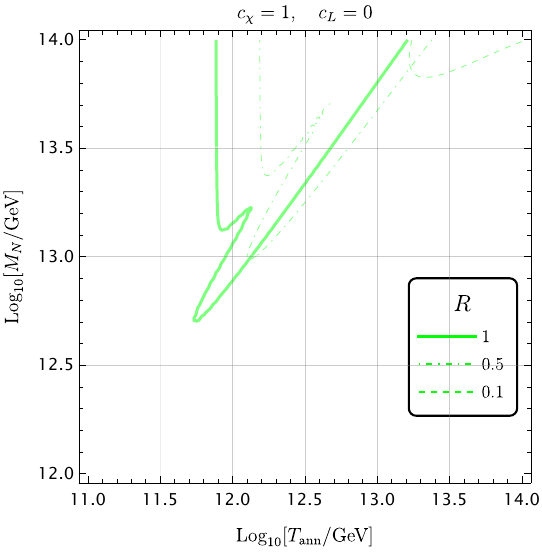}
    
    \caption{\textbf{Left panel}: Parameter space for the observed baryon asymmetry for  $c_L =1, c_\chi =0$. \textbf{Right panel}: Same for $c_L =0, c_\chi =1$. Notice that the corner of larger $M_N$ and $T_{\rm ann}$ also offers a region fitting the observed value for the case of $c_\chi =1, c_L= 0$, which is the \emph{island of the rescuing mechanism}, as discussed in the main text.   }
    \label{fig:cogenesis_para_space}
\end{figure}

On Fig.\ref{fig:cogenesis_para_space}, we display the parameter space where $Y_{\Delta L}$ can fulfill the observed abundance of baryons in the plane $T_{\rm ann}-M_N$ for different values of $R=y_D/y$. We observe that, when the source is coupled to the SM (Left panel), the parameter space is very similar to the usual DW leptogenesis model discussed in Section \ref{sec:spont_lepto} (see Fig.\ref{fig: succ_lep} for comparison). Intuitively, the parameter space is only mildly dependent on $R=y_D/y$. 

However, when the source is coupled to the DS (Right panel of Fig.\ref{fig:cogenesis_para_space}), the parameter space considerably shrinks and becomes strongly dependent on $R$. The lepton asymmetry decreases quickly with smaller $R$. In particular,  for $M_N \gtrsim 5 \times 10^{13}$ GeV, $T_{\rm ann} > M_N/10$  and $R \lesssim 0.2 $, we observe a region matching the observed abundance in the upper corner. We can refer to this region as the  ``island of the rescuing mechanism'' since it is due to this process, described around Eq.\eqref{eq:full_cogenesis_3} and Eq.\eqref{eq:saving_scaling}. 
If we increase $R \gtrsim 0.2$, the wash-out in the DS and the sharing recouples so that the abundance in the DS is then washed away, and the rescuing mechanism is not operative anymore.

In summary, we have found that there is an intricate interplay between the two sector asymmetries, depending on the ratio of the yukawa couplings $R \equiv y_D/y$, on the annihilation temperature $T_{\rm ann}/(10 M_N)$,
and on which sector is actually coupled to the CP violating source.
Two important conclusions are the following:

First, the final asymmetry in the DS and in the SM could differ by several orders of magnitude, opening the door to an asymmetric DM candidate with a mass very different than the traditional $\sim 5$ GeV (see next subsection for a concrete realisation).

Second, there is an interesting possibility of generating the SM asymmetry by a \emph{rescuing} mechanism. In this case the active source is coupled only to the DS, but the asymmetry is then transmitted to the SM via a \emph{sharing} interaction. 
This leads to the 
possibility to match the observed abundance of leptons in an unexpected corner of the parameter space: $R \sim 0.1, M_N \sim 10^{13.5} $ GeV$, T_{\rm ann} \sim 10^{14}$ GeV.

\subsection{Cogenesis and the mass of the asymmetric dark matter}
\label{sec;cog_DM}
So far we showed how to produce the abundance in the lepton and dark sector. As we can observe, remarkably, the abundance in the DS can be largely different from the abundance in the visible sector, contrarily to the common expectation that cogenesis would induce naturally $Y_{\Delta L} \approx Y_{\Delta \chi}$. This allows $m_\chi$, the mass of the candidate DM, to depart largely from $\sim 1$ GeV, as we now discuss.

We would like to refine our previous scenario in such a way that the asymmetric DS abundance accounts for the observed DM abundance. 
In this setting, the $\chi$ field is charged under lepton number and constitutes a realisation of \emph{asymmetric} DM (see \cite{Petraki:2013wwa, Davoudiasl:2012uw, Zurek:2013wia} for reviews and original papers). For this purpose, once an asymmetry has been produced in the dark sector, efficient interactions should annihilate the symmetric component, leaving only the asymmetric one. 

The annihilation of the symmetric component can be achieved through various possibilities. We here sketch one viable scenario.
For this purpose, 
with respect to the Lagrangian in Eq.\eqref{Eq:Lag_for_cogenesis}, we introduce a new scalar field $\tilde \phi$, mixing with the SM Higgs and generalise the interaction terms as 
\begin{align}
\label{eq:model_DM}
    \mathcal{L}\supset 
     y  (\tilde H \bar L)N_R+ \frac{1}{2} M_N \bar{N_R^c}N_R 
     + y_{\rm D}  (\phi \bar \chi) N_R
     + \frac{1}{2} m_{\chi} \bar \chi \chi 
     + B \tilde \phi \phi \phi +y_{\phi} \tilde \phi \bar \chi \chi 
     + \lambda |H|^2  \tilde \phi^2 +
     h.c.\, ,
\end{align}
where $B$ is a dimensionful coupling.
This Lagrangian contains a $Z_2$ symmetry which acts on $\phi$ and $\chi$,
while $\tilde \phi$ is even under it.
The dark sector (in the limit $y_D \to 0$) has an independent global $U(1)_D$ symmetry, under which $\chi$ is charged, but both $\phi$ and $\tilde \phi$ are neutral. A minimal realisation consistent with these assignments would be to consider the scalar $\phi$ to be real.
The sharing operator (i.e. when $y_{D} \neq 0$) implies that this $U(1)_D$ is identified with the lepton number.
$\tilde \phi$  mediates interactions which dilute the symmetric part of the 
particles odd under the $Z_2$ symmetry
through 2$\to$2 processes like $\phi \phi, \bar \chi \chi \to \tilde \phi \tilde \phi$, which imply a freeze-out of the symmetric part of the dark particles $\phi, \chi$ toward the $\tilde \phi$. In order for this to be efficient in removing the freeze-out symmetric abundance the coupling $y_{\phi}$ and $ B$ should be large,
and the masses of $\phi$, $\chi$, $\tilde \phi$ should be of similar order.
We consider the case in which $m_{\tilde \phi} < 2 m_{\chi}$
and $m_{\tilde \phi} < 2 m_{\phi}$ so that the new scalar only decays through the portal coupling with the Higgs, without repopulating the symmetric part of the DS (once $\tilde  \phi$ acquires a vacuum expectation value (vev)).
This decay should occur before BBN, but at the same time the $\tilde \phi$-Higgs mixing should be small enough to avoid constraints from direct detection.

The same model has been recently considered in \cite{Borah:2024wos} (there the CP violation is induced conventionally through an imaginary component in the RHN yukawas).
The viable parameter space for such a model, where the symmetric part is underabundant and the constraints coming from direct detection and late decay of $\tilde\phi$ are satisfied,
finding an upper bound of the DM mass as
$m_\chi \lesssim 500$ GeV.
We however emphasize that other realisations of the DS could open this parameter space.

Let us now study the abundance of the asymmetric DM component today. Since the DM energy fraction is observed, it leaves us with one free parameter,  the mass of the DM candidate $m_\chi$. This mass can be related to the DM fraction in the following way (again assuming that the symmetric fraction fully annihilates and that the asymmetric $\chi$ component constitutes all the DM)
\bea 
\Omega^{\rm today}_{\chi}h^2 = 0.12\bigg(\frac{m_\chi}{0.4 \text{GeV}} \bigg) \bigg(\frac{Y_{\Delta \chi}}{10^{-9}} \bigg)  \qquad \Rightarrow \qquad \frac{m_\chi}{\text{GeV}}  = \frac{4 \times 10^{-10}}{Y_{\Delta \chi}} \, , 
\eea 
In our scenario, the asymmetry abundance and its relation with the lepton asymmetry has been explored in Section \ref{sec:cogenesis_pheno}.
Among the various options, we consider here as a benchmark the case in which the visible sector is the only active one ($c_L =1$ and $c_{\chi}=0$), 
and the annihilation temperature is close to the most optimal case with $T_{\rm ann} \sim M_N/10$.
In this case, the DS abundance can be related to the lepton asymmetry with the analytic estimate in Eq.\eqref{eq:approximation_Ychi}, 
as
\bea
\frac{m_{\chi}}{\rm GeV} \approx 
50 \times R^{-2} \left(\frac{2.7 \times 10^{-10}}{Y_{\Delta L}} \right)
\left(
\frac{10^{13}\,  \rm{GeV}}{M_N}
\right)
\, . 
\eea

This provides a direct relation between the ratio of the two yukawa couplings, 
the RHN mass,
and the mass of the DM. 
As we see, depending on the value of the parameters of the model, the DM mass can span various orders of magnitude. 
As mentioned before, in this estimate it is assumed that
the entire DM abundance is made of \emph{asymmetric} DM.

\section{A model of domain wall baryogenesis}
\label{sec:spont_baryogenesis}

In the former sections, we studied in details DW leptogenesis and cogenesis via leptogenesis, in this section we switch gears and study an explicit model of \emph{baryogenesis}. In this context, the baryon number is violated and we consider the following dimension 9 operator 
\bea 
\label{eq:baryon_viol}
\mathcal{L}_{\slashed{B}}=\frac{\overline{u^cd^cd^c}udd}{\Lambda^5}+ h.c.
\eea 
which violates the baryon number by $\Delta B = 2$, while the source, which is coupled to the baryon quark current, is given by 
\bea 
\label{eq:quark coupling}
\mathcal{L}_{a-B} = 
\frac{c_q}{ f_a} \partial_{\mu} a j_{Q}^\mu , \qquad j_{Q}^\mu \equiv \frac{1}{3}\sum_{q = u,d} \bar q \gamma^{\mu} q \, .
\eea 
For the simplicity of the analysis, we assume that the operator in Eq.\eqref{eq:baryon_viol} only involves one of the SM family. 
Notice that in this case, we do not propose an explicit UV completion of the $B-$violating operator and we consider only the mechanism from the effective operator in Eq.\eqref{eq:baryon_viol}. Following the same steps as in Section \ref{sec:spont_baryo} and in agreement with\cite{Arbuzova:2016qfh}, the relevant Boltzmann equations are given by (at leading order in the expansion of the asymmetry)
\bea 
\frac{dY_{\Delta B}}{dt} = - 2 \Gamma_{\Delta B = 2} \bigg( Y_{\Delta B} + Y^{\rm eq}_{\Delta B}(t)\bigg) \, , \qquad Y^{\rm eq}_{\Delta B}(t) \equiv \frac{n_{q}}{s} \frac{2c_q\dot a}{f_a T}, \qquad Y_{\Delta B}   \approx 2 \frac{n_{q}}{s}  \xi_B \, ,
\eea 
where $Y_{\Delta B}$ is the baryon asymmetry, 
and we defined the baryon chemical potential via
\bea 
\qquad f_q = e^{-E/T + \xi_B}, \qquad f_{\bar q} = e^{-E/T - \xi_B} \, .
\eea 
As in the previous sections, one can estimate the asymmetry with an analytical expression similar to Eq.\eqref{eq:final_analytics}:
\beq
\label{eq:analytic_prod_bar}
Y_{\Delta L} = 8 \pi c_{\bar q} \frac{n^{\rm eq}_q}{s} \frac{\Gamma_{\Delta B=2}}{T}\Big|_{T=T_{\rm ann}} e^{-\beta} \,  ,\qquad 
\beta = \int_{t_{\rm ann}}^{\infty} 2\Gamma_{\Delta B =2} dt = \int_{0}^{T_{\rm ann}} \frac{2\Gamma_{\Delta B =2}}{HT} dT   \, , 
\eeq
which gives a result consistent up to 20\% with the numerical result. From now one, we will set $c_{q} =1$.

\begin{figure}[h!]
    \centering
\includegraphics[width=.56\linewidth]{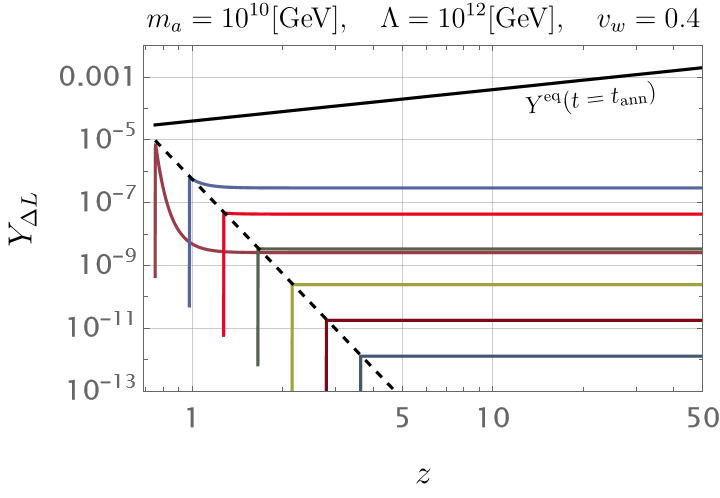}\includegraphics[width=.35\linewidth]{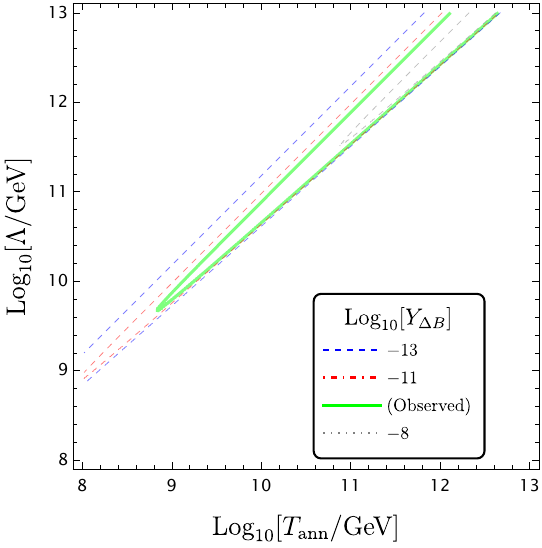}
    \caption{\textbf{Left panel}: Same figure as the Left panel of Fig.\ref{fig: T_ann}: trajectories of the asymmetry for different values of $T_{\rm ann}$, where $\Lambda = 10^{12}$ and $ m_a = 10^{10}$ GeV. \textbf{Right panel}: Value of the final baryon yield in the plane $T_{\rm ann}- \Lambda$. The Green contour is the region of the parameter space allowing for successful spontaneous leptogenesis. We observe that the tip of the Green curve lies at $ \Lambda \sim 10^{9.5}$ GeV (and $T_{\rm ann} \sim 10^9$\,GeV). }
    \label{fig: succ_baryo}
\end{figure}

Taking the phase space factors and dofs into account, we can easily estimate the rate of the baryon violating interaction,
\bea 
\Gamma_{\Delta B = 2} \sim 
\frac{T^{11}}{\Lambda^{10}}  \, ,
\eea 
which decouples at a temperature of approximately
$ T_{\rm dec} \approx \left(\Lambda^{10}/M_{\rm pl}\right)^{1/9} $.
Following the sames steps as in Section \ref{sec:para_space_spon_lepto}, one can estimate the maximal asymmetry 
\bea 
Y_{\Delta B} \Big|_{\rm max}  \approx   0.2 \, \bigg(\frac{\Lambda}{M_{\rm pl}}\bigg)^{10/9} \,  \qquad \Rightarrow \qquad \Lambda_{\rm min} \sim 10^{10} \text{ GeV.}
\eea 
where the lower bound on $\Lambda$ for the successful baryogenesis is obtained by setting $\text{Log}_{10}[Y^{\rm observed}_{\Delta B}] \approx -10 $. On the Right panel of Fig.\ref{fig: succ_baryo}, we present the value of the final baryon asymmetry in the parameter space $\Lambda - T_{\rm ann}$. One can first observe that this approximately confirms the lower bound on $\Lambda \sim 10^{10}$ GeV, 
corresponding to a lower bound on $T_{\rm ann}$ of approximately $10^{9}$ GeV.

The baryon number violating process that we used in our model, see Eq.\eqref{eq:baryon_viol}, violates the baryon number  by two units, so proton decay is not allowed, but $n-\bar n$ oscillations are present\cite{Fridell:2021gag}. The latter will be an unavoidable signature of the scenario we proposed in this section. The operator in Eq.\eqref{eq:baryon_viol} induces a neutron mixing mass of the form
\bea 
\delta m_{\bar{n}-n} \sim \frac{\Lambda_{QCD}^6}{\Lambda^5}.
\eea
 Current bounds on this mixing mass are of order $ \delta m_{\bar{n}-n}  \lesssim 10^{-33} $ GeV~\cite{Buchoff:2012bm,Syritsyn:2016ijx}, which places a bound on the typical mass scale of
$ 
\label{nnbar}
\Lambda\gtrsim 10^{6} \, {\rm GeV}
$, far from the value required for a successful baryogenesis.
As a consequence, it is unlikely that the scenario presented in this section can be testable by neutron oscillations in coming experiments~\cite{Phillips:2014fgb, Frost:2016qzt, Hewes:2017xtr}. 

\section{A model of DW-genesis using the sphalerons}
\label{Sec:with_spha}

As we have seen in the previous section, baryogenesis can occur whenever an axionic DW is sweeping through the plasma (at the moment of collapse, when the velocity has a specific direction) and, at the same time, an $L$- or $B$-violating interaction is close to decoupling. Actually, this is the case close to the EWPT cross-over at $T_{\rm EWPT} \approx 162 $ GeV, where the sphalerons are strongly active before the transition and quickly decouple after, at $T^{\rm spha}_{\rm dec} \approx 135 $ GeV\cite{Hong:2023zrf,PhysRevLett.113.141602} (see \cite{Nayak:2000pf} for a previous study involving the QCD sphaleron). In this section, we study the baryon number produced by a DW-network coupling to the baryon current and collapsing around  $T\approx T_{\rm EWPT}$.

\begin{figure}[h!]
    \centering
    \includegraphics[width=.5\linewidth]{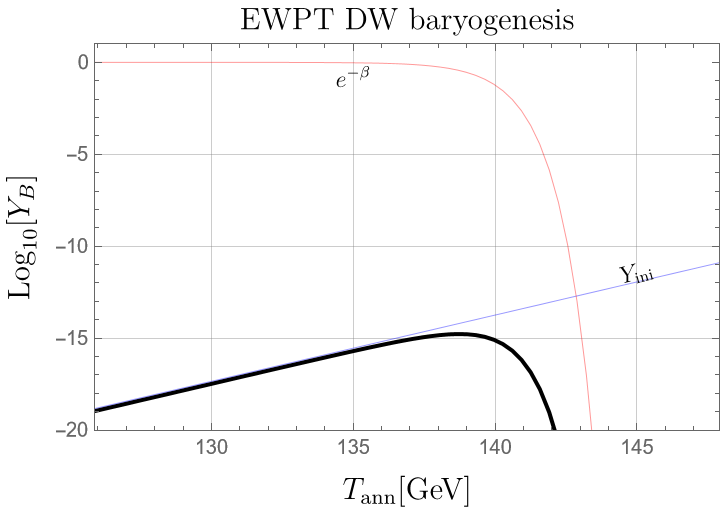}
    \caption{Yield of the baryogenesis in the scenario where the baryogenesis is catalized by DW walls at the EW phase transition. The black line is the total baryon yield, while the red shows the wash-out suppression and the blue the initial production in the form $Y_{\rm final} \approx Y_{\rm ini} e^{-\beta}$. }
    \label{fig: T_ann_shpa}
\end{figure}

The sphaleron is a 12-fermion process, violating the $B+L$ number, such that  $\Delta B = \Delta L = 3$.
The coupling between the quark current and the axion is taken to be
\bea 
\label{eq:quark coupling}
\mathcal{L}_{a-B} = 
\frac{\tilde c_q}{ f_a} \partial_{\mu} a j_{Q,5}^\mu , \qquad j_{Q}^\mu \equiv \frac{1}{3}\sum_{q = u,d, s,c, b, t} \bar q \gamma^{\mu}  q \, .
\eea 
The sphaleron rate at the EW cross-over is a sharp function of time which has been numerically simulated and is approximated by\cite{Hong:2023zrf,PhysRevLett.113.141602} 
\bea 
\Gamma_{\rm spha}(T) \approx 
\begin{cases}
8\times 10^{-7}\times T \qquad \text{if} \qquad T> T_{\rm EWPT}
    \\
   T e^{- 147.7 + 0.83 \times\frac{T}{\text{GeV}} } \qquad \text{if} \qquad T< T_{\rm EWPT}
\end{cases} \, , 
\eea 
where $T_{\rm EWPT} \approx 162$ GeV. The Boltzmann equation
controlling the generation of the baryon asymmetry and its further evolution is given by 
(the prefactor can be found in\cite{Anber:2015yca})
\begin{align}
     \frac{dY_{\Delta (L+B)}}{dt}&=-2N_c\Gamma_w \left(Y_{\Delta (L+B)} + Y^{\rm eq }_{\Delta B}(t)\right)\, ,  \qquad  Y^{\rm eq}_{\Delta B}(t)  \equiv \frac{n_{q}}{s} \frac{3\tilde c_q\dot a}{f_a T}  \, ,
\end{align}
where $N_c=3$ is the number of colors of QCD. Using the same methods as presented in the previous sections, we can approximate the produced baryon and lepton number in following way 
\beq
Y_{\Delta (B+ L)} = 6 N_c \pi c_{\bar q} \frac{n^{\rm eq}_q}{s} \frac{\Gamma_{\rm spha}}{T}\Big|_{T=T_{\rm ann}} e^{-\beta} \qquad  {\rm with}\quad 
\beta = \int_{t_{\rm ann}}^{\infty} 3\Gamma_{\rm spha} dt = \int_{0}^{T_{\rm ann}} \frac{3\Gamma_{\rm spha}}{HT} dT   \, .
\eeq
Fig.\ref{fig: T_ann_shpa} presents the baryon yield $Y_{\Delta B}$ as a function of the annihilation temperature of the DW. We however observe that the yield peaks around $T^{\rm peak}_{\rm ann} \approx 138$ GeV at a value of $Y_{\Delta B} \sim 2 \times 10^{-15}$, which is a few orders of magnitude below the observed baryon abundance. This could have been anticipated by the general upper bound 
derived in Section \ref{sec:para_space_spon_lepto}, which in this case implies
\bea 
Y_{\Delta (B+ L)}^{\rm max} = 6 N_c \pi c_{\bar q} \frac{n^{\rm eq}_q}{s} \frac{H(T)}{T}\Big|_{T=T_{\rm dec}} \approx \frac{T_{\rm dec}}{ M_{\rm pl}} \sim  10^{-16} \, ,
\eea 
which is far too small to account for the observed baryon asymmetry. Consequently, we do not pursue the study of this model further.

\section{DW dynamics, collapse and gravitational waves}
\label{sec:dynamics_DW_GW}

In the former sections, we studied several realisations of DW-genesis. We will now examine the features of the axion potential that facilitates this mechanism, along with the gravitational signatures it may produce.
Our findings are summarised in Fig.\ref{fig:ALPpara}, \ref{fig:ALPparaLim} and \ref{Fig:supp}, where the viable parameter space of the DW leptogenesis model is depicted, together with the detectability through a GW signal.
In the following we will explain the various constraints and contours which are illustrated in these plots.

\subsection{Parameter space in collapsing ALP DW}

We remind that 
the potential for the axion is given by
\bea \label{eq: axion potential with bias}
V(a) =  m_a^2 f_a^2\bigg(1- \cos \frac{ a N_{\rm DW}}{v_a}\bigg) + V_{\rm bias} (a), 
\qquad V_0 \equiv m_a^2 f_a^2 \,
\eea 
where we defined $V_0$ as the magnitude of the scalar potential breaking $U(1) \to \mathbb{Z}_{N_{\rm DW}}$ and leading to the DW formation.
The second term in Eq.\eqref{eq: axion potential with bias} is a bias that will induce a small difference in the vacuum energy of the otherwise degenerate vacua, of order $\Delta V$, resulting in the DW collapse.
A convenient parametrization for the bias is to consider the typical displacement of the axion vev which is induced by $V_{\rm bias} (a)$, in units of the axion decay constant,
which is
\footnote{In the case of the QCD axion, this would correspond to the contribution to the QCD $\theta$ term induced by the bias, see e.g. \cite{ZambujalFerreira:2021cte}.}
\bea 
\label{eq:displacement}
\Delta \theta \equiv \frac{\Delta a}{f_a N_{\rm DW}} \approx   
\frac{\Delta V}{V_0} 
\eea 
The displacement $\Delta \theta$ is also clearly a measure of the quality of the $U(1)$ ALP symmetry.
One can then relate $\Delta \theta$ to the annihilation temperature and the mass of the axion,
employing Eq.\eqref{eq:T_ann}.
With this parametrization, we can now explore the viable $f_a-m_a$ parameter space, for fixed values of $\Delta \theta$,
by imposing several consistency conditions and necessary conditions 
for successful DW-genesis.

\begin{figure}[t!]
\centering
\includegraphics[width=.46\linewidth]{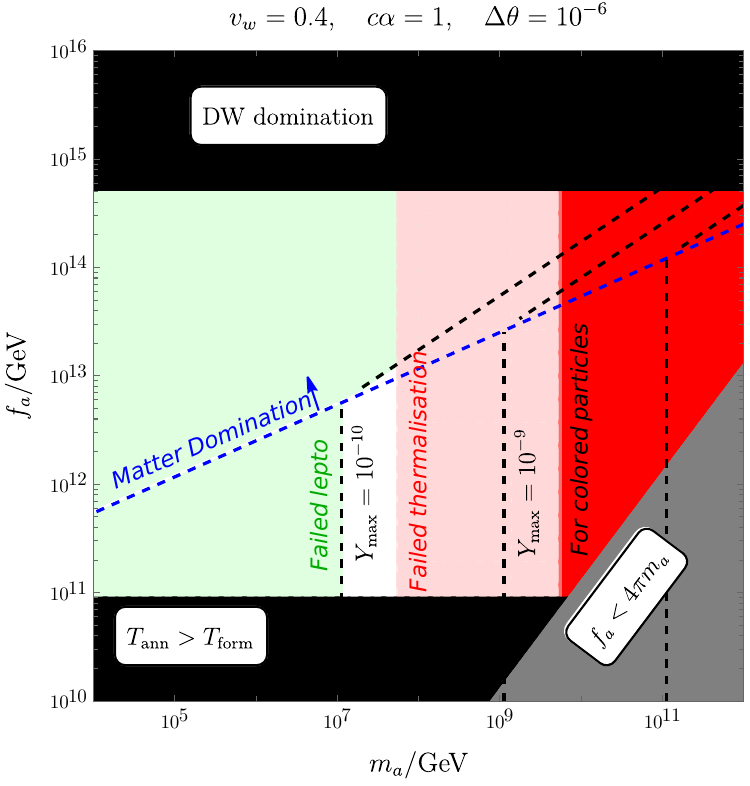}
\includegraphics[width=.48\linewidth]{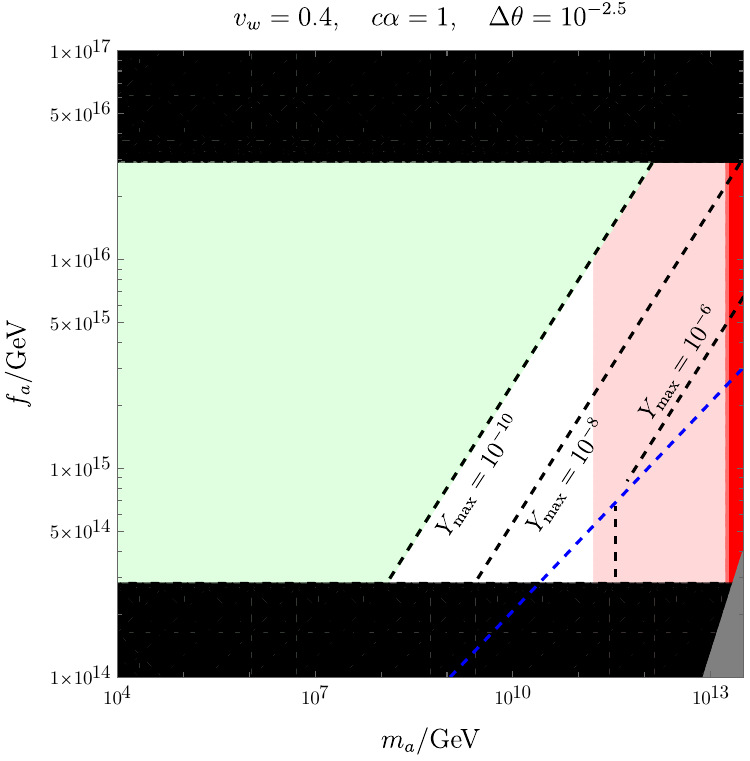}
    \caption{
    \textbf{Left panel}: Plot of the parameter space available for DW leptogenesis and detectability by the GW observer ET for $\Delta \theta = 10^{-6}$. The Black and Gray regions are inconsistent and violate the conditions in Eq.\eqref{eq:reqiALP}. The Green region does not allow possible leptogenesis because the yield is always too small to match the observed baryon number, and finally the Red regions exclude the parameter space where the leptonic plasma cannot thermalise in the DW and violates Eq.\eqref{eq:reqiALP_3}. The dashed Blue curves show the region where the axions emitted by the DWs enter in matter domination, inducing a dilution of the baryon yield, where we set $c \alpha \simeq 1$ in the ALP decay width for concreteness. The viable region for leptogenesis is displayed in White. We emphasize that no region of this plot is observable via a GW signal from the Einstein Telescope, as we will discuss in Section \ref{sec: Observational signatures and emitted gravitational waves}. The maximal value of the asymmetry can be read from the dashed Black contours. \textbf{Right panel}: Same plot for $\Delta \theta = 10^{-2.5}$.
    }
    \label{fig:ALPpara}
\end{figure}

\paragraph{Consistency conditions for our analysis}
In order to have successful DW-genesis and a viable ALP DW network we  need to require first that
\beq
\label{eq:reqiALP}
f_a > 4 \pi m_a 
\qquad
T_{\rm ann} > T_{\rm dom}  \qquad  T_{\rm form} > T_{\rm ann} \qquad (\text{consistency conditions}) \; ,
\eeq
 The regions where they
are not satisfied are displayed in Black and Grey in the plot of Fig.\ref{fig:ALPpara}. 

Secondly, 
we have seen that we need roughly $T_{\rm ann} \gtrsim 10^9$ GeV to have successful leptogenesis. The region where this is not satisfied is displayed in Green in Fig.\ref{fig:ALPpara}, and is labeled as \emph{failed leptogenesis}, considering the DW-leptogenesis as the benchmark scenario. 
The same constraint is approximately valid also for the baryogenesis model,
as discussed in Section \ref{sec:spont_baryogenesis}. 
In addition, contours of the maximal lepton asymmetry given in Eq.\eqref{eq:analytics_max} are shown with dashed Black lines.

 Furthermore, we add also the requirement of the thermalisation of the plasma inside the DW,
 which for particles interacting with the weak force is
 (the weak fine-structure constant is given by  $\alpha_w \approx 0.033$)
\beq
\label{eq:reqiALP_3}
\mathcal{O}(10)\alpha_w^2\gtrsim\frac{\gamma_w v_wm_a}{T_{\rm ann}} \qquad T_{\rm ann} \gg 10^2m_a \gamma_w v_w \quad \text{(thermalisation condition)}
\eeq 
On the other hand, for colored particles, as in the baryogenesis model, one finds
\beq
\label{eq:reqiALP_3}
\mathcal{O}(10)\alpha_s^2\gtrsim\frac{\gamma_w v_wm_a}{T_{\rm ann}} \qquad T_{\rm ann} \gg 10 m_a \gamma_w v_w \quad \text{(thermalisation condition for colored particles)}
\eeq
with $\alpha_s \approx 0.1$. 
We show in Red in Fig.\ref{fig:ALPpara}
the regions which do not satisfy this requirement (the light Red region assumes that scatterings are mediated by weak interactions, as in the leptogenesis case, while the darker Red case assumes that scatterings are mediated by gluons, as expected in the baryogenesis model).
Note that given that in the scans of Fig.\ref{fig:ALPpara}.
we are
fixing $\Delta \theta$,
the annihilation temperature 
becomes a function of $m_a$ scaling as
$T_{\rm ann} \sim \sqrt{ \Delta \theta m_a M_{\rm Pl} }$.
As a consequence, the thermalization conditions result in an upper bound on $m_a$ (and consequently on $T_{\rm ann}$).

\paragraph{Dilution due to matter domination from axions}
As DWs emit axions during their lifetime, there is the possibility that the axion energy density will dominate the universe. Assuming most of the DW energy goes into axion production such that $\rho_a \simeq \rho_\text{DW}$, the temperature at which the axions dominate the universe, $T_\text{mat}$, is given by the condition
\begin{equation}
    \rho_\text{DW}(T_\text{ann}) \left(\frac{g_*(T_\text{mat})}{g_*(T_\text{ann})}\right)\left(\frac{T_\text{mat}}{T_\text{ann}}\right)^3 = 3 H(T_\text{mat})^2 M_\text{pl}^2\,,
\end{equation}
where we assumed the axion population decays as matter after DW annihilation. This condition results in
\begin{equation}
    T_\text{mat} = \frac{30}{\pi^2 g_*(T_\text{ann})}\frac{\Delta V}{T_\text{ann}^3}\,,
\end{equation}
where we used the annihilation condition $\rho_\text{DW}(T_\text{ann}) = \Delta V$. This matter dominated epoch can however be avoided if the axions decay fast enough to photons and gluons, for which the decay rates are given by (see e.g. \cite{Mariotti:2017vtv,CidVidal:2018blh, Bauer:2017ris, Bauer:2020jbp,ParticleDataGroup:2020ssz}) 
\begin{equation}
    \Gamma_{a\rightarrow \gamma\gamma} = \frac{\alpha_\text{em}^2}{64\pi^3}\frac{m_a^3}{f_a^2}\left(\frac{E}{N_\text{DW}}\right)^2\,, \quad \Gamma_{a\rightarrow gg} = \frac{\alpha_\text{s}^2}{8\pi^3}\frac{m_a^3}{f_a^2}\left(\frac{N_\text{QCD}}{N_\text{DW}}\right)^2\,,
\end{equation}
where $E$ and $N_\text{QCD}$ are the electromagnetic and color anomaly coefficients respectively. The temperature at which the decay becomes efficient, i.e. when $\Gamma = H$, is given by
\begin{equation}
    T_\text{alp dec} = 0.04 \frac{c\ \alpha}{g_*(T_\text{alp dec})^{1/4}}\frac{\sqrt{m_a^3 M_\text{pl}}}{f_a}\,,
\end{equation}
where $\alpha = \alpha_\text{em}$ ($\alpha_\text{s}$)  and $c = E/N_\text{DW}$ ($\sqrt{8} N_\text{QCD}/N_\text{DW}$) for the decay to photons (gluons).

As soon as $T_\text{mat} < T_\text{alp dec}$, the matter dominated epoch is avoided, which leads to the condition
\begin{equation}
    \frac{f_a^3}{m_a M_\text{pl}^2} < 2.7\times 10^{-3} \left(\frac{g_*(T_\text{ann})}{g_*(T_\text{alp dec})}\right)^{1/4} c\ \alpha \ \Delta\theta^{1/2}\,,
\end{equation}
where we used Eq.\eqref{eq:displacement} and $V_0 = m_a^2 f_a^2$ to rewrite the bias $\Delta V$ in terms of the displacement $\Delta\theta$.
The region in the parameter space where the axion induces a period of matter domination is the portion above the dashed blue line in Fig.\ref{fig:ALPpara}.

In the matter domination regime, one needs to take into account the dilution from the creation of entropy due to the decay of the axion. We can define the dilution factor as the ratio of comoving entropy $S= s a^3$ after and before the decays.  
Following \cite{Scherrer:1984fd, Cirelli:2018iax, Nemevsek:2022anh}, the diluted leptonic yield is 
\begin{equation}
\label{eq:dilution}
    Y_{\Delta L} = Y^0_{\Delta L}\times D\,,\quad \text{with} \quad D = \text{Min}\left[1, 0.57 \frac{g_*(T_\text{ann})}{g_*(T_\text{alp dec})^{1/4}}\frac{\sqrt{M_\text{Pl}\Gamma}\ T_\text{ann}^3}{\Delta V}\right]\,,
\end{equation}
where $Y^0_{\Delta L}$ denotes the leptonic yield as computed in previous sections, without considering an intermediate matter dominated regime.
The inclusion of such a dilution factor in the estimate of the maximal asymmetry explains why the contours lines of $Y_{\rm max}$ in Fig.\ref{fig:ALPpara} bend above the dashed Blue line.

\paragraph{Summary on viable region for successful DW-genesis}
We finally conclude that the viable region where DW leptogenesis is possible is the one remaining in White in Fig.\ref{fig:ALPpara},
where we considered two representative values for 
$\Delta \theta$: $\Delta \theta = 10^{-6}, 10^{-2.5}$.
During this analysis, we observed that DW leptogenesis is only possible for $\Delta \theta \gtrsim 10^{-6.5}$. 
Below this value, the requirement of thermalization inside the DW is not compatible with a sufficiently large asymmetry generation.
In the case of DW-baryogenesis, where the strong interaction sets the thermalization condition,
the corresponding bound is relaxed to $\Delta \theta \gtrsim 10^{-7.3}$. 

 Consequently, since a successful solution to the strong CP problem would require $\Delta \theta \lesssim  10^{-10}$,  we conclude that the axion responsible for the DW-genesis cannot be a \emph{heavy QCD axion}.

\subsection{Observational signatures and emitted gravitational waves}
\label{sec: Observational signatures and emitted gravitational waves}

During the scaling regime of the DW network, gravitational waves are produced by the oscillations of the DWs in the network \cite{Saikawa:2017hiv,Hiramatsu:2010yz,Hiramatsu:2012sc,Hiramatsu:2013qaa} 
(see also \cite{Kitajima:2023kzu,Kitajima:2023cek,Ferreira:2023jbu,Chang:2023rll,Li:2023gil} 
for recent works). 
This emission finishes when the network collapses at $T_{\rm ann}$. The largest emission occurs at the annihilation temperature,
since the energy density of the DW network keeps on increasing, compared to the radiation one, until annihilation \cite{Hiramatsu:2013qaa}.

The gravitational wave signal is  parameterized by the energy fraction 
    \bea 
    \Omega_{\rm GW}(t, f) = \frac{1}{\rho_c(t)} \bigg(\frac{d \rho_{\rm GW}(t)}{d \ln f}\bigg) \, ,
    \eea 
where $\rho_c(t) = 3 M_{\rm pl}^2 H^2$ is the critical density of the universe at time $t$. The GW spectrum can be described by a power broken law with 
\bea 
\Omega_{\rm GW}(t, f) = \Omega^{\rm peak}_{\rm GW} (t) \times 
\begin{cases}
(f/f_{\rm peak})^{3} \qquad f < f_{\rm peak}
\\
(f/f_{\rm peak})^{-1}  \qquad f > f_{\rm peak}
\end{cases} \, ,
\eea 
where $f_{\rm peak}$ is the frequency at which the GW signal is maximal, and
the peak contribution is given by 
\bea 
\Omega_{\rm GW}^\text{peak}(t) = \frac{\tilde\epsilon_{\rm GW} G \mathcal{A}^2 \sigma^2}{\rho_c(t)} \, , 
\eea 
with $\mathcal{A} = 0.7$ and  $\tilde\epsilon_{\rm GW} = 0.7$ taken from numerical simulations performed in \cite{Hiramatsu:2013qaa}. 
After redshifing to today, the amplitude of the GW signal reads
\bea 
\label{eq:GWfromDW}
\Omega_{\rm GW}(T)\approx 2.34 \times 10^{-6} \tilde \epsilon_{\rm GW} \mathcal{A}^2\bigg(\frac{g_\star (T)}{10}\bigg) \bigg(\frac{g_{s\star} (T)}{10}\bigg)^{-4/3} \bigg(\frac{T_{\rm dom}}{T}\bigg)^4 \text{Min}\left[1, \bigg(\frac{T_\text{alp dec}}{T_{\text{mat dom}}}\bigg)^{4/3}\right] \, ,
\\
\nonumber
f_{\rm peak}(T) \approx 1.15 \times 10^{-7} \text{Hz} \times \bigg(\frac{g_\star (T)}{10}\bigg)^{1/2} \bigg(\frac{g_{s\star} (T)}{10}\bigg)^{-1/3} \bigg(\frac{T}{\text{GeV}}\bigg)\text{Min}\left[1, \bigg(\frac{T_\text{alp dec}}{T_{\text{mat dom}}}\bigg)^{1/3}\right]  \, . 
\eea 
The temperature at which we have to consider the GW emission is the annihilation temperature $T = T_{\rm ann}$.
The last factor in Eq.\eqref{eq:GWfromDW} takes into account the dilution to the gravitational wave signal (and the shift in frequency) induced by a period of matter domination
(see e.g. \cite{ZambujalFerreira:2021cte}).

\begin{figure}[t!]
\centering
\includegraphics[width=.5\linewidth]{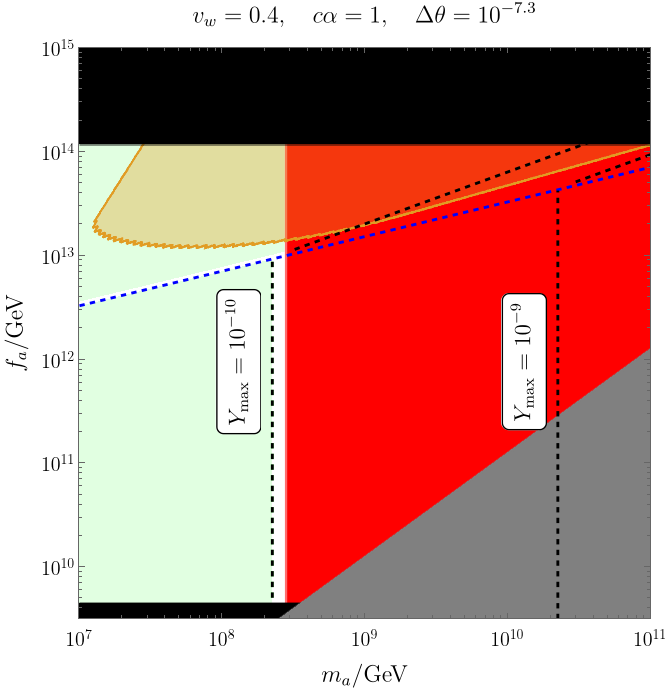}
    \caption{Same plot as in Fig.\ref{fig:ALPpara} for $\theta = 10^{-7.3}$ with a zoomed-in view of the region near the GW detection (in the Orange region). 
    }
    \label{fig:ALPparaLim}
\end{figure}

By inspection, we find that  
the GW signal emitted by the DW is strongly diluted by the matter domination 
due to axions 
and is eventually not observable at the Einstein telescope\cite{Moore:2014lga,Sathyaprakash:2012jk, Maggiore:2019uih}
in the parameter space of 
Fig.\ref{fig:ALPpara}. 
Interestingly, we also obtain that the GW signal from the DW collapse is only observable if the quality of the axion is $\Delta \theta \lesssim 10^{-7}$(similar correlation was already observed in
\cite{ZambujalFerreira:2021cte}). 
This is exemplified on Fig.\ref{fig:ALPparaLim}.

\subsection{Post-collapse average out of the asymmetry}

An important issue of the scenario under consideration is the averaging of the asymmetries after the collapse. This would come from the fact that the simultaneous collapse of different types of DWs might create asymmetries with opposite sign, which might cancel each other after the end of the collapse. 
We now describe in detail how this can lead to a suppression of the generated asymmetry and possible solutions.

We first remind that the sign of the asymmetry is determined by the change of $a/f_a$ across the DW. However, if we consider a specific $U(1) $ symmetry and $N_{\rm DW} = 2$ for simplicity, there exist in general two possible different DWs: i) $0 \to \pi$ and ii) $\pi \to 2\pi$. Assuming that the minimum in $\pi$ is favored by the bias, the collapse of the $0 \to \pi$ DW will induce a positive asymmetry while the collapse of $\pi \to 2\pi$ will induce a negative asymmetry, which will eventually cancel each other after diffusion of the baryon number. This conclusion extends naturally to any $N_{\rm DW}$ value.

Even in this case, 
we however do not expect an exact cancellation of the asymmetry, for the following reasons:    
\begin{itemize}
    \item {\textbf{Different annihilation temperatures}}: Sticking to the example above, the tension of the $0 \to \pi$ and $\pi \to 2\pi$ is naturally different due to the bias. Indeed, one can observe that $\Delta \sigma/ \sigma \sim \Delta V/V_0$, where $\Delta \sigma$ is the difference of tension between the two types of DWs. This leads to a different annihilation temperature for each type of DW so that $\Delta T_{\rm ann}/T_{\rm ann}  \sim \Delta V/V_0$. Since some of the DWs start to collapse earlier, they first 
    sweep through the plasma at different temperature, experiencing different wash-out effects, resulting in a favored sign of the asymmetry. 
    The suppression induced by the cancellation of the asymmetries is difficult to estimate precisely. 
    Taking into account the temperature dependence of the generated asymmetry on the annihilation temperature that we found in the numerical analysis of Section \ref{sec:numerics}, a small difference in annihilation temperatures will
imply a difference in the generated asymmetry of the order
\bea 
\text{Suppression} \equiv \frac{\Delta Y_{\Delta L}}{Y_{\Delta L}} \approx \mathcal{O}(1-10) \frac{\Delta V}{V_0} \,, 
\eea 
where here $\Delta Y_{\Delta L}$ denotes the difference of lepton yield from the two DW types. 
This points to a suppression of the asymmetry scaling linearly with $\Delta \theta$ (see Eq.\eqref{eq:displacement}). The dependence on $\Delta \theta$ of the final asymmetry is a reflection of the fact that the asymmetry would totally cancel in the case in which the $U(1)_{\rm PQ}$ is restored.

\item {\textbf{Different collapse velocities}}:  Due to different tensions, the velocity of the DWs will be typically different, resulting in a different amount of volume swept by the two DWs. 
One can compute this difference using the equation of motion of the DW which dictates the evolution of the boost factor
\bea 
\frac{d\gamma_w}{dz} \approx \frac{ \Delta V}{\sigma}  \qquad \Rightarrow \qquad  \frac{d\Delta\gamma_w}{dz} \approx -\frac{ \Delta V}{\sigma} \frac{\Delta \sigma}{ \sigma} \, , 
\eea 
where $\Delta \gamma_w$ is the difference in boost factor of each DW\footnote{Notice that pressure from particle reflection on the DW
\cite{Huang:1985tt,Blasi:2022ayo} and from the Chern-Simons term\cite{Ganoulis:1986rd,Huang:1985tt,Blasi:2023sej,Hassan:2024bvb} might limit the acceleration. We do not take this into account in our present estimate.} 
After the DW has travelled a distance of the order of Hubble distance, at DW annihilation, the difference in velocity has transformed into a difference in asymmetry
\bea 
\Delta \gamma^{\rm final}_w \approx \frac{\Delta V} {\sigma H_{\rm ann} }   \frac{\Delta \sigma}{\sigma} \approx \frac{\Delta \sigma}{\sigma}   \qquad \Rightarrow \frac{\Delta Y}{Y} \sim \frac{\Delta \sigma}{\sigma v_w^2} \sim \frac{\Delta V}{V_0 v_w^2} \sim \mathcal{O}(10) \frac{\Delta V}{V_0} \,, 
\eea 
where we have neglected $\mathcal{O}(1)$ numerical coefficients.

\end{itemize}
Both effects previously illustrated lead to a suppression factor proportional to $\Delta \theta$ in the asymmetry generated in DW leptogenesis. This  extra suppression has not been quantified in \cite{Daido:2015gqa}, but can significantly affect the conclusions.

We can quantify the impact of this suppression by considering the maximum 
value of the possible asymmetry in Eq.\eqref{eq:analytics_max}, and taking $T_{\rm dec} \sim T_{\rm ann} \sim M_N/10$, we obtain
\bea
\label{eq:analytics_max_suppressed}
Y_{\Delta L}\big|_{\text{max}} 
\approx
10^{-10}
\times 
\mathcal{O}(10)
\left(\frac{T_{\rm ann}}{10^{13} \, \text{GeV}} \right)
\left(\frac{\Delta \theta}{10^{-4}}\right) D
\eea
where 
$D$ is the dilution factor in Eq.\eqref{eq:dilution}, 
and we normalized $T_{\rm ann}$ around the highest possible value.
Indeed, assuming that the reheating temperature after inflation is around $T_R \sim 10^{14} - 10^{15}$ GeV, RHNs with such high masses can become thermal, and the annihilation temperature should be one order of magnitude smaller.
In conclusion, matching the observed abundance of baryons allows for $\Delta \theta  $ as low as 
$\sim 10^{-3}- 10^{-4}$.

\begin{figure}[t!]
\centering
\includegraphics[width=.46\linewidth]{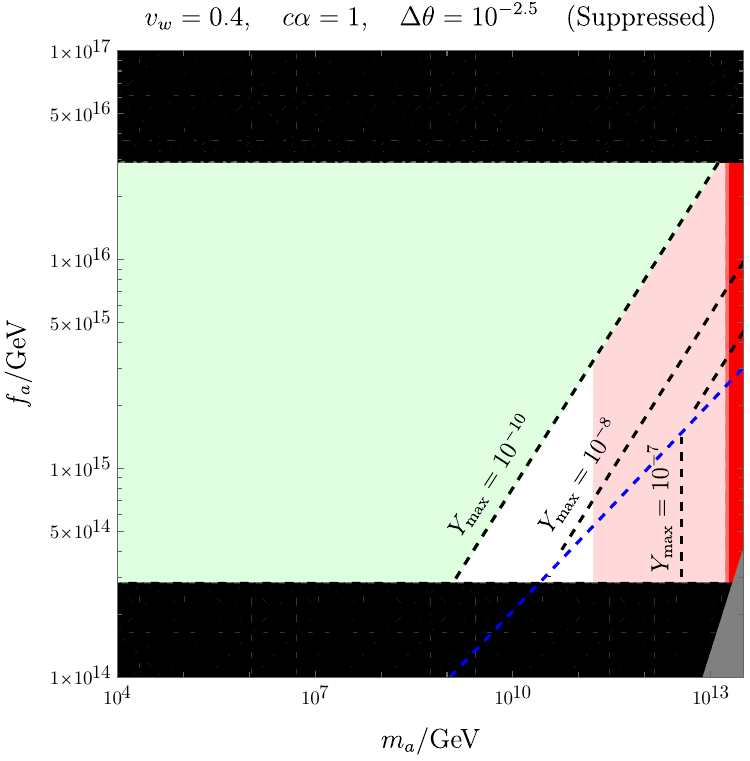}
\includegraphics[width=.45
\linewidth]{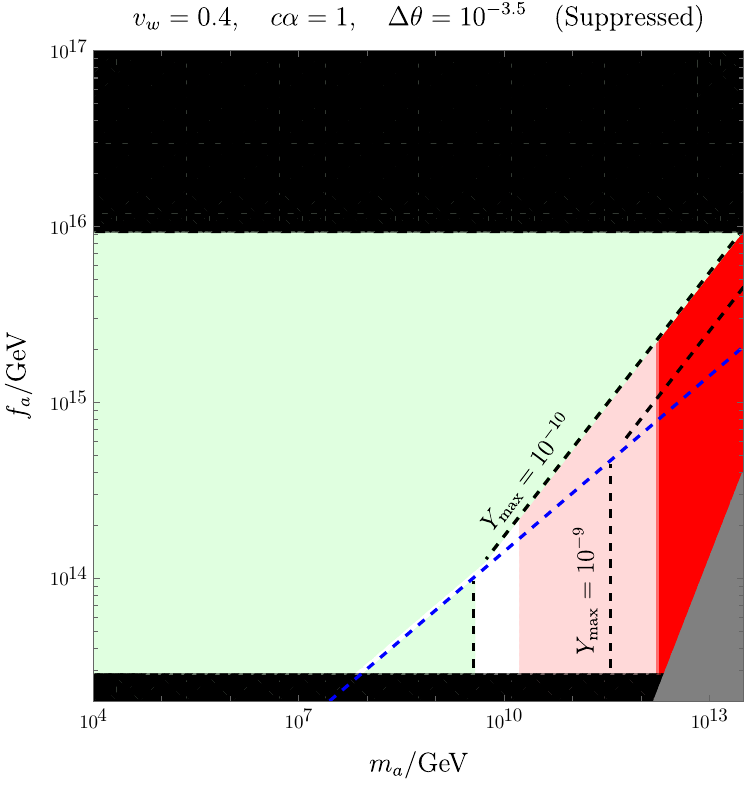}
    \caption{Similar Plots that in Fig.\ref{fig:ALPpara} \emph{now including the $\Delta \theta$ suppression} in Eq.\eqref{eq:analytics_max_suppressed} for $\Delta \theta = 10^{-2.5} $ and  $\Delta \theta = 10^{-3.5} $. 
    \label{Fig:supp}
    }
   \end{figure}

One can now wonder what is the typical numerical value of $\Delta \theta\approx\Delta V/V_0$ appearing in axion models. This turns out to be very model dependent. In particular it depends on the assumptions associated with the origin of the potential forming the axion domain walls and of the potential bias.
We can try to relate $\Delta \theta$ to other quantities characterizing the DW network dynamics under realistic assumptions.

In this paper, we considered the axion potential $V_0$ to be produced by a dynamical mechanism like a confining sector, with a temperature-dependent potential. 
This implies that the formation temperature of the DW is at $T_{\rm form} \sim V_0^{1/4}$.
In addition, we assumed that the bias is generated by a temperature-independent potential (a typical example is a Planck-suppressed operator).
In this case the 
annihilation temperature is given  by Eq.\eqref{eq:T_ann},
\be
T_{\rm ann}^2 \approx  \frac{ \Delta V M_{\rm Pl}}{24 \, m_a f_a^2} 
\qquad ~
\Rightarrow
\quad
\text{T-independent bias}
\ee
where we assumed the same number of DoF as in the SM.
This expression implies a relation between $\Delta \theta$ and the formation and annihilation temperatures as
\be
\Delta \theta \approx \frac{ 24 \, f_a}{ M_{\rm Pl}} \left( \frac{T_{\rm ann}}{T_{\rm form}} \right)^2
\qquad \Rightarrow
\quad
\text{T-independent bias}
\ee
The minimal requirement that $T_{\rm ann} < T_{\rm form}$, for a fixed $\Delta \theta$, indeed excludes the Black region in the bottom part of the Fig.\ref{fig:ALPpara}. 
Demanding $f_a$ to be $\lesssim$ than the maximal reheating temperature $\sim 10^{15}$GeV, 
one can expect the displacement to be of order $\Delta \theta \lesssim 10^{-2}$, corresponding  
to an annihilation temperature of $T_{\rm ann} \gtrsim 10^{11}$ GeV (see Eq.\eqref{eq:analytics_max_suppressed})
for successful leptogenesis. This would select only the regions with large 
$T_{\rm ann}$ as viable ones 
in Fig.\ref{fig:ALPpara}.
We eventually find out that the viable parameter space, taking into account the aforementioned $\Delta \theta$ suppression (and the matter domination dilution), is very limited in our scenario and the results are illustrated in Fig.\ref{Fig:supp}.

Alternatively, 
one can consider the possibility that 
the bias is induced by an explicit breaking term generated by another strong dynamics.
In this case, the potential is generated at temperatures of order $ \sim \Delta V^{1/4}$.
Following a similar line of reasoning as above, we find in this different scenario the scaling to be
\be
\Delta \theta \approx \left( \frac{T_{\rm ann}}{T_{\rm form}} \right)^4 \qquad
\qquad
\Rightarrow 
\quad
\text{T-dependent bias}  \; .
\ee
Finally, the formation of DWs could also be due to a T-independent potential, which we did not consider in this paper. 
This option could lead to even more interesting cases:
for instance, if in addition the bias is induced by a strong sector at the confining temperature, the axion displacement parameter $\Delta \theta$ could be essentially unsuppressed.
We leave this investigation for future work.

\medskip

All the previous conclusions extend naturally to the case of larger $N_{\rm DW}$. 
Alternatively, 
the suppression proportional to $\Delta \theta$ can be fully circumvented in other scenarios, possibly involving pre-inflationary conditions\footnote{Note that in pre-inflationary models of DW-genesis, further constraints could arise from isocurvature perturbations, as discussed in  \cite{Daido:2015gqa}.}, that we quickly sketch here:
\begin{itemize}
    \item The case in which there is a population bias 
    (see for example
\cite{Coulson:1995nv,Gonzalez:2022mcx,Kitajima:2023kzu,Larsson:1996sp,Correia:2014kqa,Correia:2018tty}), possibly coming from the dynamics during inflation, that favors one type of DW over the other. This can come from large $U(1)$ breaking terms only active at large field values\cite{Enqvist:2003gh} as it is required in  other baryogenesis mechanisms like \emph{axiogenesis} and Affleck-Dine baryogenesis which also need high energy $U(1)$ breaking.  In this case, one type of DW is favored, the axion domains do not span the entire $2 \pi$ range, and there is no suppression of the type described above. 
    \item In the context of a dynamical collapse during which one of the DW, say $0 \to \pi$, collapses long before the other, the one from $\pi \to 2\pi$. 
    This could for example be induced by the mixing of different axions and different biases with hierarchical scales. We leave a detailed analysis of this mechanism for future investigations.
    
    \item In multi-axion models, 
    a possibility 
    is that the axionic DW is induced by a mixing between two (or more) axions, see \cite{Lee:2024xjb,Lee:2024toz}. In this case, the resulting axion DWs do not have to spam an entire winding range from $0$ to $2\pi$.

\end{itemize}

In such scenarios, no suppression is expected and the estimates presented in the previous sections (and Fig.~\ref{fig:ALPpara}) are applicable without modifications.

\section{Conclusion and outlook}
\label{sec:conclusions}

When a (axionic) domain wall with a derivative coupling to the lepton/baryon current $\propto\partial_\mu a/f_a j^\mu_{L/B}$ passes through the plasma, interactions violating the conservation of the lepton/baryon number are biased and produce a lepton/baryon asymmetry, via the familiar mechanism of \emph{spontaneous baryogenesis}. We call this scenario DW-genesis. During the oscillations of the wall in the scaling regime, the lepton/baryon number is washed out by the symmetry of oscillations, as the yield from a wall with velocity $v_w$ is the opposite of the one produced by a wall with a velocity $-v_w$. The symmetry is however broken when the DW network collapses towards a preferred direction, and a burst of lepton/baryon asymmetry is produced. 
Then, after the passage of the collapsing DW, some wash-out can occur if the lepton/baryon violating interaction is still active. The final lepton/baryon asymmetry observed today is thus controlled by the \emph{annihilation temperature} of the DW and by the rate of the baryon/lepton violating interaction at this temperature.

The mechanism of leptogenesis with axion DW has been previously proposed in \cite{Daido:2015gqa}.
In this paper, we have significantly extended the previous study in several directions:
i) we have derived the Boltzmann equations describing the mechanism in generality, and systematically studied several new concrete realisations;
ii) in the case of DW-leptogenesis, we have included the possibility of RHN in thermal equilibrium, which can boost the asymmetry production and hence enlarge the viable parameter space; iii) we focused mainly on the post-inflationary scenario, where the Peccei-Quinn symmetry is restored after reheating, and we have quantified the 
suppression due to cancellations between different axionic DWs at collapse.

Some properties of the DW-genesis mechanism appeared to be model-independent: 
the asymmetry is typically maximised when the lepton/baryon-number violating interaction is close to decoupling when the DW network collapses. At this peak, the asymmetry scales like $Y_{\Delta B, L} \sim H/T\big |_{T = T_{\rm ann}}$, yielding a lower bound on the annihilation temperature to match the observed abundance, which is $T_{\rm ann} \gtrsim 10^9 $ GeV for both the leptogenesis and baryogenesis realisations of the mechanism.

As mentioned, the mechanism of DW-genesis is versatile enough to accommodate different realisations. We study four scenarios: i) A type I see-saw realisation with heavy RHNs. At the moment of collapse, the DW coupling to the lepton number produces the observed asymmetry if the mass of the RHN lies in the region $M_N \in [10^9, 10^{14}]$ GeV. We emphasize that such mechanism does not require any CP violation in the see-saw sector. ii) A dark sector charged under lepton number and coupled to the RHN, leading to a  scenario of cogenesis where the fermion in the dark sector is the asymmetric dark matter candidate. We find that
 the DW motion will populate simultaneously the two sectors but the relative asymmetries can differ by orders of magnitude. This allows the DM mass to depart from $\sim 1$ GeV by few orders of magnitude. We find $m_{\rm DM} \in [0.1, 10^5]$ GeV can explain the observed DM fraction. iii) A model with a dimension 9 operator breaking the baryon number by two units, where a DW coupled to a quark current could induce the observed asymmetry if the scale of the effective operator $\Lambda \gtrsim 10^{10}$\,GeV. This scale is however too high to be observable at neutron-anti-neutron oscillation experiments. iv) Finally, the attractive idea that DWs collapsing during the EWPT could bias the sphalerons and produced the observed asymmetry. We however show that the asymmetry produced in this way is always five orders of magnitude below the observed one.  

We also anticipate that our analysis does not necessarily exhaust the possible models and regimes of DW-genesis.
For instance, it would be interesting to study the case of Dirac leptogenesis\cite{Dick:1999je,Murayama:2002je,Barrie:2024yhj} where the lepton number is conserved and produced with opposed sign in the visible and the dark sector respectively.

In this paper, we focused on the interesting case when the plasma could efficiently thermalise inside the domain wall, $T_{\rm ann} \gg  \gamma_w v_w m_a/\alpha^2$ which is the spontaneous baryogenesis regime and imposes strong constraints on the parameter space. However, one might wonder if, like in electroweak baryogenesis, the baryon asymmetry could come from CP-violating reflections on the domain wall itself. We leave the exploration of this interesting idea to future studies.

We eventually observed that the presence of different types of DWs might yield to a partial cancellation of the asymmetry during the collapse. We quantify this cancellation in minimal models and propose some mechanisms to avoid the corresponding suppression. We argue that it is not universal and will not plague all the realisations of the DW-genesis scenario. We nevertheless leave for future studies a complete classification of the viable possibilities.

We showed that the DW-genesis mechanism pinpoints to a specific annihilation temperature for the DW, as high as $T_{\rm ann} \gtrsim 10^9$ GeV, for successful lepton/baryogenesis,
corresponding to a GW signal peaked at $O(100)$ Hz frequencies (or higher). However, due to the relative low quality of the axion and the matter domination induced by the axion produced at the collapse, the GW signal results not observable 
at the future Einstein Telescope (see Fig.\ref{fig:ALPparaLim})
in the DW-genesis realisations explored in this paper.
Our results nevertheless motivate further the characterization of the gravitational wave signal from the collapse of the DW network.

\section*{Acknowledgements}
It is a pleasure to thank Simone Blasi, Graham White, Valerie Domcke, Fuminobu Takahashi for useful discussions. We also thank Tae Hyun Jung and Eung Jin Chun for useful discussions about the biasing in the decay of the RHN.

AM, XN, AR and MV are supported
by the "Strategic Research Program High-Energy Physics” of the Vrije Universiteit Brussel, by the
iBOF “Un-locking the Dark Universe with Gravitational Wave Observations: from Quantum Optics to Quantum Gravity” of the Vlaamse Interuniversitaire Raad,
and by the 
the ``Excellence of Science - EOS" - be.h project n.30820817. AR is supported
by FWO-Vlaanderen through grant number 1152923N.

\appendix

\section{Axion couplings in the presence of a RHN mass}
\label{app:axionEFT}
In the SM plus the RHN neutrino with a Majorana mass, the axion effective Lagrangian 
in the leptonic sector can be written in terms of derivative couplings
\footnote{See e.g. \cite{Bauer:2020jbp,Bonilla:2021ufe} for the generic form of the axion effective Lagrangian in the SM.}
as
(for one flavour) 
\be
\label{eq:axionEFT}
\mathcal{L} \supset \frac{\partial_{\mu} a}{f_a}
\left( c_L \bar L \gamma^\mu L + c_R 
\bar e_R \gamma^\mu e_R + c_N \bar N_R \gamma^{\mu} N_R
\right) \, . 
\ee
By performing an axion dependent rotation of the leptons (LH and RH) and the RHN, we eliminate all these couplings and we 
generate the following non derivative interactions
\bea 
\label{eq:L_violating_1_bis}
\mathcal{L}^{} \supset 
e^{-i(c_L - c_R) a/f_a } y_{\rm{SM}} \bar L  H e_R +
e^{i(c_N - c_L) a/f_a } y_N (  \tilde H \bar L)N_R +\frac{1}{2} e^{2 c_N a/f_a }M_N \bar N^c_R N_R + \text{h.c.}  \; , 
\eea 
In addition, this transformation  will generically induce extra
contributions to possible non-derivative axion couplings to the hypercharge and to $SU(2)_L$ field strengths.
Thus, besides such anomaly terms, the three relevant independent combinations of axion couplings in the presence of a RHN mass term are $c_L - c_R, c_L- c_N, c_N$. Only the coefficient $c_L$ is important for generating the lepton asymmetry through spontaneous DW leptogenesis,
and it will be the one considered in the main body of the paper.

\section{Computing the average Yield}
\label{app:finite_time_collapse}

In the main text, we assumed that the temperature is constant over the entire duration of the DW collapse. This however cannot be the case as the typical duration of the collapse of a spherical DW enclosing an Hubble volume will be 
\bea 
t_{\rm collapse} \approx \frac{1}{v_w \gamma_w H(t_{\rm ann})} \, .
\eea 

To this end, we assume a collapsing closed DW of size equal to the Hubble radius at annihilation, i.e. $R_{\rm initial} \sim H^{-1}(T_\text{ann})$. As the DW collapses, it acts as a source of a lepton asymmetry for each point enclosed by the wall.

During the collapse of the DW, the universe keeps cooling and the temperature decreases. Assuming that the DW collapse spherically, the temperature is related to the radius of the circular DW 
\bea 
\frac{dr}{dT} =  \frac{v_w}{HT} =  \frac{v_w}{aT^3}, \qquad a \equiv  \sqrt{g_\star\pi^2/90}/M_{\rm pl}\qquad \Rightarrow \int^{R_{\rm ini}}_{r} dr = -\frac{v_w}{2 a} \bigg[\frac{1}{T^2}\bigg]^{T_{\rm ann}}_{T_{}} \, . 
\eea 
where $r$ is the distance from the origin to the considered point within the Hubble radius, as depicted in Fig.\ref{fig:TotalYield_method}. We also imposed the boundary conditions $T(r = R_{\rm ini}) = T_{\rm ann}$. 
Finally, we obtain 
\bea 
R_{\rm ini} - r = -\frac{v_w}{2} \bigg(\frac{1}{aT^2_{\rm ann}}-\frac{1}{aT^2}\bigg) \, , 
\eea 
where we can read that at $r = R_{\rm ini}, T= T_{\rm ann}$ and when $r = 0, T= T_{\rm end}$. We will take
\bea 
\label{eq:T_vs_r}
R_{\rm ini} = \frac{1}{H(T= T_{\rm ann})}\, , \qquad \frac{1}{T_{\rm ann}^2} \bigg(\frac{2}{v_w}+1\bigg) - \frac{2ra}{v_w} = T^{-2} \, , \qquad T= \frac{1}{\sqrt{a}\sqrt{(2/v_w +1)R_{\rm ini} - \frac{2r}{v_w}}}
\eea 
Consequently, we obtain 
\bea 
T_{\rm end} \approx T_{\rm ann} \frac{1}{\sqrt{2/v_w + 1}}\,  \qquad \Delta T \approx T_{\rm ann} \bigg(1- \frac{1}{\sqrt{2/v_w + 1}} \bigg) \, . 
\eea 
As a consequence, the temperature and thus the rates can change sizably over the timescale of the collapse. In this appendix, we estimate the change in baryon yield due to this change of temperature. 

\subsection{Analytical considerations}

\begin{figure}
    \centering
    \includegraphics[scale=0.25]{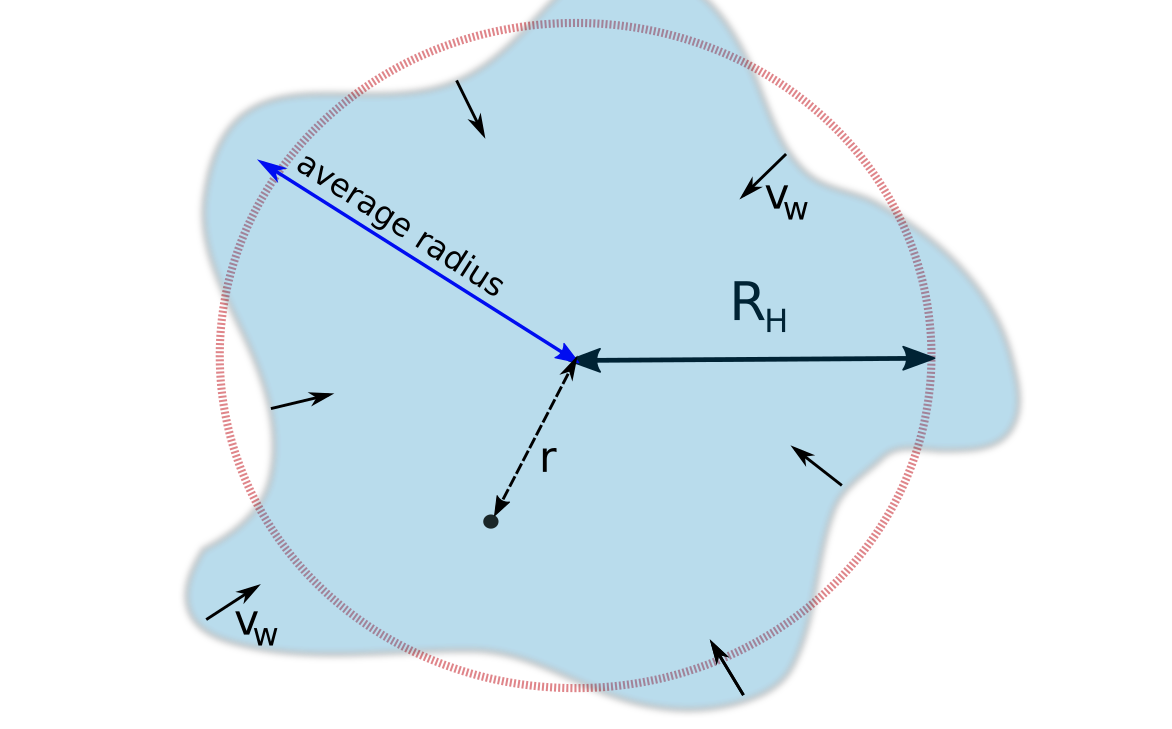}
    \caption{Sketch of a collapsing closed DW. For the purpose of this study, we assume that the initial radius of the wall is roughly given by the Hubble radius at annihilation, i.e. $\text{\emph{average radius} of the DW} \sim H^{-1}(T_\text{ann})$. On the other hand, $r$ is the distance of some point in the plasma to the centre of the enclosed Hubble region. $r_\text{max}$ coincides with the distance of the most outer part of the wall to the centre.}
    \label{fig:TotalYield_method}
\end{figure}

  For each point of the plasma (including the ones already inside the wall), we compute the yields and take the average. A sketch of the situation is depicted in Fig.\ref{fig:TotalYield_method}. 
  Assuming the wall is collapsing at a \emph{constant} velocity $v_w$, the time it takes for the DW to reach a point at a distance $r$ from the center is $ t \sim \frac{R_{\rm ini}-r}{v_w}$. Therefore, the effective chemical potential has the form
\begin{equation}
    \mu_\text{eff} \equiv \frac{\Dot{a}}{f_a} = \frac{2 m_a \gamma_w v_w}{\cosh\left[m_a \gamma_w v_w \left(t-t_\text{ann} - \frac{r}{v_w}\right)\right]}\,,
\end{equation}
where $t_\text{ann}$ is the time of annihilation. The average yield over the sphere enclosed by the DW is then
\begin{equation}\label{eq: average yield}
    \langle Y_L \rangle = \frac{3}{R_{\rm ini}^3}\int_0^{R_{\rm ini}}r^2 Y_{\Delta L}(r)\mathrm{d}r\,,
\end{equation}
where the radius $r$ can be parameterized using the temperature $r(T)$. On the other hand, one can  estimate  
the behaviour of the $Y_{\Delta L}(T)$ to be $Y_{\Delta L}(T) \propto T^2$ in the regime of $T \ll M_N/10$. 

Using now Eq.\eqref{eq:T_vs_r} to solve Eq.\eqref{eq: average yield}, we can compute the 
following ratio
\bea 
R \equiv\frac{\langle Y_L \rangle^{\rm integrated}}{\langle Y_L \rangle^{\text{non-integrated }}} \approx - \frac{3v_w}{8}  \bigg(2 (3 + v_w) + (2 + v_w)^2\text{Log}[v_w/(2 + v_w)]\bigg)
\quad \text{for} \quad Y_{\Delta L}(T) \propto T^2 \, . 
\label{eq:ratio}
\eea 

\begin{figure}
    \centering 
     \includegraphics[scale=0.7]{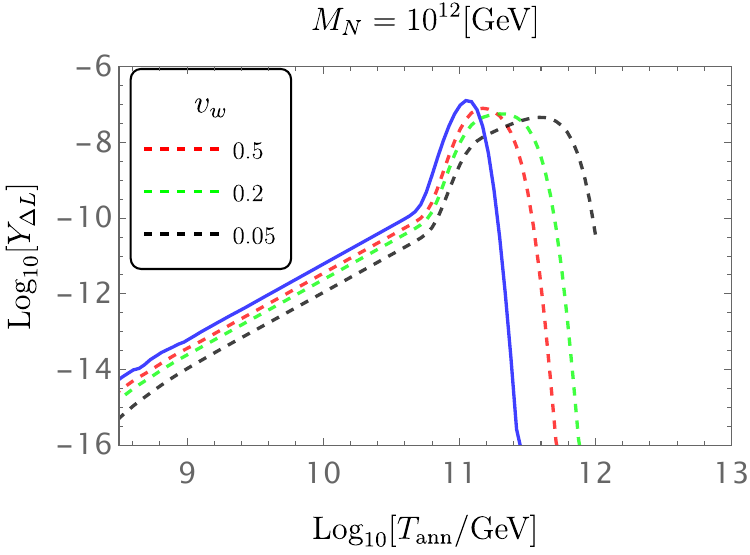}
    \includegraphics[scale=0.9]{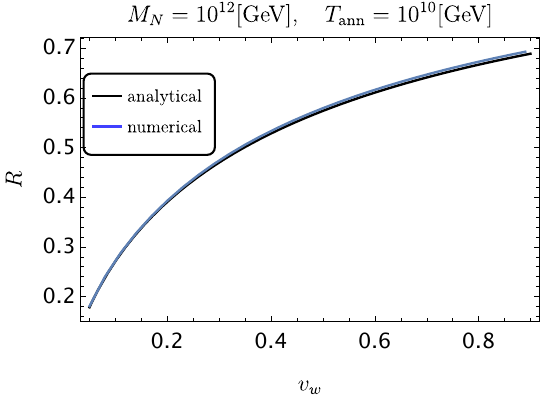}
    \caption{\textbf{Left Panel}: Modification in the $Y_{\Delta L}$ as a function of $v_w$, the dashed lines are obtained from the integrated procedure, while the solid line is the unintegrated procedure. 
    \textbf{Right panel}: The ratio of $R$ given by Eq.\eqref{eq:ratio} as a function of $v_w$, obtained in the case of the ``scattering dominated regime''. The Blue curve is the numerical solution while the Black one is the approximation from Eq.\eqref{eq:ratio}.   }
    \label{fig:yieldperTann}
\end{figure}

It is interesting to notice that, even though the non-integrated yield $\langle Y_L \rangle^{\text{non-integrated }}$ \emph{does not} depend on the velocity of the wall $v_w$, the integrated version $\langle Y_L \rangle^{\text{integrated }}$ does depend on the velocity. 
The ratio in Eq.\eqref{eq:ratio} goes to $0.6$ when $v_w \to 1$ while for $v_w \to 0.1$, it goes to $\sim 0.2$. This can be understood intuitively by the fact that when the wall moves slower (smaller $v_w$), the temperature changes more over the full collapse, and thus induces a larger suppression. 
Note that in a realistic spherical collapse, the velocity will also evolve during the collapse. We however do not expect very large boost factors at the end.

 \subsection{Numerical considerations}

We now turn to the numerical study of the shift. 
The final average lepton asymmetry is then calculated using Eq.\eqref{eq: average yield} and plotted as a function of the annihilation temperature on the Left panel of Fig.\ref{fig:yieldperTann}. We compare the result from the integrated (dashed)  and the unintegrated (solid Blue) for different values of the velocity of the wall $v_w$. As expected from Eq.\eqref{eq:ratio} and intuition, the separation increases with smaller velocities $v_w$.

\begin{figure}
    \centering 
     \includegraphics[scale=0.6]{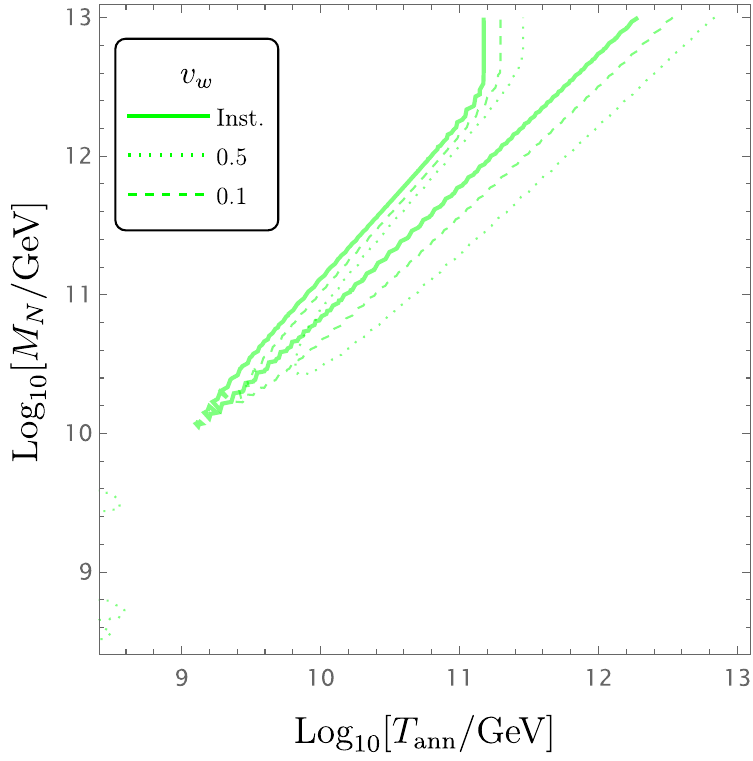}
    \caption{Modification of the region matching the observed abundance when the collapse is not instantaneous. The solid curve corresponds to instantaneous collapse.  }
    \label{fig:modif_para}
\end{figure}

We also observe a \emph{shift} of the position  and a \emph{broadening} of the peak.  Another important information for the phenomenology of our model is the region of the parameter space matching the observed asymmetry, as in Fig.\ref{fig: succ_lep}. We show the modification to the region in Fig.\ref{fig:modif_para} and observe non-trivial shift of the green region, with a larger shift for slower velocities of the DW. 

\section{Computation of the Boltzmann equations in the shift formalism}
\label{app:spon_baryo}

In this Appendix, we present the computation of the asymmetry produced inside the domain wall using the dispersion relation approach, which is complementary to the approach we followed in the main text.

\subsection{Shift in energy approach}

As mentioned above, the interaction between the DW and the LH leptons takes the form
\bea \label{eq: L(a-L)}
\mathcal{L}_{a-L} = \frac{c_L}{f_a} \partial_{\mu} a \bar L \gamma^{\mu}  L    \, . 
\eea
In the main text, we rotated away the couplings between the axion and the leptonic current. The phase then reappeared in the vertices of the $L$-violating interaction. It is however equivalent to keep the coupling between the axion and the leptonic current and eliminate the phases in the vertices. This will be the method we illustrate in this Appendix. 
In fact, the addition of the interaction between the non-trivial background and the lepton current
would imply the following modification of the energy momentum relation (by inserting the interaction in the Dirac equation),
\bea
\label{eq:dynsplit}
E = \sqrt{\vec p^2+m_L^2} \pm c_L \frac{\dot a}{2f_a} \, , 
\eea
where $\pm$ distinguish the energy eigenstates of particle and anti-particle \cite{Ibe:2015nfa} and $m_L$ is the mass of the particle under consideration. This is the so called \emph{dynamical splitting}\cite{Arbuzova:2016qfh} induced by the non-trivial background. This can be seen by noting that the interaction in Eq.\eqref{eq: L(a-L)} simply modifies the Dirac equation, obtained via the Euler-Lagrange equations, to
\be
\label{eq:splitted_Dirac}
\bigg( i\gamma^{\mu} \partial_{\mu}  - m_L - c_L \frac{\dot a}{f_a} \gamma^0 \bigg) L = 0 \, .
\ee

Then, one can look at the plane wave solutions for particle and anti-particles, and realize that the extra piece proportional to $\dot a$ simply implies a shift of the type in Eq.\eqref{eq:dynsplit}
(this is done in QFT in \cite{Ibe:2015nfa} and in relativistic QM in \cite{Chun:2023eqc}). Notice that this solution assumes that the chemical potential is constant in time: $\dot a \neq 0$ while $\ddot a =0$. This means that, as already discussed in Section \ref{sec:spont_baryo}, the solutions considered above are only valid for a time $\Delta t < \dot a /\ddot a$.

We now want to see how this dynamical splitting is taken into account in the Boltzmann equation governing the abundance of leptons and anti-leptons.

\subsection{Solving the Dirac equation}

We now move on to solve the Dirac equation in Eq.\eqref{eq:splitted_Dirac}. It admits the well known solutions in the limit of vanishing chemical potential $\dot a /f_a \to 0$, 
\bea 
L(x) = \int \frac{d^3p}{(2\pi)^3} b_{\vec p} u_{\vec p}(t) e^{i \vec p \cdot \vec x} +  \int \frac{d^3p}{(2\pi)^3} d^\dagger_{\vec p} v_{\vec p}(t) e^{-i \vec p \cdot \vec x}  \, ,  
\eea 
where $ b_{\vec p}$ ($d_{\vec p}$) is the  annihilation operator of the particles (anti-particles) and $p \cdot x = Et - \vec p \cdot \vec x$. $u_{\vec p}(t)$ and $v_{\vec p}(t)$ are the time-dependent wave functions which are the solutions of the Dirac equation with eigenstates $\pm E= \pm\sqrt{p^2 + m^2}$, 
\bea 
u_{\vec p}(t) = \frac{u_{\vec p}^0}{2E} e^{-i Et} \qquad v_{\vec p}(t) = \frac{v_{\vec p}^0}{2E} e^{i Et} \, . 
\eea 
 Writing the complete Dirac field 
\bea 
L(x) = \int \frac{d^3p}{(2\pi)^32E} b_{\vec p} u_{\vec p}^0 e^{-i  p \cdot  x} +  \int \frac{d^3p}{(2\pi)^3 2E} d^\dagger_{\vec p} v_{\vec p}^0 e^{i  p \cdot  x}  \, , 
\eea 
then yields the usual conservation of energy and momentum in the scattering rates: 
\bea 
\mathcal{S} = i (2\pi)^4 \delta(\sum p_i - \sum p_f) \mathcal{M} \, . 
\eea 

This solution is however modified in the presence of the background because of the different Dirac equation for particle and anti-particle. The solution of the Dirac equation is identical except for the energy eigenstates, which are 
\bea 
E_1= \sqrt{p^2 + m^2} + \frac{c_L \dot a}{2 f_a}, \qquad E_2= -\sqrt{p^2 + m^2} + \frac{c_L \dot a}{2 f_a} \, , 
\eea 
where we assumed that $\dot a$ is a constant. Consequently the Dirac field becomes 
\bea 
L(x) = \int \frac{d^3p}{(2\pi)^32E} b_{\vec p} u_{\vec p}^0 e^{-i  p \cdot  x} e^{-i\frac{c_L \dot a}{2 f_a}  t} +  \int \frac{d^3p}{(2\pi)^3} d^\dagger_{\vec p} v_{\vec p}^0 e^{i  p \cdot  x } e^{-i\frac{c_L \dot a}{2 f_a} t}  \, . 
\eea 
Going through the computation of the matrix element as before gives 
\bea 
\mathcal{S}^n = i (2\pi)^4 \delta^{(3)}\bigg(\sum \vec p_i - \sum \vec p_f \bigg) \delta\bigg(\sum E_i - \sum E_f + n \frac{c_L \dot a}{2 f_a} \bigg) \mathcal{M} \, ,
\eea 
where $n$ is some number that relates to the number of particles that couple to the axion in the initial and final states. Two reactions are important for the purpose of DW leptogenesis: the scatterings $HL \to H^cL^c$ and the decays $N \to HL$. It is easy to see that 
\bea 
N \to HL: \qquad n = -1, \qquad \text{and} \qquad  HL \to H^cL^c: \qquad n = 2 \, . 
\eea

This permits us to recover the result that we have found in Section \ref{sec:spont_baryo}, as it allows to reconstruct directly Eq.\eqref{eq:collision_int_1} and \eqref{eq:collision_int_2}.  The rest of the computation then follows exactly as in Section \ref{sec:spont_baryo}.

In \cite{Chun:2023eqc}, with similar reasoning  for the axion coupling $c_N$ to the Majorana fermion $N_R$, it was shown that the background induces an asymmetry in the \emph{helicity} eigenstates of the heavy neutrino. Since each helicity of the RHN has the same probability as the decay to $LH$ and $L^cH^c$, this asymmetry is not relevant for the lepton asymmetry.

\section{Validity of the constant $\dot a$}
\label{app:validity_app}

The procedure described in the main text and in Appendix \ref{app:spon_baryo} assumed that $\dot a$ is constant. This is of course approximately fulfilled for the Affleck-Dine mechanism and axiogenesis, but this is much more disputable for DW baryogenesis, where inside the DW $\ddot a \neq 0$. In this Appendix, we show that this brings subleading corrections if $T\gg m_a v_w\gamma_w$, which is already implied by the thermalization regime we impose
\bea 
\label{eq:thermalisation_cons}
T \gg  m_a v_w \gamma_w /\alpha^2 \qquad \text{(thermalisation)} \, . 
\eea 
We follow \cite{Arbuzova:2016qfh} where a detailed analysis of this issue has been performed.
Let us begin by expanding $a/f_a \equiv \theta$, 
\bea 
\label{eq:taylor exp}
\theta(t) = \theta_0(t_0) + \dot \theta_0(t_0)(t-t_0) + \frac{\ddot \theta_0 (t_0)}{2}(t-t_0)^2+..., \qquad 
\eea 
where we can identify $t_0 \equiv t_{\rm passage}$. 
 Using that $m_a\sim L_w^{-1}$, we have $\Dot{\theta}/\Ddot{\theta}\sim L_w/v_w\gamma_w$.
The approximation that we used in the main text, i.e. that we can truncate the expansion in Eq.\eqref{eq:taylor exp} at leading order, holds for a given period of time of order $L_w/v_w \gamma_w$,
which is the timescale for a particle to cross the domain wall. 

In the Boltzmann equations, the collision term  can be considered as a function of two variables $\mathcal{C}(t_1, t_2) \equiv \mathcal{C}(\Delta t,  t) $, 
$\Delta t$ being the local variable varying rapidly and 
$ t \equiv (t_1+t_2)/2$ being a global variable, varying slowly. This means that also the phase is a function of those two variables
\bea 
\label{eq:taylor_exp}
\theta(\Delta t, t) = \theta_0(t) + \dot \theta_0(t)\Delta t + \frac{\ddot \theta_0 (t)}{2}\Delta t^2+..., \qquad 
\eea 
and we can now perform the Fourier transform over $\Delta t$.
From splitting the time-integration in time steps of $ t_{\rm max} < L_w/ \gamma_w v_w$, one can build a coarse-grained amplitude
\begin{align}
\label{eq:M_in_fourier}
\mathcal{M}(p, t) \propto \int^{t+ t_{\rm max}/2}_{t-t_{\rm max}/2} d\Delta t  e^{2i c_L \theta(\Delta t, t)} e^{i (E_1 + E_2 - E_3 -E_4) \Delta t}  (2\pi)^3 \delta^3 (\vec p_1 + \vec p_2 - \vec p_3 -\vec p_4) \mathcal{M}\, . 
\end{align} 
where $t_{\rm max}$ is a free-time parameter on which we perform a Fourier transform. 
We expect this to be a good approximation as long as the typical momentum $p \sim T$ is larger than the inverse of the time-scale on which we perform the integration, which corresponds to $T \gg  1/t_{\rm max} > \gamma_w v_w/L_w$, as argued above.

\begin{figure}
    \centering
\includegraphics[width=0.7\linewidth]{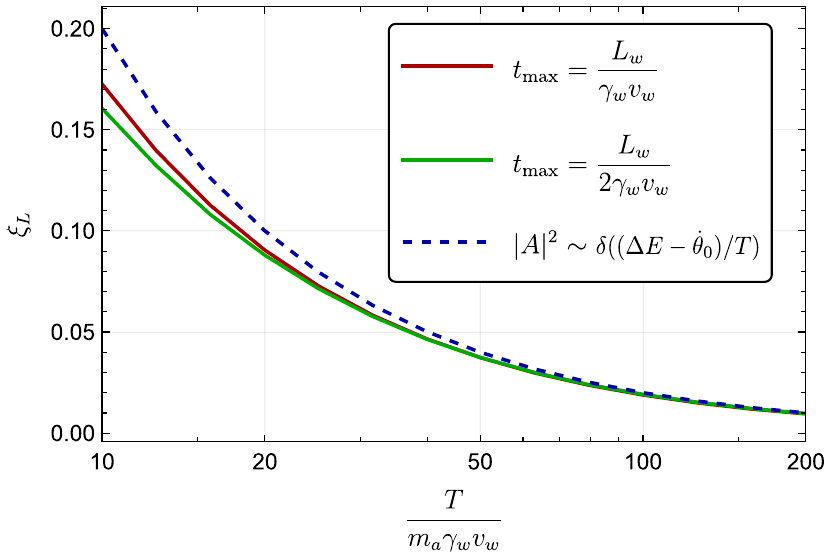}
    \caption{Equilibrium chemical potential computed from numerically solving Eq.\eqref{eq: A with tmax} and \eqref{eq:new_bolt} with $t_{\rm max} = L_wv_w \gamma_w$ (red) and $t_{\rm max} = L_wv_w \gamma_w/2$ (green) and computed from the constant $\dot \theta$ approximation (dashed blue). The $\xi_L$ is defined in Eq.\eqref{eq:distributions}.}
    \label{fig:ComparisonV}
\end{figure}

As a concrete example where to numerically check this estimation, we can study the BE for the production of the lepton number via scatterings.  The BE depends on the vertex function as follows, following \cite{Arbuzova:2016qfh},
\begin{align}
\label{eq:new_bolt}
    \dot n_{\Delta L} + 3H n = -\frac{(2\pi)^3}{t_{\rm max}} \int \prod_i \frac{d^3p_i^3}{(2\pi)^3 2E_i} \delta\left(\sum_{a,b} {\bf p_i}-\sum_{c,d} \bf p_f\right) |A(t)|^2 (f_af_b-f_cf_d) \; ,
\end{align}
The product of the phase spaces is performed over all incoming and outgoing particles.
%
Here, $A$ is the amplitude of the interaction:
\begin{align}
\label{eq: A with tmax}
    A(t)=\int_{t-t_{\rm max}/2}^{t+t_{\rm max}/2} d\Delta t e^{i\left( \Delta E \Delta t -\theta(\Delta t,t) \right)}\mathcal{M}_{HL \to H^cL^c} \; ,
\end{align}
with $\Delta E$ the difference in energy of the asymptotic initial and final states. 
One can numerically solve Eq.\eqref{eq: A with tmax} and \eqref{eq:new_bolt} and compare with the 
delta-function approximation used in Section~ \ref{sec:derivation_BE}.
We show the results of such comparison
in Fig.\ref{fig:ComparisonV}, where we fixed $t_{\rm max} = L_wv_w \gamma_w$ (Red) and $t_{\rm max} = L_wv_w \gamma_w/2$ (Green). Comparing with the constant $\dot \theta$ result (dashed Blue), one observes that the approximation becomes better in the limit $ t_{\rm max} T \to \infty$. At $t_{\rm max} T \sim 100$, the agreement is at the percent level.

\section{The rates and the Boltzmann equations}
\label{app:un_sub}
In this Appendix, we explicitly provide the rates that we used for the computations in the unsubtracted scheme and we discuss the resulting BEs for leptogenesis and then for cogenesis.

\paragraph{Leptogenesis}

The BEs for DW leptogenesis, ignoring the CP-violating contributions and considering only the $\Delta L=2$ scatterings, take the simple form\cite{Buchmuller:2002rq, Giudice:2003jh}
\bea 
\frac{dY_{\Delta L}}{dz} = \frac{1}{H s z} \bigg(D - \bar D - 2 N_s -N_t + \bar N_t\bigg) \, , 
\eea 
where we used schematic notations as in \cite{Giudice:2003jh} for the thermal rates
\bea 
D = [N \to HL]\, , \qquad N_s =[HL \to H^cL^c ], \qquad N_t = [LL \to H^cH^c] \,.
\eea 
In the BEs, two contributions dominate the washout: the decay and inverse decay $D$, and the $2 \to 2$ scatterings $N_s$ and $N_t$. For the decays, one obtains immediately 
\bea 
D - \bar D = -\gamma_D \frac{Y_{\Delta L}}{Y_{\rm eq}}\,. 
\eea 
 For the scatterings, the $s$-channel exchange of $N_s$ must be computed by subtracting the contribution due to the on-shell $N_R$ exchange, because in the BEs
 this effect is already taken into account by successive decays, $LH \to N \to L^c H^2$\cite{Giudice:2003jh}.
Using that $\gamma^{\rm on-shell}(HL \to H^cL^c) = \gamma^{\rm eq}(LH \to N)\text{Br}_{N\to \text{SM}}/2$, where SM designates the leptons of the Standard Model, and where $\text{Br}_{N\to \text{SM}} = 1$ (assuming no decay toward a dark sector) and $\gamma^{\rm eq}(LH \to N)$ is the thermal rate for the production of an on-shell RHN from the plasma $HL$, one obtains 
\bea 
\gamma^{\rm on-shell}(HL \to H^cL^c) = \gamma_D/2 \, .
\eea 
Notice the factor of two different with\cite{Giudice:2003jh}, which comes from our definition of $\Gamma_D = \Gamma_{N \to HL} = y_N^2\frac{M_N}{16 \pi}$. The subtracted scattering rates then read 
\bea 
\gamma(HL \to H^cL^c) = \gamma_{N_s} - \gamma^{\rm on-shell}(HL \to H^cL^c) = \gamma_{N_s} - \gamma_D/2 \, .
\eea 

After this procedure, the \emph{subtracted} $N'_s$ piece becomes 
\bea 
N'_s = - \frac{Y_{\Delta L}}{Y_{\rm eq}}(\gamma_{N_s} - \gamma_D/2) \, , \qquad N_s = N'_s - \frac{Y_{\Delta L}}{Y_{\rm eq}}\gamma_D/2= - \frac{Y_{\Delta L}}{Y_{\rm eq}}\gamma_{N_s}  \, ,
\eea 
 where the prime denotes the subtraction procedure. 
From this discussion, we understand that, in general, the BEs can be solved in two equivalent ways. Firstly, we can separate the on-shell decay and inverse decay, and we can write the BEs as  in Eq.\eqref{eq:lepto_equation}
\bea 
\frac{dY_{\Delta L}}{dz} = - \frac{1}{szH}\bigg( 
\frac{\gamma_D}{Y_L^{\rm eq}} (Y_{\Delta L} + Y^{\rm eq}_{\Delta L}(t)) + 2 \frac{\gamma'_{2 \to 2} }{Y_L^{\rm eq}} \bigg( Y_{\Delta L} + Y^{\rm eq}_{\Delta L}(t)\bigg)\bigg) \qquad \text{(subtracted scheme)}\,,
\eea 
in which the rate for the $2 \to 2$ scatterings should be understood as the subtracted rate $\gamma'_{2 \to 2}$ (primed) in the narrow width approximation\cite{Buchmuller:2004nz, Davidson:2008bu}: $\gamma'_{2 \to 2} \equiv \gamma_{N_s} + \gamma_{N_t} - \gamma_D/2$. This is in the spirit of integrating out the heavy RHN. 

Secondly, one might stay in the UV-theory and compute the full rate, where the BEs become
\bea 
\frac{dY_{\Delta L}}{dt} =  - 
 2 \frac{\gamma_{2 \to 2} }{szH Y_L^{\rm eq}} \bigg( Y_{\Delta L} + Y^{\rm eq}_{\Delta L}(t)\bigg)
\qquad \text{(un-subtracted scheme)}\,,
\eea 
where now $\gamma_{2 \to 2}$ (unprimed) denotes the full (unsubtracted) rate, $\gamma_{2 \to 2} \equiv \gamma_{N_s} + \gamma_{N_t}$. The \emph{subtracted} $\gamma'_{2 \to 2}$ can be computed in the following way\cite{Buchmuller:2002rq}\footnote{It has been noticed in \cite{Giudice:2003jh} that the rates defined in Eq.\eqref{eq:rates_buch} are only partially subtracted, 
a further $\gamma_D/8$ needs to be subtracted from Eq.\eqref{eq:rates_buch}.}, 
\bea 
\label{eq:rates_buch}
\gamma_i (z) = \frac{M_N^4}{64\pi^4}\frac{1}{z} \int^{\infty}_{0} dx \sqrt{x} \hat\sigma_i(x) K_1(z\sqrt{x}) \, , \qquad \Gamma_i (z) = \frac{M_Nz^2}{128\pi^2} \int^{\infty}_{0} dx \sqrt{x} \hat\sigma_i(x) K_1(z\sqrt{x}) - \gamma_D/8
\eea 
where $K_1$ is the Bessel function of the first kind and $\hat\sigma$ the reduced cross section in the hierarchical approximation, i.e. $M_1 \ll M_2$, which takes the form 
\bea 
\hat \sigma_i = \frac{(y^\dagger y)^2}{2\pi} f^i(x) \, , \qquad \frac{(y^\dagger y)^2}{M_N^2} = \frac{m_\nu^\dagger m_\nu}{v_{\rm EW}^4}, \qquad v_{\rm EW} = 174 \text{ GeV} \, . 
\eea 
In the hierarchical regime, $m_\nu^\dagger m_\nu \approx \Delta m^2_{\rm atm}$. The $f^i$ functions take the form 
\begin{align}
f^{HL \to HL}(x) &= 1 + \frac{(x-1)}{P(x)}  + \frac{x(x-1)^2}{2P(x)^2}- \frac{1}{x}\bigg(1+ \frac{(1+x)(x-1)}{P(x)}\bigg)\log (1+x) 
\notag \\
f^{LL \to HH}(x)& = \frac{x}{x+1} + \frac{1}{x+2} \log (1+x) \, , 
\end{align} 
where $P(x) \equiv (x-1)^2 + (2b)^2$, where $b\equiv y^\dagger y/16\pi$. For more details about the computation of the scattering rates following reference and its appendices might be useful \cite{Hahn-Woernle:2009jyb}.

\begin{figure}[t!]
\centering
\includegraphics[width=0.5\linewidth]{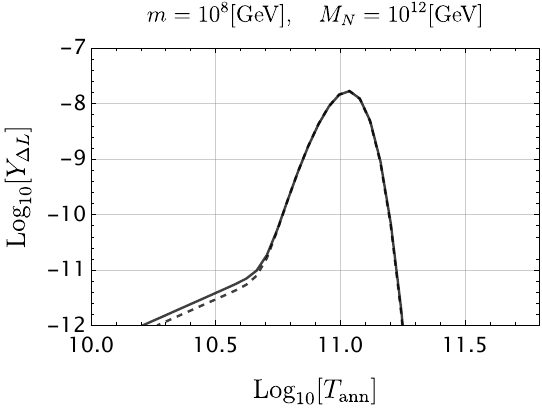}
    \caption{Comparison between the subtracted scheme (solid line) and un-subtracted scheme (dashed line).     }
    \label{fig:comparison_scheme}
\end{figure}

We have verified that both schemes bring the same final evolution and final asymmetry within a very small error, see Fig.\ref{fig:comparison_scheme} for the comparison. 
The unsubtracted scheme has however the virtue to clarify the role of the RHN in the source term, as we explain now. Indeed in the main text, we have decided to rotate the lepton $L \to e^{ic_L a/f_a} L$ to eliminate the coupling between the axion and the RH leptonic current,
\bea 
\label{eq:L_viol_final}
\mathcal{L}^{\rm IR}_{\slashed{L}} = e^{2i c_L a/f_a} y_N^2\frac{(\bar L^c \tilde H^*)(\tilde H^\dagger L)}{M_N} + h.c.
\qquad \mathcal{L}^{\rm UV}_{\slashed{L}} = e^{-i c_L a/f_a} y_N ( \tilde H \bar L)N_R + \frac{1}{2}M_N \bar N^c_R N_R + \text{h.c.}  \,. 
\eea 

In the unsubtracted scheme, only the UV Lagrangian is relevant
\bea 
\mathcal{L}^{\rm UV}_{\slashed{L}} = e^{-i c_L a/f_a} y_N (\tilde  H \bar L)N_R + \frac{1}{2}M_N \bar N^c_R N_R + \text{h.c.} \,.
\eea 
We argued in the main text that a further rotation of the RHN fields, $N \to e^{ic_L a/f_a}N$ should be irrelevant, as the shift in the Dirac equations of Majorana particles only distinguishes helicities. We now verify that one can rotate the RHN without any impact on the source terms in the BEs. Let us apply the rotation $N \to e^{ic_L a/f_a}N$ to the former lagrangian. One obtains
\bea 
\mathcal{L}^{\rm UV}_{\slashed{L}} =  y_N (\tilde  H \bar L)N_R +\frac{1}{2} e^{2i c_L a/f_a}M_N \bar N^c_R N_R + \text{h.c.} \, ,
\eea 
which removes the phase from the vertex and brings it to the mass term. In the unsubtracted scheme, only the $\gamma_{2 \to 2}$ rate matters, which is built from two vertices $yN(\tilde H \bar L)$ and one insertion of the Majorana mass $M_N$. It is now clear that the diagram before or after the RHN rotation yields the same phase in the diagram: in the former case, it comes from each of the two vertices and in the latter case it comes from the mass insertion. This conclusion then naturally also applies to the subtracted scheme. 

\paragraph{Cogenesis}

In the main text, we decided to study the model of cogenesis within the unsubtracted scheme. In the unsubtracted scheme, we saw that the BEs for the cogenesis read 

\begin{align} 
\label{eq:full_cogenesis_app}
\frac{dY^{L}_{\Delta L}}{dt} & = - 
 2\text{Br}_{N \to LH}^2 \Gamma^{\Delta L = 2}_{(HL)(LH)} \bigg( Y_{\Delta L} + Y^{\rm eq}_{L}(t)\bigg)
 - \text{Br}_{N \to LH}\text{Br}_{N \to \phi \chi}\Gamma^{\Delta L = 2}_{(HL)(\phi \chi)} \bigg( Y_{\Delta L} +  Y_{\Delta \chi} + Y^{\rm eq}_{L+\chi}(t)\bigg)
 \notag
 \\
 &-   \text{Br}_{N \to LH}\text{Br}_{N \to \phi \chi}\Gamma^{\Delta L = 0}_{(HL)(\phi \chi)} \bigg( Y_{\Delta L} -  Y_{\Delta \chi} + Y^{\rm eq}_{L-\chi}(t)\bigg) \, ,
  \notag
 \\ 
 \frac{dY^{\chi}_{\Delta \chi}}{dt}& =
- 
 2 \text{Br}_{N \to \phi \chi}^2\Gamma_{(\phi \chi)(\phi \chi)}^{\Delta L = 2} \bigg( Y_{\Delta \chi} + Y^{\rm eq}_{\chi}(t)\bigg) 
 -  \text{Br}_{N \to LH}\text{Br}_{N \to \phi \chi} \Gamma_{(HL)(\phi \chi)}^{\Delta L = 2} \bigg( Y_{\Delta L}+ Y_{\Delta \chi} + Y^{\rm eq}_{L+\chi}(t)\bigg) 
 \notag
 \\
&  -   \text{Br}_{N \to LH}\text{Br}_{N \to \phi \chi}\Gamma^{\Delta L = 0}_{(HL)(\phi \chi)} \bigg( -Y_{\Delta L} +  Y_{\Delta \chi} - Y^{\rm eq}_{L-\chi}(t)\bigg) 
 \, ,
\end{align} 

where the rates are given by\cite{Falkowski:2017uya}
\bea 
\label{eq:rates_F}
\Gamma^{ij \to kl}(z) = \frac{ M_N z^2}{  256\pi^2} \int^{\infty}_0 dx \sqrt{x} K_1(z \sqrt{x}) \hat{\sigma}(x) + \frac{\gamma_D/4}{n^{\rm eq}_{L}}\,. 
\eea 
Notice that we  added the term $\gamma_D/4/n^{\rm eq}_{L}$ to restore the subtracted piece from the expressions given in \cite{Falkowski:2017uya}. The reduced cross-sections read
\begin{align}
\hat\sigma_{HL \to HL}(x) &= \frac{(y^2(1+R^2))^2}{2\pi}\bigg[\frac{x/2}{D(x)}+ \frac{x-\log (1+x)}{x}- \frac{2(x-1)}{D(x)}\frac{(x+1)\log(x+1) -x}{x}
\notag 
\\
&+ \frac{x/2}{(x+1)}+ \frac{\log (x+1)}{x+2}\bigg] \, ,
\notag 
\\
\hat\sigma^{\Delta L = 2}_{HL \to \phi \chi}(x) &= \frac{(y^2(1+R^2))^2}{2\pi}\bigg[\frac{x/2}{D(x)}+ \frac{x}{x+1} + \frac{x-\log (x+1)}{x} \bigg] \, ,
\notag 
\\
\hat\sigma^{\Delta L = 0}_{HL \to \phi \chi}(x) &= \frac{(y^2(1+R^2))^2}{2\pi}\bigg[\frac{x^2/2}{D(x)}+ \frac{(x+1)\log(x+1) -x}{x+1} + \frac{(x+2)\log(x+1) -2x}{x}\bigg] \, ,
\end{align} 
where we defined again
\bea 
D(x) \equiv (x-1)^2 + b^2, \qquad b \equiv \frac{\Gamma^{D}_{\rm tot}}{M_N}=  \frac{y^\dagger y(1+R^2)}{16\pi} \, .
\eea 
In the on-pole limit where $x\to 1$, the terms proportional to $x/D(x)$ and $x^2/D(x)$ will dominate in the reduced cross-section. As a consequence, the integral at the pole reads
\begin{align}
    \int_0^\infty dx\sqrt{x}K_1(z\sqrt{x})\frac{x}{2D(x)} \to \frac{ K_1(z)}{2} \int_0^\infty dx \frac{1}{(x-1)^2+b^2} \; ,
\end{align}
where we used that in the vicinity of the pole, only 
$D^{-1}(x)$ varies rapidly, allowing the other functions in the integrand to be approximated as constants with their arguments set to 
$x=1$. Using that $b \ll 1$, the integral becomes equal to
\begin{align}
 \frac{z^2}{4}K_1(z)\Gamma_D=\frac{1}{4}\frac{\gamma_D}{n_L^{\rm eq}} \: .
\end{align}

Turning to the BEs in Eq.\eqref{eq:full_cogenesis_app}, one can now combine all the on-pole pieces to obtain 
\begin{align} 
\frac{n^{\rm eq}_L}{\gamma_D} \frac{dY_{\Delta L}}{dt} & \stackrel{\text{on pole}}{=}  - 
 \text{Br}_{N \to LH}^2  \bigg( Y_{\Delta L} + Y^{\rm eq}_{L}(t)\bigg)-   \frac{1}{2}\text{Br}_{N \to LH}\text{Br}_{N \to \phi \chi}\bigg( Y_{\Delta L} +  Y_{\Delta \chi} + Y^{\rm eq}_{L+\chi}(t)\bigg) 
 \notag
 \\
 &
 -   \frac{1}{2}\text{Br}_{N \to LH}\text{Br}_{N \to \phi \chi} 
 \bigg( Y_{\Delta L} -  Y_{\Delta \chi} + Y^{\rm eq}_{L-\chi}(t)\bigg) \, ,
  \notag
 \\ 
 \frac{n^{\rm eq}_\chi}{\gamma_D} \frac{dY_{\Delta \chi}}{dt}& \stackrel{\text{on pole}}{=}- 
  \text{Br}_{N \to \phi \chi}^2
 \bigg( Y_{\Delta \chi} + Y^{\rm eq}_{\chi}(t)\bigg) -  \frac{1}{2}\text{Br}_{N \to LH}\text{Br}_{N \to \phi \chi} 
 \bigg( Y_{\Delta L}+ Y_{\Delta \chi} + Y^{\rm eq}_{L+\chi}(t)\bigg) 
 \notag
 \\
& 
 -  \frac{1}{2} \text{Br}_{N \to LH}\text{Br}_{N \to \phi \chi}
 \bigg( -Y_{\Delta L} +  Y_{\Delta \chi} - Y^{\rm eq}_{L-\chi}(t)\bigg) 
 \, ,
\end{align} 
leading, after a bit of simplification, to 
\begin{align} 
\label{eq:BEs_sub_bis}
\frac{n^{\rm eq}_L}{\gamma_D}\frac{dY_{\Delta L}}{dt}  \stackrel{\text{on pole}}{=} - 
\text{Br}_{N \to LH}  \bigg( Y_{\Delta L} + Y^{\rm eq}_{L}(t)\bigg)
 \qquad 
\frac{n^{\rm eq}_\chi}{\gamma_D} \frac{dY_{\Delta \chi}}{dt} \stackrel{\text{on pole}}{=}- 
  \text{Br}_{N \to \phi \chi}
 \bigg( Y_{\Delta \chi} + Y^{\rm eq}_{\chi}(t)\bigg)  \, ,
\end{align} 
where we used Eq.\eqref{eqn:Yeqsource} and the fact that 
\bea 
Y^{\rm eq}_{L-\chi}(t) + Y^{\rm eq}_{L+\chi}(t) \approx 2Y^{\rm eq}_{L}(t) \qquad -Y^{\rm eq}_{L-\chi}(t) + Y^{\rm eq}_{L+\chi}(t) \approx 2Y^{\rm eq}_{\chi}(t)  ,\qquad \text{Br}_{N \to HL} + \text{Br}_{N \to \phi \chi} = 1 \, . 
\eea

As expected, taking all the cross-sections into account, the on-shell rate in the BE becomes the one of the inverse decay. In this limit, the two sets of equations decouple and evolve separately. The consequence of this fact is intuitive and important: \emph{the passive sector (the sector having no source), can be only produced via the off-shell interactions, while the active sector is produced via on-shell and off-shell interactions}. We will observe a few examples of this fact in Appendix \ref{app:moreplots}.

Let us now turn back to the full system of equations in Eq.\eqref{eq:full_cogenesis_app}. 
This system of equations need to fulfill some consistency requirements which apply separately on each operator.

\begin{itemize}
    \item \textbf{$\Delta L =2$ Sharings}: The $\Delta L =2$ cross-terms fulfill the following constraints 
\bea 
 \frac{dY_{\Delta L-\Delta \chi}}{dt} = 0\quad \xrightarrow{\text{equilibrium}} Y_{\Delta L-\Delta \chi}  =  0 
 \\
 \frac{dY_{\Delta L+\Delta \chi}}{dt} =  -   2\text{Br}_{N \to LH}\text{Br}_{N \to \phi \chi}\Gamma^{\Delta L =2}_{(HL)(\phi \chi)} \bigg( Y_{\Delta L+\Delta \chi} + Y^{\rm eq}_{L+\chi}(t)\bigg)  \quad \xrightarrow{\text{equilibrium}} Y_{\Delta L+\Delta \chi} \neq 0 \, ,
\eea 
or in other words: only the combination $\chi+ L$ has a $\Delta L =2$ source term while the combination  $\chi- L$ is conserved. This can be understood from Eq.\eqref{Eq:Lag_for_cogenesis_IR}, where we observe that the vertex violates $\chi+ L$ but not $\chi- L$. 
\item \textbf{$\Delta L =0$ Sharings}: On the other hand, the $\Delta L =0$ cross term fulfils the following constraints 
\bea 
 \frac{dY_{\Delta L-\Delta \chi}}{dt} = -2\text{Br}_{N \to LH}\text{Br}_{N \to \phi \chi}\Gamma^{\Delta L =0}_{(HL)(\phi \chi)}  \bigg( Y_{\Delta L -\Delta \chi} + Y^{\rm eq}_{L-\chi}(t)\bigg)\quad \xrightarrow{\text{equilibrium}} Y_{\Delta L-\Delta \chi} \neq 0 
 \\
 \frac{dY_{\Delta L+\Delta \chi}}{dt} =  0 \quad \xrightarrow{\text{equilibrium}} Y_{\Delta L+\Delta \chi}  =  0 \, ,
\eea 
which is also what we expect from considering that the $\Delta L =0$ conserves $L+\chi$. 
\end{itemize}

\section{Cogenesis discussion}
\label{app:moreplots}

\begin{figure}[t!]
\centering
\includegraphics[width=0.8\linewidth]{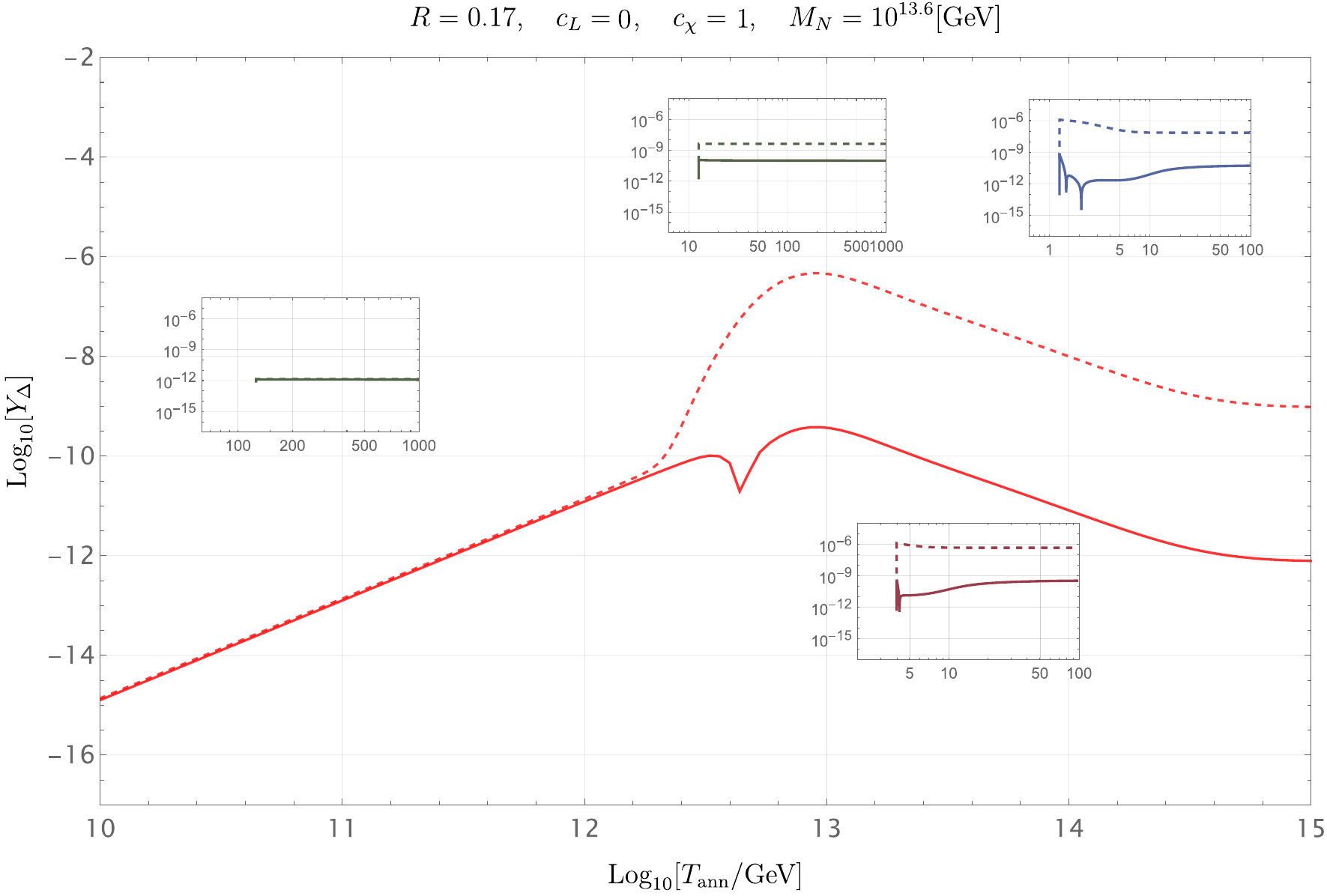}
\includegraphics[width=0.8\linewidth]{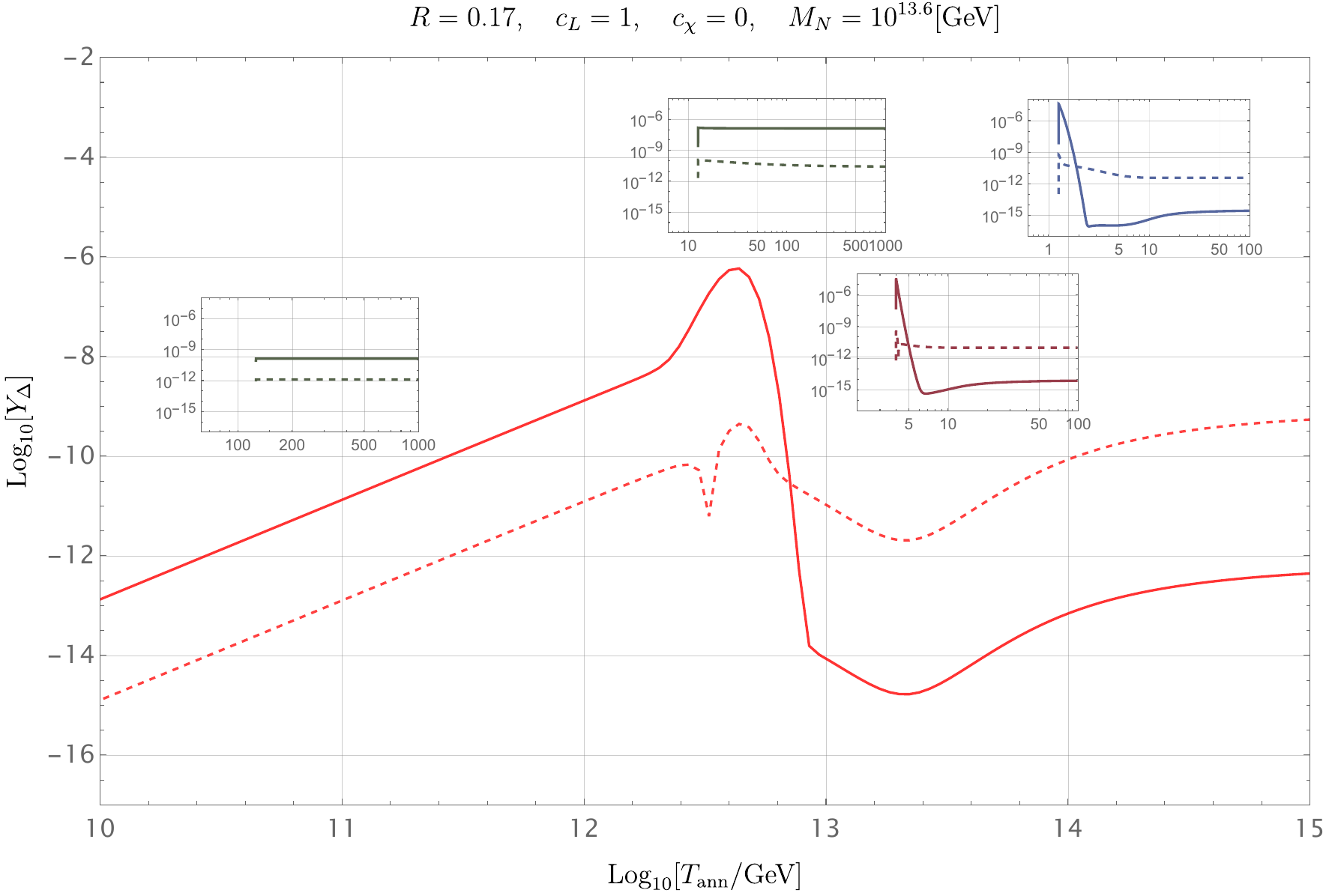}
    \caption{\textbf{Top panel}: Final asymmetry in $L$, $Y_{\Delta L}$ (solid) and $\chi$, $Y_{\Delta \chi}$ (Dashed) as a function of the annihilation temperature for $R = 0.17, c_L=0, c_\chi =1$ and $M_N = 10^{13.6}$ GeV.  We also display on the miniature plots the trajectory of $Y_{\Delta L}$ (solid line) and $Y_{\Delta \chi}$ (dashed line)  as a function of $z$, for different $T_{\rm ann}$. \textbf{Bottom panel}: Same as the top plot with $c_L = 1, c_\chi=0$. } 
    \label{fig:CogenesisMoneyPlot}
\end{figure}

In this Appendix, we elucidate the different characteristics appearing in Fig.\ref{fig:TrajCog_On} and focus on the case where the DS is weakly coupled with $R = 0.17$. We first present multiple zoomed-in plots depicting the $z$-trajectories of the asymmetries in the upper panel of Fig.\ref{fig:CogenesisMoneyPlot} for the DS being active, and in the lower panel of Fig.\ref{fig:CogenesisMoneyPlot} for the SM being active. 

For $R = 0.17$, the SM sector interacts with the heavy neutrino more strongly than the DS and one has the following hierarchy
    \bea 
    \text{Br}_{N \to \phi \chi}^2\Gamma_{(\phi \chi)(\phi \chi)} \ll \text{Br}_{N \to HL}\text{Br}_{N \to \phi \chi}\Gamma_{(HL)(\phi \chi)} \ll \text{Br}_{N \to HL}^2\Gamma_{(HL)(HL)} \qquad R \ll 1 \, ,
    \eea 
    as we can observe in  Fig.\ref{fig:TrajCog_On} (for $c_L = 1, c_\chi =0$, and $c_L = 0, c_\chi =1$).     
    
\begin{figure}[t!]
\centering
\includegraphics[width=0.5\linewidth]{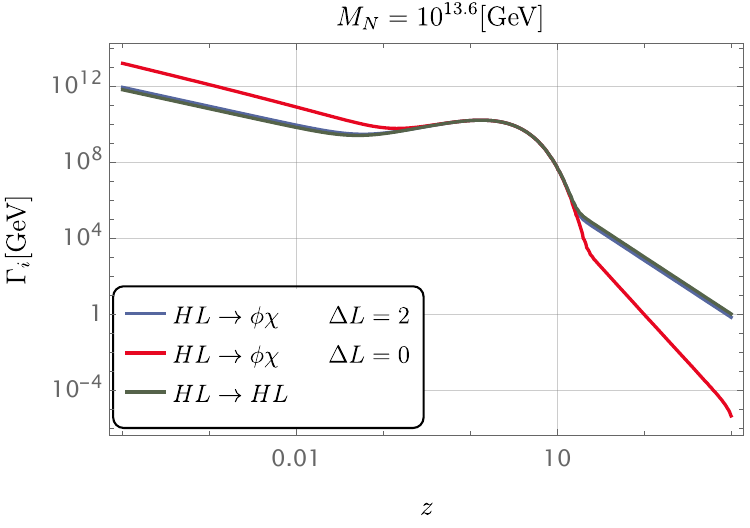}
    \caption{Rates for the three relevant operators as a function of $z$. We observe that the $\Delta L =0$ sharing is dominant for $z \lesssim 0.1$ and subdominant for $z \gtrsim 10$.}
    \label{fig:RRates}
\end{figure}

    We begin by explaining the case $c_L = 0, c_\chi =1$ (DS active) with the aid of Fig.\ref{fig:CogenesisMoneyPlot}.   We will study the values of the final asymmetries from high $T_{\rm ann}$ to low $T_{\rm ann}$:\\
    \newline
    \underline{$T_{\rm ann}>4\times 10^{14}$ GeV: the plateau} \\
     For $T_{\rm ann}>4\times 10^{14}$ GeV, numerically, we observe that the ratio of $Y_{\Delta L}/Y_{\Delta \chi}$ scales like 
    \bea 
    \frac{Y_{\Delta L}^{\rm final}}{Y_{\Delta \chi}^{\rm final} }\propto R^2 \, , \qquad Y^{\rm final}_{\Delta \chi} \approx  Y^{\rm initial}_{\Delta \chi} \propto R^2  \qquad \Rightarrow  \qquad Y^{\rm final}_{\Delta L} \propto R^4 \, . 
    \eea 
     We can understand this result in the following way. First of all,  $HL \leftrightarrow \phi \chi$ serves as a source for both the lepton- and the $\chi$-number and $N \to \phi \chi$ serves only as a source for the $\chi$-number. Since $Y_{\Delta \chi}$ is produced via the sharing  with $\Delta L = 0$ (see Fig.\ref{fig:RRates} to compare the rates) which scales like $R^2$, the DS asymmetry also scales like $R^2$. One can also understand the fact that the initial asymmetry in the SM is lower than in the DS because the passive sector can be only produced via the \emph{off-shell} interactions. 
     
     Secondly, in the limit $R \ll 1$ and long time after the production, the BEs simplify to
    \begin{align} 
\label{eq:full_cogenesis_b}
\frac{dY_{\Delta L}}{dt} & \approx - 
 2  \Gamma^{\Delta L = 2}_{(HL)(LH)}   Y_{\Delta L} 
-  R^2\Gamma^{\Delta L = 2}_{(HL)(\phi \chi)}   Y_{\Delta \chi} 
 +   R^2\Gamma^{\Delta L = 0}_{(HL)(\phi \chi)}  Y_{\Delta \chi}   \, ,
  \notag
 \\ 
 \frac{dY_{\Delta \chi}}{dt}& = 0 \qquad \Rightarrow \qquad Y_{\Delta \chi} = Y^{\rm initial}_{\Delta \chi} \, .
\end{align}

    When $N \to HL$ becomes efficient, the abundance of $L$ is largely depleted by washout, while the abundance of $\chi$ mostly remains, since $N \to \phi \chi$ is always decoupled in the regime $R\approx 0.17$. After the decoupling of the decay and inverse decay, the weak coupling between the SM and the DS at a temperature order $T \sim M_N/10$  recreates a very small asymmetry in the SM sector. This ``saved'' asymmetry can be  estimated by solving algebraically the LHS of Eq.\eqref{eq:full_cogenesis_b} and evaluate it at decoupling of the inverse decay,  
\bea 
\label{eq:simplified}
\frac{dY_{\Delta L}}{dt} = 0 \bigg|_{\rm dec} \quad  \Rightarrow
\quad Y^{\rm final}_{\Delta L} \approx - R^2\bigg[\frac{ \Gamma^{\Delta L = 2}_{(HL)(\phi \chi)}- \Gamma^{\Delta L = 0}_{(HL)(\phi \chi)}}{2  \Gamma^{\Delta L = 2}_{(HL)(LH)} }    \bigg]\bigg|_{\text{dec}} Y^{\rm initial}_{\Delta \chi} \propto -0.1R^2 Y^{\rm initial}_{\Delta \chi}    \,, 
\eea 
where the further suppression comes from a partial cancellation coming from $\Gamma^{\Delta L = 2}_{(HL)(\phi \chi)}- \Gamma^{\Delta L = 0}_{(HL)(\phi \chi)}$. Notice that the sign of the two asymmetries are \emph{opposed}. This is what we call the "\emph{rescuing mechanism}", which hides some asymmetry in the weakly coupled DS and then send it back to the SM, after the wash-out decoupled. 
Finally, both the $\Delta L$ and $\Delta \chi$ display a plateau in $T_{\rm ann}$. This is because the final asymmetry is dictated by the initial asymmetry in the DS, which is 
\bea 
Y^{\rm final}_{\Delta \chi} \propto \frac{\Gamma_{\rm prod}}{T} \approx 
 \frac{\Gamma^{\Delta L = 0}_{(HL) (\phi \chi)}}{T}  \propto T_{\rm ann}^0 \; , 
\eea 
since the asymmetry is produced via $\Delta L =0$ sharing operators for $z< 0.1$, as we can observe in Fig.\ref{fig:RRates}. 

\underline{$T_{\rm ann} \sim 4\times 10^{14}$ GeV : the ankle}

 We also observe an ankle at $T_{\rm ann } \sim 4 \times 10^{14}$ GeV. 
 The trajectory can be understood in the same way as for the plateau, the only crucial difference is that now the production is dominated by on-shell interactions (decay and inverse decay) instead of off-shell scatterings. The peak which follows the plateau at lower $T_{\rm ann}$ comes thus from on-shell scatterings (decay and inverse decays). 
 \\
\newline
\underline{$T_{\rm ann} \sim 5\times 10^{12}$ GeV : the dip}

The dip occurring around $T_{\rm ann} \sim 5\times 10^{12}$ GeV comes from the change in the relative sign of the asymmetries $Y_{L}, Y_{\Delta \chi}$. For $T_{\rm ann} > 5\times 10^{12}$ GeV, the production is dictated by the sharing, and as we can observe in Fig.\ref{fig:RRates}, the $\Delta L =0$ sharing becomes subdominant for $z \gtrsim 10$, around $T_{\rm ann} \sim 5\times 10^{12}$  GeV. At $T_{\rm ann} \sim 5\times 10^{12}$ GeV, however, the production is dominated by the $\Delta L= 2$ operator and the sign of the asymmetry is identical. Exactly at the position of this dip, the production in the $L$-sector is suppressed by the cancellation between the two aforementioned operators. 
\\
\newline
\underline{Low $T_{\rm ann}$ behaviour} \\
At $T_{\rm ann} \ll M_N/10$, we observe that $Y_{\Delta L}\to Y_{\Delta \chi}$. This can be again understood from Fig.\ref{fig:RRates} where we observe that in this regime the dominant operator is the sharing $HL \leftrightarrow \phi \chi$ with $\Delta L=2$, which naturally produces equal asymmetry in the SM and the DS \emph{with the same sign}. \\
\newline
The case of $c_L=1, c_\chi=0$ (SM active) displayed in the bottom panel of Fig.\ref{fig:CogenesisMoneyPlot} can be understood following a very similar analysis.

\bibliographystyle{JHEP}
{\footnotesize
\bibliography{biblio}}
\end{document}